\documentclass[journal,comsoc]{IEEEtran}
\usepackage{amsmath,amsfonts}
\usepackage[caption=false,font=normalsize,labelfont=sf,textfont=sf]{subfig}
\usepackage{textcomp}
\usepackage{stfloats}
\usepackage{url}
\usepackage{verbatim}
\usepackage{graphicx}
\def\BibTeX{{\rm B\kern-.05em{\sc i\kern-.025em b}\kern-.08em
    T\kern-.1667em\lower.7ex\hbox{E}\kern-.125emX}}
\usepackage{balance}

\usepackage{tikz}
\usepackage{pgfplots}
\usepackage{pgfplotstable}
\pgfplotsset{compat=newest}
\usepackage[noadjust]{cite}

\usepackage[detect-all]{siunitx}
\usepackage{booktabs}
\usepackage{circuitikz}
\usepackage{mathtools}
\usepackage{hyperref}
\hypersetup{unicode=true,pdfborder = {2 0 0.1}}
\usepackage{relsize}

\usetikzlibrary{arrows,shapes.misc,chains,scopes,shadows,arrows.meta}
\usetikzlibrary{calc}
\usetikzlibrary{fit}
\usetikzlibrary{intersections}
\usetikzlibrary{shapes}
\usetikzlibrary{scopes}
\usetikzlibrary{backgrounds}
\usetikzlibrary{positioning,decorations.pathreplacing,calc}
\usetikzlibrary{decorations.text}

\DeclareSIUnit\bpcu{bpcu}
\DeclareSIUnit\bpQs{bpQs}

\newcommand{\bm}[1]{\mathbf{#1}}

\newcommand{\sicindex}{t} 
\newcommand{\siclength}{N}

\newcommand{\lay}{\ell}
\newcommand{\maxlay}{L}

\newcommand{\osidx}{u}

\newcommand{\ok}[1]{\textcolor{black}{#1}}

\pgfmathsetmacro{\blockwidth}{30}
\pgfmathsetmacro{\blockheight}{17}

\tikzstyle{comblock} = [drop shadow={shadow xshift=0.1em,shadow yshift=-0.1em},thick,black,fill=white,draw=black, rectangle,minimum height=\blockheight pt,minimum width=\blockwidth pt]

\tikzstyle{sumstyle} = [thick,draw, circle,inner sep=-1pt,outer sep=0pt,font=\normalsize,fill=white,drop shadow={shadow xshift=0.1em,shadow yshift=-0.1em},]

\tikzstyle{comtriag} = [regular polygon, regular polygon sides=3,
              draw=black, black,  text width=0.90em,
              inner sep=0.9mm, outer sep=0mm,
              shape border rotate=-90]

\tikzstyle{input} =  [coordinate]
\tikzstyle{output} = [coordinate]

\DeclareMathOperator*{\T}{T}

\DeclareMathOperator{\nonl}{\xi}
\DeclarePairedDelimiter\parens{\lparen}{\rparen}
\newcommand{\nonlp}[1]{\nonl\parens*{#1}}

\DeclareMathOperator*{\argmin}{argmin}

\DeclareMathOperator*{\sinc}{sinc}
\DeclareMathOperator*{\vstack}{cat}

\renewcommand{\tt}{\text{t}}

\newcommand{\dimC}[1]{\ensuremath{\in \mathbb{C}^{#1}}}
\newcommand{\dimR}[1]{\ensuremath{\in \mathbb{R}^{#1}}}

\usetikzlibrary{intersections}
\usetikzlibrary{calc}
\usetikzlibrary{shapes}
\usetikzlibrary{scopes}
\usetikzlibrary{backgrounds}
\usetikzlibrary{positioning,decorations.pathreplacing,spy}

\definecolor{TUMBlack}{cmyk}{0,0,0,1}     %
\definecolor{TUMWhite}{cmyk}{0,0,0,0}     %

\definecolor{TUMBlue} {cmyk}{1,0.43,0,0}  %

\definecolor{TUMDarkBlue}   {cmyk}{1,0.57,0.12,0.7}      %
\definecolor{TUMDarkerBlue} {cmyk}{1,0.54,0.04,0.19}     %
\definecolor{TUMMediumBlue} {cmyk}{0.9,0.48,0,0}         %
\definecolor{TUMLighterBlue}{cmyk}{0.65,0.19,0.01,0.04}  %
\definecolor{TUMLightBlue}  {cmyk}{0.42,0.09,0,0}        %

\definecolor{TUMDarkGray}  {cmyk}{0,0,0,0.8}  %
\definecolor{TUMMediumGray}{cmyk}{0,0,0,0.5}  %
\definecolor{TUMLightGray} {cmyk}{0,0,0,0.2}  %

\definecolor{TUMGreen} {cmyk}{0.35,0,1,0.2}         %
\definecolor{TUMOrange}{cmyk}{0,0.65,0.95,0}        %
\definecolor{TUMIvory} {cmyk}{0.03,0.04,0.14,0.08}  %

\definecolor{TUMBeamerYellow}    {rgb}{1.00,0.71,0.00}  %
\definecolor{TUMBeamerOrange}    {rgb}{1.00,0.50,0.00}  %
\definecolor{TUMBeamerRed}       {rgb}{0.90,0.20,0.09}  %
\definecolor{TUMBeamerDarkRed}   {rgb}{0.79,0.13,0.25}  %
\definecolor{TUMBeamerBlue}      {rgb}{0.00,0.60,1.00}  %
\definecolor{TUMBeamerLightBlue} {rgb}{0.25,0.75,1.00}  %
\definecolor{TUMBeamerGreen}     {rgb}{0.57,0.67,0.42}  %
\definecolor{TUMBeamerLightGreen}{rgb}{0.71,0.79,0.51}  %

\definecolor{I8LogoRed}         {rgb}{0.51,0,0.08}
\definecolor{I8LightBlue}       {rgb}{0.725,0.812,0.882}
\definecolor{I8DarkBlue}        {rgb}{0.490,0.573,0.667}
\definecolor{I8Blue}            {rgb}{0.576,0.624,0.718}

\definecolor{mycolor1bright}{HTML}{E9D8A6}%
\definecolor{mycolor7bright}{HTML}{EE9B00}%
\definecolor{mycolor6bright}{HTML}{AFCBFF}%
\definecolor{mycolor5dark}{HTML}{166337}%
\definecolor{mycolor6}{HTML}{FF0000}%

\definecolor{mycolor6dark}{HTML}{961717}%
\definecolor{mycolorlightorange}{HTML}{7A7A7A}%
\definecolor{mycolorcube}{HTML}{C2C2C2}%
\definecolor{mycolor5bright}{HTML}{FFFFFF}%

\DeclareFontEncoding{LS1}{}{}
\DeclareFontSubstitution{LS1}{stix}{m}{n}
\DeclareSymbolFont{stixletters}{LS1}{stix}{m}{it}
\DeclareMathAccent{\cev}{\mathord}{stixletters}{"91}
\DeclareMathAccent{\vec}{\mathord}{stixletters}{"92}

\newcommand{\fw}[1]{\vec{#1}} 
\newcommand{\bw}[1]{\cev{#1}}

\newcommand{\Wfw}[2]{\fw{W}_{#1}^{#2}} %
\newcommand{\Cfw}[2]{\fw{C}_{#1}^{#2}} %
\newcommand{\Cbw}[2]{\bw{C}_{#1}^{#2}} %
\newcommand{\hfw}[2]{\fw{\mathbf{h}}_{#1}^{#2}} %
\newcommand{\hbw}[2]{\bw{\mathbf{h}}_{#1}^{#2}} %

\newcommand{\mufwsub}[2]{\fw{\mu}_{#1}{(#2)}} %
\newcommand{\mubwsub}[2]{\bw{\mu}_{#1}{(#2)}} %

\newcommand{\St}{H} %
\newcommand{\st}{h} %

\newcommand{\fI}{\psi} %
\newcommand{\fII}{\psi'} %

\newlength\nntablevspace
\makeatletter
\DeclareRobustCommand{\rvdots}{%
  \vbox{
    \baselineskip4\p@\lineskiplimit\z@
    \kern-\p@
    \hbox{.}\hbox{.}\hbox{.}
  }}
\makeatother

\begin{document}

\title{
Neural Network-Based Successive Interference Cancellation for Non-Linear Bandlimited Channels
\thanks{Date of current version \today. %
\emph{(Corresponding author: Daniel Plabst.)}}
\thanks{Daniel Plabst, Tobias Prinz, Francesca Diedolo, Thomas Wiegart, Norbert Hanik and Gerhard Kramer are with the Institute for Communications Engineering, School of Computation, Information, and Technology, Technical University of Munich, 80333 Munich, Germany (e-mail: daniel.plabst@tum.de; tobias.prinz@tum.de; francesca.diedolo@tum.de; thomas.wiegart@tum.de, norbert.hanik@tum.de; gerhard.kramer@tum.de). Georg Böcherer is with the Munich Research Center, Huawei Technologies Düsseldorf GmbH, 80992 Munich, Germany (e-mail: georg.bocherer@huawei.com).}}

\author{\IEEEauthorblockN{
Daniel Plabst, 
Tobias Prinz, 
Francesca Diedolo, 
Thomas Wiegart, 
Georg Böcherer,\\ 
Norbert Hanik and Gerhard Kramer
}

}

\maketitle
\thispagestyle{plain} %
\pagestyle{plain}

\begin{abstract}
Reliable communication over bandlimited and nonlinear channels usually requires equalization to simplify receiver processing. Equalizers that perform joint detection and decoding (JDD) achieve the highest information rates but are often too complex to implement. 
To address this challenge, model-based neural network (NN) equalizers that perform successive interference cancellation (SIC) are shown to approach JDD information rates for bandlimited channels with a memoryless nonlinearity and additive white Gaussian noise. The NNs are chosen to have a periodically time-varying and recurrent structure that imitates the forward-backward algorithm (FBA) in every SIC stage. 
Simulations for short-haul fiber-optic links with square-law detection show that NN-SIC nearly doubles current spectral efficiencies, and bipolar or complex-valued modulations achieve energy gains of up to $\SI{3}{dB}$ compared to state-of-the-art intensity modulation. Moreover, NN-SIC is considerably less complex than equalizers that perform JDD, mismatched FBA processing, and Gibbs sampling.
\end{abstract}


\section{Introduction}
\label{sec:introduction}

\IEEEPARstart{C}{ommunication} links with low transmit power are often modeled as linear intersymbol interference (ISI) channels with additive white Gaussian noise (AWGN). Orthogonal frequency division multiplexing (OFDM) converts such channels into parallel memoryless channels~\cite[Chap.~3.4.4]{tse2005fundamentals}, providing a practical method to approach capacity~\cite[Eq.~5.39]{tse2005fundamentals}. However, device constraints or high transmit power may introduce nonlinearities at the transmitter or receiver, e.g., through a power amplifier (PA)~\cite{eriksson2019nonlinear,mollen2016massive}, or a square-law detector (SLD)~\cite{chagnon_optical_comms_short_reach_2019}. Nonlinearities degrade OFDM performance~\cite{merchan1998OFDM} in general, and the receiver may need to apply joint detection and decoding (JDD) to approach capacity. Unfortunately, JDD is often too complex to implement, especially for higher-order modulation~\cite[Sec.~IV]{muller_capacity_separate2004}.

A pragmatic method to reduce complexity is separate detection and decoding (SDD) for each channel input symbol; see~\cite{sheik_achievable_2017,liga_information_2017}. SDD rates may be significantly lower than JDD rates~\cite{muller_capacity_separate2004,prinz2023successive}. Two methods that use SDD to approach JDD performance are turbo detection and decoding (TDD)~\cite{douillard1995iterative,alexander-ETT98,wang1999iterative} and multi-level coding (MLC) with successive interference cancellation (SIC)~\cite{prinz2023successive,wachsmann_multilevel_1999,PfisterAIRFiniteStateChan2001,soriaga_determining_2007}.
TDD requires a dedicated code design to approach capacity while MLC-SIC permits using off-the-shelf codes; see~\cite{PfisterAIRFiniteStateChan2001,soriaga_determining_2007,ten2004design}.

\subsection{Equalizers to Compute APPs}
We focus on MLC-SIC with detectors/equalizers that output \emph{a posteriori} probabilities (APPs) or approximations of APPs. Consider the following equalizers.
\begin{itemize}
    \item A forward-backward algorithm (FBA)~\cite{bcjr_1974} that computes symbol-wise APPs. A related approach is hard- or soft-output Viterbi equalization~\cite{hagenauer1989SOVA,kschischang2001factor}.
    \item Gibbs sampling (GS)~\cite[Ch.~29]{mackay2003information} as a Markov-chain, Monte-Carlo method to approximate APPs.
    \item Linear equalizers~\cite{muller_capacity_separate2004} followed by a demapper (a memoryless APP calculator).
    \item Nonlinear equalizers with a demapper, such as decision feedback equalizers
    (DFEs)~\cite{cioffi1995mmsedfe}, Volterra filters~\cite{taiho1985secondvolterra} and neural networks (NNs).
\end{itemize}
We compare the methods and discuss simplifications.
\begin{itemize}
    \item The FBA outputs exact APPs but is too complex for large channel memory or modulation alphabets. One can trade off complexity and performance via mismatched models with small memory~\cite{{plabst2022achievable}} or channel shortening filters~\cite{rusek2012optimal}.
    \item GS simplifies FBA calculations but may converge slowly at high signal-to-noise ratio (SNR), requiring many parallel samplers or iterations~\cite{senst2011rao,prinz2023successive}.
    \item Linear equalizers cannot cancel nonlinearities and lose significant rate at high SNR~\cite[Fig.~7-9]{muller_capacity_separate2004}. For example, minimum mean square error (MMSE) filtering often assumes Gaussian inputs~\cite{cioffi1995mmsedfe,yin2021soft,shoumin2004near} and is suboptimal for discrete inputs.
    \item Nonlinear equalizers such as DFEs exhibit error propagation~\cite[Fig.~10]{muller_capacity_separate2004} and require decoding in the feedback path. Even with perfect feedback, MMSE-DFE~\cite{cioffi1995mmsedfe} approaches the JDD rates only at low SNRs~\cite[Fig.~9]{muller_capacity_separate2004}.
    Volterra filters are too complex~\cite{taiho1985secondvolterra} in general.
\end{itemize}

The main limitation of the above methods is that, for high rates, either the complexity is high or residual interference reduces rates significantly compared to JDD. We focus on model-based NN equalizers and show they can efficiently achieve high rates for large-memory models. 

\subsection{Review of NN Receivers}
We review the literature on NN receivers. For decoding, recurrent NNs (RNNs) were used to approximate the Viterbi algorithm~\cite{xiao1996viterbiNN,shlezinger2020viterbi} and the FBA~\cite{kim2020physical,jiang2019learn,haykin2000adaptive}. NN autoencoders were used for both turbo encoding and decoding in~\cite{gruber2017ondeep,nachmani2016learning,bruck1989neural,jiang2019deepturbo,jihang2019turbo,clausius2021serial}.
For equalization, we categorize papers based on whether they use JDD, SDD, TDD, or SIC.

\subsubsection{JDD}
The paper~\cite{ye2017initial} performs JDD for ISI channels with two memory taps and a memoryless amplifier nonlinearity. A large receiver NN demaps and decodes polar-coded BPSK, but the approach does not scale to longer codes, and JDD cannot account for interblock interference. 
The works~\cite{zhang2022design,han2022sparse} consider JDD for ISI channels with three memory taps and sparse code multiple access, respectively. The paper~\cite{doerner2021bitwise} proposes a JDD autoencoder for memoryless MIMO channels with few antennas. The approach is limited to small MIMO and modulation alphabets; extending to frequency-selective or massive MIMO is too complex.

\subsubsection{SDD}
The paper~\cite{xu2018joint} treats ISI with separate NNs for symbol-wise APP detection and decoding.
The paper~\cite{liao2019deep} estimates symbol-wise APPs using NNs for millimeter wave wireless channels. 
The paper~\cite{honkala2021deeprx} studies OFDM with a convolutional NN for channel estimation and symbol-wise APP detection.
OFDM is also considered in~\cite{fischer2022adaptive}, but with feedback between the decoder and NN detector to adapt the NN to the channel state information (CSI) during a retraining phase.
The paper~\cite{samuel2019learning} uses NNs for SDD for MIMO channels, and~\cite{uhlemann2020deep,haeger2020model,buetler2021modelextended} applies NNs for SDD and long-haul fiber links.
The authors of~\cite{karanov2018end,karanov2019recurrent} study short-reach fiber-optic systems with intensity modulation, SLD, and NN autoencoders.

The papers~\cite{shlezinger2020viterbi,shlezinger2020deep} approximate sequence-MAP detection by NN-Viterbi detection and NN-based soft interference cancellation~\cite{choi2000iterativesoft} (IC), respectively. 
Sequence-MAP detection is suboptimal with SDD, as the detector passes hard decisions to the decoder. For time-varying channels, both approaches may be improved by ``online'' retraining via a feedback path between the detector and decoder~\cite{fischer2022adaptive}. 
The paper~\cite{baumgartner2022softSIC} uses~\cite{shlezinger2020deep} for single-carrier frequency-domain equalization, where NNs are trained across different Rayleigh fading channels. 

\subsubsection{TDD}
The paper~\cite{tsai2020neural} applies NN-TDD for ISI channels with CSI uncertainty; see~\cite{jiang2019deepturbo} that also learns the component decoders. 
The authors of~\cite{xu2022tdd} propose TDD for OFDM and replace the modulator and demapper with a NN. For frequency-selective Rayleigh fading channels, autoencoding outperforms conventional OFDM due to better frequency diversity.
The paper~\cite{doerner2023joint} studies TDD autoencoders for short packet communication and synchronization.

\subsubsection{SIC}
NNs for soft-SIC were applied to code division multiple access~\cite{minghuang2000successive} by subtracting interference estimates between stages. NN post-processing improves the bit-error rate. 
NN-based SIC receivers were also applied to non-orthogonal multiple access (NOMA).
The paper~\cite{lin2019deep} studies the NOMA downlink with perfect CSI and replaces a soft-SIC receiver with a single NN. 
The paper~\cite{kang2020deep} optimizes an NN-based precoder and SIC detector for NOMA with imperfect CSI. In each SIC stage, an NN estimates the data and interference from other users to subtract interference; see ~\cite[Fig.~2]{kang2020deep}. 
Instead, the papers~\cite{aref2020deep,kim2023enhancedsoft} process the estimated interference and received channel output symbols jointly.
The paper~\cite{vanluong2022deep} performs NN-SIC for broadcast channels with imperfect CSI; each user treats the other users' signals as noise. The APPs are computed using soft information from prior SIC stages, similar to soft-decision feedback equalization~\cite{yang2013soft}.

The above works apply soft IC with no channel decoding between the stages, i.e., significant error propagation occurs if the interference is not entirely suppressed. The information rate is that of SDD.

\subsection{Contributions and Organization}
We propose a periodically time-varying recurrent NN equalizer that is inspired by the FBA and combine it with MLC-SIC and one NN-equalizer per stage\footnote{A conference version of this manuscript is available at~\cite{plabst2024neuralISIT}.}.
In contrast to the NN literature, the NN-SIC receiver decodes between stages and approaches JDD performance as the number of SIC stages increases. As a showcase, we study oversampled short-reach fiber links with long ISI and an SLD and compare the results to using mismatched FBA, SDD, and Gibbs sampling receivers~\cite{prinz2023successive}. Remarkably, the NN-SIC receiver outperforms and is substantially simpler than existing mismatched receivers that approximate JDD. For example, by applying the NN-SIC equalizer to bipolar and complex-valued modulation, one gains up to \SI{3}{dB} in energy efficiency over classic intensity modulation, making the technique a candidate for future short-range links with direct detection.

This paper is organized as follows. Sec.~\ref{sec:system-model} introduces the system model and Sec.~\ref{sec:detection_and_decoding} reviews SDD and SIC. Sec.~\ref{sec:fba_with_ic} and~\ref{sec:app_nn} discuss FBA-equalization and NN-equalization with IC, respectively. Sec.~\ref{sec:numerical} computes information rates for SDD and SIC receivers for short-reach fiber-optic links with a SLD and compares with~\cite{prinz2023successive,plabst2022achievable}. We find that NN-SIC substantially increases current spectral efficiencies for short-haul fiber links with SLD~\cite{tasbihi2021direct,prinz2023successive}. Sec.~\ref{sec:conclusion} concludes the paper. 

\subsection{Notation}
\label{subsec:notation}
Column vectors and matrices are written using bold letters, e.g., $\bm{a}$. The transpose of $\bm{a}$ is $\bm{a}^{\T}$ and $\vstack{(\mathbf{a},\mathbf{b})} = [ \mathbf{a}^{\T}, \mathbf{b}^{\T}]^{\T}$ stacks $\mathbf{a}$ and $\mathbf{b}$. 
The real and imaginary parts of the complex number $z$ are $\mathfrak{R}\{z\}$ and $\mathfrak{I}\{z\}$, respectively. The phase of $z$ is $\angle z$. We denote strings of scalars and vectors by $x_\kappa^n=(x_\kappa,\ldots,x_n)$ and $\mathbf{X}_\kappa^n= (\mathbf{X}_\kappa,\ldots,\mathbf{X}_n)$, respectively, and omit the subscript if $\kappa = 1$. For positive integers $a,b,c,d$, we use the notation $(x_{\kappa,u})_{a\leq \kappa \leq b, c\leq u \leq d  }$ to denote the string $(x_{\kappa,u} \mid a\leq \kappa \leq b,\, c\leq u \leq d)$.

Random variables (RVs) are written in uppercase letters, and their realizations in lowercase. The probability mass function (PMF) and density of a vector of discrete and continuous RVs $\bm{X}$ is written as $P_{\bm{X}}$ and $p_{\bm{X}}$, respectively. We often discard subscripts on PMFs or densities if the arguments are uppercase or lowercase versions of their RVs. 
The conditional PMF of a discrete RV $X$ given $Y$ is $P_{X\lvert Y}$. Similarly, we use $P_{X\lvert Y}( \cdot |y)$ for the PMF of $X$ given $Y=y$.

The sinc function is $\sinc(t) = \sin(\pi t)/(\pi t)$. 
The  convolution of $g(t)$ and $h(t)$ is $g(t)*h(t)$ and the energy of $a(t)$ is $\|a(t)\|^2 = \int_{-\infty}^{\infty} \big|\, a(t) \big|^2 \mathrm{d}t $.
Entropy, conditional entropy, and mutual information are defined as in~\cite[Chap.~2]{cover1991elementsofIT}, and we measure the quantities in bits.

\section{System Model}
\label{sec:system-model}
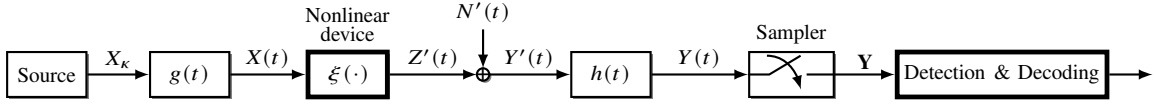
\begin{figure*}[t!]
    \centering
    \usetikzlibrary{decorations.markings}
\tikzset{node distance=2.3cm}

\pgfdeclarelayer{background}
\pgfdeclarelayer{foreground}
\pgfsetlayers{background,main,foreground}
\tikzset{boxlines/.style = {draw=black!20!white,}}
\tikzset{boxlinesred/.style = {densely dashed,draw=red!50!white,thick}}

\pgfmathsetmacro{\samplerwidth}{32}

\tikzset{midnodes/.style = {midway,above,text width=1.5cm,align=center,yshift=-0.1em}}
\tikzset{midnodesRP/.style = {midway,above,text width=1.5cm,align=center,yshift=-1.4em}}

\begin{tikzpicture}[]
    \footnotesize
    \node[comblock] (dms) {Source};
    \node[comblock,node distance=1.9cm,right of=dms] (txfilter) {$g(t)$};
    \node [comblock,right of=txfilter,node distance=2.1cm,line width=1.8pt] (sld) {$\nonlp{\cdot}$};
    \node [sumstyle,right of=sld,node distance=1.8cm] (sumnode) {$+$};
    \node [input, name=noise,above of=sumnode,node distance=0.6cm] {Input};
    \node [comblock,right of=sumnode,node distance=1.7cm] (rxfilter) {$h(t)$};
    \node [comblock,right of=rxfilter,minimum width=\samplerwidth pt,node distance=2.4cm] (sampler) {};
    \node [comblock,right of=sampler,node distance=2.8cm,line width=1.8pt] (dsp) {Detection \& Decoding};
    \node [input, name=output, right of=dsp,node distance=2cm] {Output};
    \draw[thick] (sampler.west) -- ++(\samplerwidth/4 pt,0) --++(\samplerwidth/2.7 pt,\samplerwidth/4.5 pt );
    \draw[thick] (sampler.east) -- ++(-\samplerwidth/3pt,0);

    \draw ($(sampler.west) + (\samplerwidth/4.5 pt,0.25)$)edge[out=0,in=100,-latex,thick] ($(sampler.east) + (-\samplerwidth/2.5 pt,-0.2)$);

    \draw[-latex,thick] (dms) -- node[midnodes](){$X_\kappa$} (txfilter);
    \draw[-latex,thick] (txfilter) -- node[midnodes](s_ti){$X(t)$  } (sld);

    \draw[-latex,thick] (sld) -- node[midnodes](){$Z^\prime(t)$\\[0.2em] } (sumnode);
    \draw[-latex,thick] (sumnode) -- node[midnodes](){$Y'(t)$\\[0.2em] } (rxfilter);
    \draw[-latex,thick] (rxfilter) -- node[midnodes](){$Y(t)$\\[0.2em] } (sampler);
    \draw[-latex,thick] (sampler) --  node[midnodes](){$\mathbf{Y}$\\ [0.2em] } (dsp)  ;
    \draw[-latex,thick] (dsp) --  node[midnodes,xshift=0cm,align=center](){ }(output); %

    \draw[-latex,thick] (noise) -- (sumnode);
    \node[above] () at  (noise) {$N'(t)$};

    \node[above,yshift=1.1em] () at (sampler) {Sampler};
    \node[below,yshift=-1.3em, text width=1.9cm, align=center] () at (sampler) {}; %
    
    \node[below,yshift=-1.3em] () at (rxfilter) {};

    \node[above,yshift=+1.2em,text width=1.6cm,align=center,] () at (sld) {Nonlinear\\[-0.2em] device};

\end{tikzpicture}
    \caption{Bandlimited channel with a memoryless nonlinear device and additive noise.}
    \label{fig:continuous_detailed_system_model}
\end{figure*}

We study bandlimited channels with a memoryless nonlinear device $\nonlp{\cdot}$ and additive noise, as shown in Fig.~\ref{fig:continuous_detailed_system_model}. For example, this model applies to wireless communications with nonlinearities at the transmitter due to PAs, mixers, and digital-to-analog converters (DACs)~\cite{eriksson2019nonlinear,mollen2016massive}, or at the receiver due to low-noise amplifiers (LNAs), mixers, and analog-to-digital converters (ADCs). The model also describes fiber-optic communications with nonlinearities at the transmitter due to a driver amplifier, DAC, Mach-Zehnder modulator, and optical amplifiers, or at the receiver with a square-law detector (SLD), a single photodiode~\cite[Fig.~2]{plabst2022achievable}, LNA, and ADC. 

Fig.~\ref{fig:continuous_detailed_system_model} assumes additive noise $N'(t)$ after the nonlinearity. For instance, $N'(t)$ might model the amplified thermal noise of a radio-frequency amplifier~\cite{friis1944noise,rapp1991effects,saleh1981frequency,ghorbani1991effect}, or the amplified spontaneous emission noise of an erbium-doped fiber amplifier,
or the lumped noise of photo-detection~\cite[Sec.~II]{wiener_filter_plabst2020}, the LNA, and the ADC.

\subsection{Continuous Time Model}
\label{sec:time-continuous-model}
The source in Fig.~\ref{fig:continuous_detailed_system_model} generates uniformly, independently and identically distributed (u.i.i.d.) symbols $(X_\kappa)_{\kappa \in \mathbb{Z}} = \left( \ldots, X_1, X_2, X_3,\ldots \right)$ from the symbol alphabet $\mathcal{A} = \{a_1,\ldots,a_M \}$ where $M = 2^m$. After filtering with $g(t)$, the baseband waveform is
\begin{align}
    X(t)  = \sum\limits_{\kappa} X_\kappa \cdot g(t-\kappa T_\text{s})
\end{align}
where $B=1/T_\text{s}$ is the symbol rate and $g(t)$ collects all linear effects, including the bandwidth limitations of the pulse, nonlinear device, and DAC. 
The nonlinear device puts out $Z'(t) = \nonlp{X(t)}$ to which noise $N'(t)$ is added. The noise is modeled as a complex white Gaussian random process with two-sided power spectral density (PSD) $N_0/2$ Watts per Hertz per dimension.

The signal and noise are filtered by $h(t)$, which may include a receiver bandwidth limitation and anti-aliasing filter matched to the ADC. Fig.~\ref{fig:continuous_detailed_system_model} can also be extended to colored noise using an additional whitening filter as the front-end of $h(t)$.  
The filtered noise $N(t)=N'(t)*h(t)$ is a stationary circularly-symmetric complex Gaussian process with autocorrelation function (ACF)
\begin{align}
\varphi_\text{NN}(\tau) =  N_0 \cdot (h^*(-\tau) *  h(\tau)).
\end{align}
Next, we describe the effect of the nonlinear device on $X(t)$.

\subsection{Memoryless Nonlinear Device}
\label{sec:memoryless_nl}
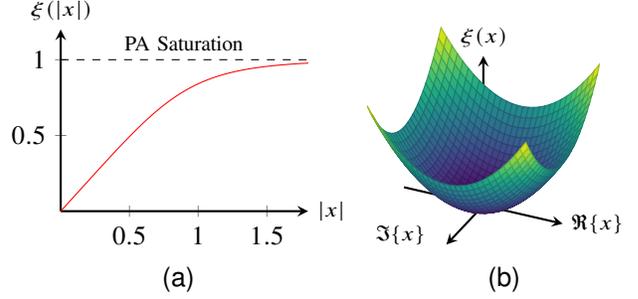
\begin{figure}[!t]
    \begin{minipage}[b]{0.55\columnwidth}
    \subfloat[\label{fig:sspa}]{
    \begin{tikzpicture}
\begin{axis}[width=1\textwidth,
          height=4.0cm,axis x line=middle,
          axis y line=middle,
          x axis line style={thick},
          y axis line style={thick},
          xlabel={\footnotesize $|x|$},
          ylabel={\footnotesize $\nonlp{|x|}$},
          xmin=0,xmax=1.8,
          ymin=0,ymax=1.2,
          xlabel style={anchor=west},
          ylabel style={anchor=south},]
\addplot [domain=0:2, red, samples=200]{x/(1+x^(2*2))^(1/(2*2))}; %
\draw[dashed] (axis cs: 0,1) -- node[midway,above,font=\footnotesize]{PA Saturation} (1.8,1); 
\end{axis}
\end{tikzpicture}}
    \end{minipage}
    \hfill
    \begin{minipage}[b]{0.440\columnwidth}
    \subfloat[\label{fig:absq}]{%
     \begin{tikzpicture}
\begin{axis}[axis x line=middle,
          axis y line=middle,
          axis z line=middle,
          x axis line style={thick},
          y axis line style={thick},
          z axis line style={thick},
          y dir=reverse,
          ylabel style={at={(current axis.left of origin)},anchor=south west},
          width=1.2\textwidth,height=5.1cm,colormap/viridis,
          xlabel={\footnotesize $\mathfrak{R}\{x\}$},
          ylabel={\footnotesize $\mathfrak{I}\{x\}$},
          zlabel={\footnotesize $\nonlp{x}$},
          xlabel style={anchor=west},
          ylabel style={anchor=north west,yshift=-0.5cm},
          zlabel style={anchor=south},
          xmin=-5,xmax=+5,
          ymin=-5,ymax=+5,
          zmax=62,
          ticks=none,
          ]
 
\addplot3 [surf,
    domain =-5:5,
    domain y = -5:5,
    samples = 30,
    samples y = 30,
    draw = green!70!black,opacity=0.8] {x^2 + y^2};
\end{axis}
 
\end{tikzpicture}}
    \end{minipage}
    \caption{Nonlinear functions for a (a) PA: $\nonlp{|x|} = |x|/\sqrt[\leftroot{1}\uproot{2}4]{(1+|x|^{4})}$ and (b) SLD: $\nonlp{x} = \lvert x\rvert^2$.}
    \label{fig:appexamples}
\end{figure}

The nonlinearity expands the bandwidth, e.g., if $\nonlp{\cdot}$ is a polynomial of degree $d_\xi$ then $Z'(t)$ occupies $d_\xi$ times the bandwidth of $X(t)$~\cite[Sec.~3.1]{zhou2004spectral},~\cite[Thm.~1]{wise1977effect}. 
More generally, the bandwidth of $Z'(t)$ may be unbounded but is bandlimited via $h(t)$. We consider the two applications with the nonlinear functions shown in Fig.~\ref{fig:appexamples}.

\subsubsection{Wireless Transmitter with a PA} The model in~\cite[Fig.~1]{ochiai2013analysis} has
\begin{align}
    Z'(t) = \nonlp{|X(t)|} \cdot \exp(\mathrm{j} \angle X(t))
    \label{eq:WT-PA}
\end{align}
with a nonlinear real-valued function $\nonlp{\cdot}$ that models a solid-state PA that distorts the magnitude; see~\cite[Sec.~3.2]{rapp1991effects}.

\subsubsection{Optical Fiber Receiver with a SLD}
\label{sec:fiber_optic_application}
The model in~\cite[Fig.~2]{plabst2022achievable} has a single polarization and $g(t) \propto \mathrm{sinc}(B t) * g_\text{SSMF}(t)$ where $g_\text{SSMF}(t)$ is the linear response (dispersion) of a standard single-mode fiber (SSMF). The SLD outputs $\nonlp{\cdot} = \lvert \,\cdot\, \rvert^2$ and the front-end of the ADC is a brickwall filter $h(t) \propto \mathrm{sinc}(2 B t)$ with twice the transmit filter bandwidth. 

\subsection{Discrete Time Model}
\label{sec:time-discrete_model}
One obtains a discrete time model by collecting samples $Y_k = Y( k T_\text{s}')$, $k\in\mathbb{Z}$, where $T_\text{s}' = 1/(B N_\text{os})$ corresponds to sampling at rate $B N_\text{os}$ with oversampling factor $N_\text{os}$.
The Nyquist-Shannon sampling theorem is met if $Y(t)$ is bandlimited and $N_\text{os}$ is sufficiently high. The $k^\text{th}$ sample is
\begin{equation}
\begin{aligned}
    Y_k =  Z_k + N_k
     \label{eq:y_time_discrete}
\end{aligned}
\end{equation}
where
\begin{align}
    Z_k 
    &= \left[ h(t) * \nonlp{X(t)}
    \right]_{t=kT_\text{s}'} \\
    N_k & = [h(t) * N'(t)]_{t = k T_\text{s}'} .
\end{align}
The discrete-time noise $N_k$ is stationary, circularly symmetric, and complex Gaussian with ACF
\begin{align}
    \varphi_\text{NN}[k] = N_0 \cdot \,\varphi_\text{NN}(\tau = k T_\text{s}')
    .
\end{align} 

\subsection{Approximation via Simulation}
\label{sec:approximation_via_simulation}
We use oversampling to address the bandwidth expansion. Let $T_\text{sim} = T_\text{s}/N_\text{sim}$ be the simulation sampling period where $N_\text{sim}$ is the simulation oversampling factor and $d = N_\text{sim}/N_\text{os}$ is a positive integer. One can approximate
\begin{align}
   Z_k
   \approx \sum\limits_{\osidx'} h_{\osidx'}  \nonl\!\bigg(\sum\limits_{\osidx} g_{\osidx}  X'_{\left(d\cdot k-\osidx'\right)-\osidx}\bigg)
   \label{eq:disc_approx_z_k}
\end{align}
where $(X'_\osidx)_{\osidx\in\mathbb{Z}}=((0,\ldots,0,X_\kappa))_{\kappa \in\mathbb{Z}}$ is a $N_\text{sim}$-fold upsampled string, and $g_\osidx = g(\osidx T_\text{sim})$, $h_\osidx = h(\osidx T_\text{sim})$ are the oversampled filters. 
The SLD example above has $d_{\nonl} = 2$ and $N_\text{os}=N_\text{sim}=2$ results in sufficient statistics, and~\eqref{eq:disc_approx_z_k} is an equality. However, if the PA model in \eqref{eq:WT-PA} is not a polynomial, then one may need to choose a large $N_\text{sim}$ so that~\eqref{eq:disc_approx_z_k} is a good approximation.

\begin{figure}
    \centering
    \pgfplotstableread{
X Y
-70	0.00034395434005692
-69	0.000642365211434637
-68	0.00133391168177826
-67	0.00220024042649002
-66	0.00294201956522927
-65	0.00329020207732367
-64	0.00311548950555488
-63	0.00247931334016457
-62	0.00163197379064498
-61	0.000927402146557727
-60	0.000694168258640714
-59	0.00111109847691925
-58	0.0021151091533114
-57	0.00341447479467743
-56	0.00458341341668048
-55	0.00521686444570312
-54	0.00509551432266411
-53	0.00426519110678262
-52	0.00306581558131275
-51	0.00202043346264178
-50	0.00165107748689623
-49	0.00228265268580782
-48	0.00388522439305358
-47	0.00606885976703243
-46	0.00819457281410886
-45	0.00959538180018348
-44	0.00983097723439321
-43	0.00887078533655091
-42	0.00717131366504125
-41	0.00555250141982348
-40	0.00493176120395989
-39	0.00598697212990415
-38	0.00889646561070438
-37	0.0132108773578828
-36	0.0179526541951616
-35	0.0219203814546114
-34	0.0240902067599131
-33	0.0240866799773607
-32	0.0223555394615533
-31	0.0201349244462534
-30	0.0191296161619762
-29	0.0209759161612851
-28	0.0267326394544616
-27	0.036375119435545
-26	0.0487220982666968
-25	0.0617188360641336
-24	0.0731091441594863
-23	0.0809931865340418
-22	0.0846966652826184
-21	0.0851077381650468
-20	0.084590258384194
-19	0.0863999945870284
-18	0.0937795524825601
-17	0.108981931178333
-16	0.13245983982992
-15	0.162529999586048
-14	0.195597250226608
-13	0.227072193597959
-12	0.252393867460207
-11	0.268214873555745
-10	0.273239988198279
-9	0.268532332501445
-8	0.257072926037829
-7	0.242949369010836
-6	0.230191100748319
-5	0.221704709051301
-4	0.218605590968675
-3	0.220120597480464
-2	0.22409825949458
-1	0.227908773767205
0	0.229449918286105
1	0.227908773767205
2	0.22409825949458
3	0.220120597480464
4	0.218605590968675
5	0.221704709051301
6	0.230191100748319
7	0.242949369010836
8	0.257072926037829
9	0.268532332501445
10	0.273239988198279
11	0.268214873555745
12	0.252393867460207
13	0.227072193597959
14	0.195597250226608
15	0.162529999586048
16	0.13245983982992
17	0.108981931178333
18	0.0937795524825601
19	0.0863999945870284
20	0.0845902583841939
21	0.0851077381650467
22	0.0846966652826183
23	0.0809931865340417
24	0.0731091441594864
25	0.0617188360641336
26	0.0487220982666969
27	0.036375119435545
28	0.0267326394544616
29	0.0209759161612851
30	0.0191296161619762
31	0.0201349244462534
32	0.0223555394615533
33	0.0240866799773607
34	0.0240902067599131
35	0.0219203814546114
36	0.0179526541951616
37	0.0132108773578828
38	0.00889646561070437
39	0.00598697212990415
40	0.00493176120395988
41	0.00555250141982347
42	0.00717131366504125
43	0.00887078533655091
44	0.00983097723439321
45	0.00959538180018348
46	0.00819457281410886
47	0.00606885976703243
48	0.00388522439305358
49	0.00228265268580782
50	0.00165107748689623
51	0.00202043346264179
52	0.00306581558131275
53	0.00426519110678262
54	0.00509551432266412
55	0.00521686444570312
56	0.00458341341668048
57	0.00341447479467743
58	0.0021151091533114
59	0.00111109847691925
60	0.000694168258640716
61	0.000927402146557729
62	0.00163197379064499
63	0.00247931334016457
64	0.00311548950555488
65	0.00329020207732368
66	0.00294201956522927
67	0.00220024042649002
68	0.00133391168177826
69	0.000642365211434638
70	0.000343954340056921
}{\CDCIR}

\begin{tikzpicture}[scale=1,]
\begin{axis}[width=.52\textwidth,height=4.1cm,
          xlabel={\footnotesize $t/T$},
          ytick={},
          yticklabels={},
          axis x line=middle,
          axis y line=middle,
          x axis line style={thick},
          y axis line style={thick},
          ymin=-0.03,ymax=0.37,
          xmin=-69,xmax=69,
          ytick style={draw=none},
          xtick style={font=\footnotesize},
          tick label style={font=\footnotesize},
          xtick={-20,-10,0,10,20},
          xticklabels={$-2$,$-1$,$0$,$1$,$2$},
          xlabel style={anchor=west},
          ]
\addplot[red,restrict x to domain=-69:69] table []{\CDCIR}; 
\addplot+[ycomb,blue,mark options={blue,mark=o,fill=blue},restrict x to domain=-69:69] table [each nth point =5]{\CDCIR}; 

\legend{$|g(t)|^2$,$|g_\osidx|^2$}

\end{axis}
\end{tikzpicture}
    \caption{Magnitude-squared response of an SSMF channel with parameters in Tab.~\ref{tab:simparams}, length \SI{30}{\kilo\meter} and $g(t) \propto \mathrm{sinc}(Bt) * g_\text{SSMF}(t)$; circles show samples with $N_\text{sim} = 2$.}
    \label{fig:combined_filter_taps}
\end{figure}
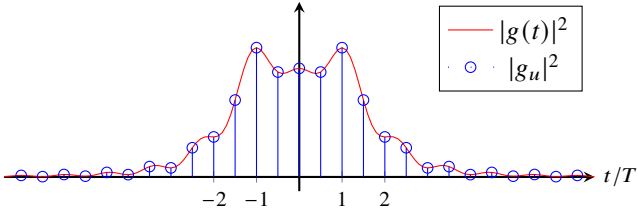

To illustrate, suppose the filters $g_\osidx$ and $h_\osidx$ in~\eqref{eq:disc_approx_z_k} have odd lengths $K_g$ and $K_h$, respectively, and
the filter taps are zero outside the intervals
$[-\lfloor K_g/2\rfloor,\lfloor K_g/2\rfloor]$ and $[-\lfloor K_h/2\rfloor,\lfloor K_h/2\rfloor]$.
The filters have symbol memories $\widetilde{K}_g = \lfloor (K_g-1)/N_\text{sim} \rfloor$ and $\widetilde{K}_h = \lfloor (K_h-1)/N_\text{sim} \rfloor$, and the total system memory is 
\begin{align}
    \widetilde{K} = \widetilde{K}_g + \widetilde{K}_h.  
    \label{eq:total_memory}
\end{align}
For example, Fig.~\ref{fig:combined_filter_taps} shows a bandlimited SSMF channel $g(t)$ with memory $\widetilde{K}_g = 13$.

We use block transmission with $\lfloor K_g/2\rfloor + \lfloor K_h/2\rfloor$ zeros at the beginning and end of each block. We collect a block of $n$ transmit symbols in the vector
\begin{align}
    \mathbf{x} &= [x_1,\; x_2,\; x_3, \quad \dots \quad x_{n} ]^\mathrm{T}  \dimC{n}.
    \label{eq:x_length_n_vector_stacking}
\end{align}
The receiver collects $N_\text{os}$ channel output symbols per transmitted symbol in the vector
\begin{align}
    \mathbf{y} &= [y_1,\; y_2,\; y_3, \quad \dots \quad y_{N_\text{os}  n} ]^\mathrm{T}  \dimC{N_\text{os} n}.
    \label{eq:y_length_n_vector_stacking}
\end{align}

\section{SDD and SIC Rates}
\label{sec:detection_and_decoding}
This section derives information rates with SDD and SIC; see~\cite[Sec. IV A-B]{prinz2023successive}.
\subsection{SDD Rates}
SDD computes the symbol-wise APPs $P_{X_\kappa | \mathbf{Y}}(\cdot \rvert \mathbf{y})$, $\kappa \in \lbrace 1, \ldots, n \rbrace$, and the decoder uses these to estimate the data. Consider the entropy rates
\begin{align}
    & H_n(\bm{X}) = \frac{1}{n}  H(\bm{X}) \label{eq:single_letter_h} \\
    & H_n(\bm{X}\lvert \bm{Y}) = \frac{1}{n}  H(\bm{X} |\bm{Y}).
    \label{eq:single_letter_mi}
\end{align}
A lower bound on the information rate of $\mathbf{X}$ and $\mathbf{Y}$ is:
\begin{align}
    I_n(\bm{X};\bm{Y}) &= H_n(\bm{X}) - H_n(\bm{X}|\bm{Y}) \label{eq:joint_detection_decoding_rate}
    \\
    &\ge H_n(\bm{X}) - \frac{1}{n}\sum\limits_{\kappa=1}^n H(X_\kappa|\bm{Y})\; := I_{n,\text{SDD}} \label{eq:separate_detection_decoding_rate} 
\end{align}
with equality if and only if $X_\kappa$ and $X^{\kappa-1}$ are conditionally independent given $\mathbf{Y}$. Define the limiting rates as
\begin{align}
    I(\mathcal{X};\mathcal{Y}) := \lim_{n \rightarrow \infty} I_n(\bm{X};\bm{Y}), \quad
     I_\text{SDD} := \lim_{n\rightarrow \infty} I_{n,\text{SDD}}
\end{align}
and note that the bound \eqref{eq:separate_detection_decoding_rate} gives $I_\text{SDD} \le I(\mathcal{X};\mathcal{Y})$.

JDD achieves the rate $I(\mathcal{X};\mathcal{Y})$, but is usually too complex to implement~\cite{huber1992trelliscodierte,schuh_reduced_2013,giallorenzi1996multiuser}. In practice, one is often limited to SDD and the rate $I_\text{SDD}$, but~\cite[Fig.~7-9]{muller_capacity_separate2004} and~\cite[Fig.~6]{prinz2023successive} show that SDD loses significant rate for ISI channels at medium to high SNRs. This is because SDD neglects stochastic dependencies when computing APPs of systems with memory.

\subsection{SIC Rates}
\label{sec:sic}
SIC increases the SDD rates and is implemented with $S$ stages and a different forward error control (FEC) code for each stage~\cite{PfisterAIRFiniteStateChan2001}.

\subsubsection{SIC Encoding}
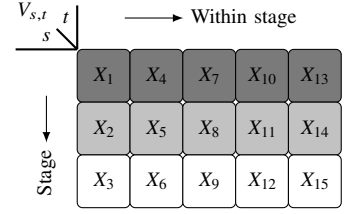
\begin{figure}
    \centering
    \usetikzlibrary{decorations.markings}
\tikzset{node distance=0.9cm}

\pgfdeclarelayer{background}
\pgfdeclarelayer{foreground}
\pgfsetlayers{background,main,foreground}

\pgfmathsetmacro{\samplerwidth}{30}

\tikzset{redbox/.style = {rounded corners=2pt,draw,font=\normalsize,minimum height=0.9cm,minimum width=0.9cm,fill=mycolorlightorange,}}
\tikzset{bluebox/.style = {rounded corners=2pt,draw,font=\normalsize,minimum height=0.9cm,minimum width=0.9cm,fill=mycolorcube,}}
\tikzset{bbox/.style = {rounded corners=2pt,draw,font=\normalsize,minimum height=0.9cm,minimum width=0.9cm,fill=mycolor5bright,}}
\tikzset{dot/.style = {anchor=base,fill,circle,inner sep=1pt}}

\begin{tikzpicture}[scale=0.78, transform shape]
\renewcommand{\baselinestretch}{1}

\node[redbox] (u1) {$X_1$};
\node[bluebox,below of=u1] (u2) {$X_2$};
\node[bbox,below of=u2] (u3) {$X_3$};

\node[redbox,right of=u1] (uM1) {$X_4$};
\node[bluebox,below of=uM1] (uM2) {$X_5$};
\node[bbox,below of=uM2] (uM3) {$X_6$};

\node[right of=uM1,redbox] (u2M1) {$X_7$};
\node[below of=u2M1,bluebox] (u2M2) {$X_{8}$};
\node[bbox,below of=u2M2] (u2M3) {$X_{9}$};

\node[redbox,right of=u2M1,minimum width=0.9cm] (u3M1) {$X_{10}$};
\node[bluebox,below of=u3M1,minimum width=0.9cm] (u3M2) {$X_{11}$};
\node[bbox,below of=u3M2] (u3M3) {$X_{12}$};

\node[redbox,right of=u3M1,minimum width=0.9cm] (u4M1) {$X_{13}$};
\node[bluebox,below of=u4M1,minimum width=0.9cm] (u4M2) {$X_{14}$};
\node[bbox,below of=u4M2] (u4M3) {$X_{15}$};

\draw[thick ] (u1) -- (-0.8,+0.8) node[above left]{$V_{s,\sicindex}$};
\draw[thick ] ($(u1) + (-0.45,+0.45)$) -- node[above](m){$s$} (-1.5,+0.45);
\draw[thick ] ($(u1) + (-0.45,+0.45)$) -- node[left,yshift=0.16cm](r){$\sicindex$} (-0.45,+1.2);

\draw[-latex] ($(m) + (0,-1.0)$) -- ($(m) + (0,-1.8)$)  node[left,rotate=90]{Stage}; 
\draw[-latex] ($(r) + (+1.0,0)$) -- ($(r) + (2,0)$) node[right]{Within stage};

\end{tikzpicture}
    \caption{SIC with $S=3$ stages and $n=15$ input symbols.}
    \label{fig:SP_conversion}
\end{figure}

To encode, downsample $\bm{X}$ by a factor of $S$ to create $S$ sequences of length $\siclength = n/S$, where we assume $\siclength$ is an integer. The symbols in the $s^\text{th}$ SIC stage are
\begin{align}
    \bm{V}_{s} &= (V_{s,t})_{t=1}^N = \big(X_{\kappa(s,1)}, X_{\kappa(s,2)} \ldots X_{\kappa(s,N)}   \big)
    \label{eq:subsampling}
\end{align}
where $\kappa(s,t) = s + (t-1)S$ converts a parallel indexing $(s,t)$ to a serial indexing $\kappa(s,t)$. For example, consider Fig.~\ref{fig:SP_conversion} with $S=3$, $n=15$, $N=5$, and thus
\begin{align}
    \bm{V}_{1} = \left(X_{1}, X_{4}, X_{7}, X_{10}, X_{13} \right).
    \label{eq:downsample-example}
\end{align}

\subsubsection{SIC Receiver}
\begin{figure}
    \centering
    \usetikzlibrary{decorations.markings}
\tikzset{node distance=1cm}

\pgfdeclarelayer{background}
\pgfdeclarelayer{foreground}
\pgfsetlayers{background,main,foreground}

\pgfmathsetmacro{\samplerwidth}{30}

\tikzset{redbox/.style = {draw,minimum height=1.9em,minimum width=3.6em,fill=mycolorlightorange}}
\tikzset{bluebox/.style = {draw,minimum height=1.9em,minimum width=3.6em,fill=mycolorcube}}
\tikzset{greenbox/.style = {draw,minimum height=1.9em,minimum width=3.6em,, opacity=1,fill=mycolor5bright},
mydash/.style={rectangle,fill=white}}
\tikzset{dot/.style = {anchor=base,fill,circle,inner sep=1pt}}
\tikzstyle{point}=[fill,shape=circle,minimum size=3pt,inner sep=0pt]

\begin{tikzpicture}[]
\footnotesize

\node[] (y) {};
\node[redbox,right of=y,align=center,node distance=1.5cm,font=\footnotesize,label={[xshift=+0.5cm,yshift=-0.05cm]above:{}}] (app1) {APP 1: $P(v_{1,t} |\, \mathbf{y})$};
\node[redbox,right of=app1,align=center,node distance=1.75cm] (dec1) {Dec. 1};

\node[bluebox,below of=app1,yshift=-0.10cm,xshift=1.55cm,align=center,label={[xshift=0.72cm,yshift=-0.05cm]above left:{}}] (app2) {APP 2: $P( v_{2,t} |\, \mathbf{y}, \hat{\mathbf{v}}_1)$};
\node[bluebox,right of=app2,node distance=1.98cm,align=center] (dec2) {Dec. 2};

\node[below of=app2,node distance=0.70cm,xshift=-0.2cm,yshift=0.1cm] (recDots) {$\ddots$};
\node[greenbox,below of=app2,yshift=-0.15cm,xshift=1.9cm,align=center,label={[xshift=0.7cm,yshift=-0.05cm]above left:{}}] (appM) {APP $S$: $P( v_{S,t} |\, \mathbf{y}, \hat{\mathbf{v}}^{S-1})$};
\node[greenbox,right of=appM,node distance=2.19cm,align=center] (decM) {Dec. $S$};

\draw[-latex] (app1 -|  y) -- (app1);
\draw (app1) --(dec1);

\draw[-latex] (app1 -| y) |- (appM);
\draw[-latex] (app1 -| y) |- (app2);

\node[xshift=-0.3cm,yshift=-0.2cm](y_desc) at(y |- app2) {$\mathbf{y}$}; 

\draw[-latex] (dec1) -| node[mydash,pos=0.87]{$\rvdots$} (appM.20) node[midway,above](v1label) {$\hat{\mathbf{v}}_1$};
\draw[-latex] (dec1) -| (app2.16);
\draw[-latex] (app2) --  (dec2) -| node[mydash,pos=0.70]{$\rvdots$} (appM.15) node[midway,above] {$\hat{\mathbf{v}}_2$};

\node[xshift=1.2cm,font=\footnotesize](out1) at(dec1 -| decM) {$\hat{\mathbf{v}}_1$}; 
\node[xshift=1.2cm,font=\footnotesize](out2) at(dec2 -| decM) {$\hat{\mathbf{v}}_2$};
\node[xshift=1.2cm,font=\footnotesize](outM) at(decM -| decM) {$\hat{\mathbf{v}}_S$};

\draw (appM) -- (decM); 
\path[] (app1) -- (appM);

\draw[-latex,] (dec1 -| appM.21) -- (out1);
\draw[-latex,] (dec2) -- (out2);
\draw[-latex,] (decM) -- (outM);

\draw [decorate, decoration = {mirror,brace}] ($(app1.west) + (0,-0.35)$) -- node[midway,below,font=\footnotesize] {Stage 1} ($(dec1.east) + (0,-0.35)$);
\draw [decorate, decoration = {mirror,brace}] ($(app2.west) + (0,-0.35)$) -- node[midway,below,font=\footnotesize] {Stage 2} ($(dec2.east) + (0,-0.35)$);
\draw [decorate, decoration = {mirror,brace}] ($(appM.west) + (0,-0.35)$) -- node[midway,below,font=\footnotesize] {Stage $S$} ($(decM.east) + (0,-0.35)$);

\end{tikzpicture}
    \caption{SIC receiver with SDD for each stage.}
    \label{fig:sic}
\end{figure}
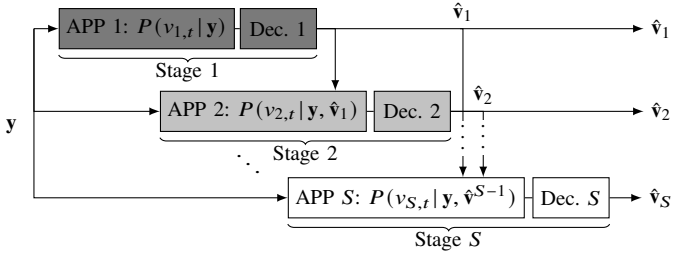

We detect and decode using $S$ stages; see Fig.~\ref{fig:sic}. To illustrate, consider again $S=3$ and partition $\bm{X}$ into $\mathbf{V}_1$, $\mathbf{V}_2$, $\mathbf{V}_3$. 
The first stage performs SDD and calculates the APPs $P_{V_{1,t} |\mathbf{Y}}(\cdot \rvert \mathbf{y})$ for $t = 1,\ldots,N$. A decoder estimates $\hat{\mathbf{V}}_1$.
The second stage uses $\hat{\mathbf{V}}_1$ as prior information and calculates the APPs $P_{V_{2,t} | \mathbf{V}_1, \mathbf{Y}}(\cdot \rvert \hat{\mathbf{v}}_1,\mathbf{y})$ for $t = 1,\ldots,N$. Correct prior information increases the information rate in the second SIC stage. The final stage computes the APPs $P_{V_{3,t} | \mathbf{V}_1, \mathbf{V}_2, \mathbf{Y}}(\cdot \rvert \hat{\mathbf{v}}_1,\hat{\mathbf{v}}_2,\mathbf{y})$ for $t = 1,\ldots,N$.
\subsubsection{SIC Rates}
Define the long string $\bm{V} = (\bm{V}_s)_{s=1}^S$. Since $\mathbf{V}$ is a reordered version of $\mathbf{X}$, we have $I_n(\bm{X};\bm{Y}) =I_n(\bm{V};\bm{Y})$. 
The rate of stage $s$ is
\begin{align}
    I^{s}_{N\!,\text{SIC}} =   \frac{1}{\siclength} \sum\limits_{t = 1}^N I(V_{s,t};  \bm{Y},\bm{V}^{s-1})
    \label{eq:per_stage_rate}
\end{align}
and $I^{s}_{N\!,\text{SIC}}$ is non-decreasing in $s$ and is at most $m$. The average SIC rate is
\begin{align}
    I_\text{$n$,SIC} = \frac{1}{S} \sum\limits_{s=1}^S   I^{s}_{N\!,\text{SIC}}
\end{align}
and satisfies
\begin{align}
    I_\text{$n$,SDD} \le I_\text{$n$,SIC}
    \le  I_n(\bm{V};\bm{Y}).
    \label{eq:sic_inequality_alt}
\end{align}

Define $I^{s}_{\text{SIC}} := \lim_{N \rightarrow \infty } I^{s}_{N\!,\text{SIC}}$ and $I_{\text{SIC}} := \lim_{n \rightarrow \infty } I_{n,\text{SIC}}$.
We encode $\mathbf{V}_s$ with a code rate less than $I^{s}_{\text{SIC}}$ to ensure reliable decoding as the block length grows, i.e., we may assume  $\hat{\mathbf{V}}_s = \mathbf{V}_s$; see~\cite{PfisterAIRFiniteStateChan2001}. 
Comparing the limiting rates, we obtain
\begin{align}
     I_\text{SDD} \le I_\text{SIC} \le I(\mathcal{X};\mathcal{Y}) .
    \label{eq:sdd_jdd_inequality}
\end{align}
SIC can approach the JDD performance by increasing $S$~\cite[Fig.~6]{prinz2023successive}. 
Also, $S$ should increase with the total memory $\widetilde{K}$~\eqref{eq:total_memory}, modulation alphabet size, and ISI magnitude; see~\cite[Sec.~IV~B and Fig.~2]{prinz2023successive} on how the effective memory is reduced at higher SIC stages.

\section{FBA Equalizer}
\label{sec:fba_with_ic}
A SIC receiver requires symbol-wise APP calculations at each stage. 
We extend the APP calculations in~\cite[Sec. III-IV]{plabst2022achievable} from SDD  to SIC. In particular, we show how the APPs of Fig.~\ref{fig:sic} are computed via the FBA. The FBA structure is later used to design model-based NNs.

\begin{figure*}
    \centering
    \tikzset{
factor/.style = {drop shadow={shadow xshift=0.1em,shadow yshift=-0.1em,}, draw, thick, fill=white, fill opacity=1,rectangle, minimum height=1.0em, minimum width=1.0em,font=\footnotesize,outer sep=0,inner sep=0},
larger/.style = { minimum height=1.8em, minimum width=1.8em,},
var/.style  = {anchor=center,fill=white,draw,ellipse,minimum size=0.3cm,inner sep=1pt},
input/.style = {coordinate},
output/.style= {coordinate},
pinstyle/.style = {pin edge={to-,thin,black}},
mydash/.style={anchor=center,fill=white,rectangle},
graystyle/.style={draw=black,opacity=1,text=black}
}

\pgfdeclarelayer{background}
\pgfdeclarelayer{foreground}
\pgfsetlayers{background,main,foreground}   %

\begin{tikzpicture}[auto, scale=1.11, transform shape, black, draw=black, node distance=2.5cm,>=stealth,line width=0.2mm,font=\footnotesize]
        
    \node[input](inputl){}; 
    \node[factor,draw=,larger,right of=inputl](F2){}; 
    \node[factor,larger,right of=F2,node distance=3.2cm,](F3){}; 
    \node[factor,larger,right of=F3](F4){}; 
    \node[factor,larger,right of=F4](F5){}; 
    \node[factor,larger,right of=F5](F6){}; 

    \node[input,right of=F6](inputr){}; 

    \node[above of=F2,node distance=1.05cm,](xs){}; 
    \node[,above of=F3,node distance=1.05cm,](xsp1){}; 
    \node[,above of=F4,node distance=1.05cm,](xsp2){}; 
    \node[,above of=F5,node distance=1.05cm,](xsp3){}; 
    \node[,above of=F6,node distance=1.05cm,](xsp4){}; 
    
    \node[input,below of=F2,node distance=0.85cm](ys){}; 
    \node[input,below of=F3,node distance=0.85cm,graystyle](ysp1){}; 
    \node[input,below of=F4,node distance=0.85cm,graystyle](ysp2){}; 
    \node[input,below of=F5,node distance=0.85cm,graystyle](ysp3){}; 
    \node[input,below of=F6,node distance=0.85cm,graystyle](ysp4){}; 

    \draw[] (inputl) node[left,mydash,pos=1]{$\ldots$} -- node[var, pos=0.5, label=below:{$\St_{S,t-1}$}]{} (F2); 
    \draw[] (F2) -- node[var, pos=0.5,  label=below:{$\St_{s,t}$}]{} (F3); 
    \draw[] (F3) -- node[var, pos=0.5, label=below:{$\St_{s+1,t}$}]{} (F4); 
    \draw[] (F4) -- node[mydash,pos=0.5]{$\ldots$}  (F5); 
    \draw[] (F5) -- node[var, pos=0.5, label=below:{$\St_{S-1,t}$}]{} (F6); 
    \draw[] (F6) -- node[var, pos=0.5, label=below:{$\St_{S,t}$}]{} (inputr) node[right,mydash,pos=1]{$\ldots$};

    \draw[] (xs) -- node[,var, pos=0, fill=black, label=above:{$V_{s,t}$}]{} (F2); 
    \draw[] (xsp1) -- node[graystyle,var, pos=0, label=above:{$V_{s+1,t}$}]{} (F3); 
    \draw[] (xsp2) -- node[graystyle,var, pos=0, label=above:{$V_{s+2,t}$}]{} (F4); 
    \draw[] (xsp3) -- node[graystyle,var, pos=0, label=above:{$V_{S-1,t}$}]{} (F5); 
    \draw[] (xsp4) -- node[graystyle,var, pos=0, label=above:{$V_{S,t}$}]{} (F6); 
    
    \draw[] (ys) -- node[,var, pos=0, label={[below,align=center,yshift=-0.30cm]$(\breve{\mathbf{y}}_{\kappa(j,t)})_{j=1}^{s}$, $h^c_{S,t-1}$, $(v_{j,t})_{j=1}^{s-1}$}]{} (F2);  %
    \draw[] (ysp1) -- node[graystyle,var, pos=0, label=below:{$\breve{\mathbf{y}}_{\kappa(s+1,t)}$, $h^c_{s,t}$}]{} (F3); 
    \draw[] (ysp2) -- node[graystyle,var, pos=0, label=below:{$\breve{\mathbf{y}}_{\kappa(s+2,t)}$, $h^c_{s+1,t}$}]{} (F4); 
    \draw[] (ysp3) -- node[graystyle,var, pos=0, label=below:{$\breve{\mathbf{y}}_{\kappa(S-1,t)}$, $h^c_{S-2,t}$}]{} (F5); 
    \draw[] (ysp4) -- node[graystyle,var, pos=0, label=below:{$\breve{\mathbf{y}}_{\kappa(S,t)}$, $h^c_{S-1,t}$}]{} (F6);

\end{tikzpicture}
     \caption{Factor graph for the FBA in stage $s$. The string $(v_{j,t})_{j=1}^{s-1}$ represents known symbols. Due to previous SIC levels, the $h^c_{j,t}, j \in \lbrace s,\ldots, S\rbrace$, are known parts of the channel state. The first state index takes on values in $\{s,\ldots,S\}$. The filled variable node marks a symbol for which an APP is calculated.}
    \label{fig:fba}
\end{figure*}
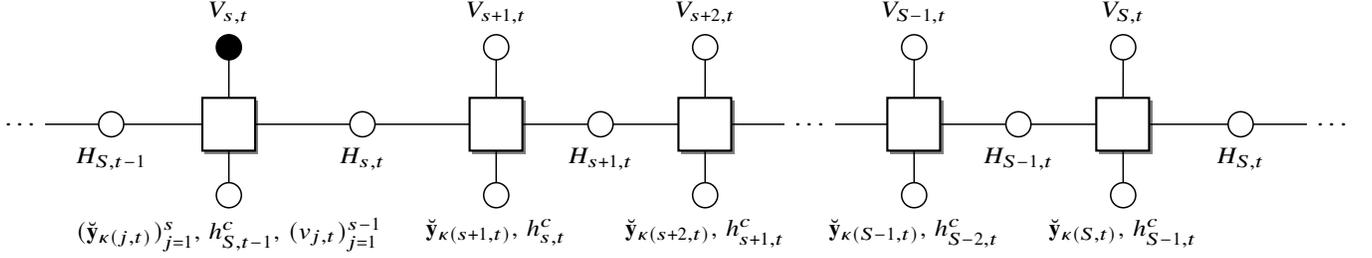

Consider the channel outputs $\breve{y}_k := y_{k-k_0}$, where $k_0 = \lfloor \lfloor K_g / 2 \rfloor/d + \lfloor K_h/ 2 \rfloor/d \rfloor$ and $d = N_\text{sim}/N_\text{os}$; see Sec.~\ref{sec:approximation_via_simulation}. We collect $N_\text{os}$ output symbols per input $X_\kappa$ in the string
\begin{align} 
    \mathbf{\breve{y}}_{\kappa} 
    := \big(\breve{y}_{\kappa + N_\text{os}(\kappa-1) + \ell}\big)_{\ell = 0}^{N_\text{os}-1}
    .
    \label{eq:single_input}
\end{align} 
For SIC stage $s$, $s \in \lbrace 1,\ldots, S\rbrace $, the FBA computes the APPs
\begin{align}
    P_{V_{s,t}|\breve{\mathbf{Y}},\mathbf{V}^{s-1}}( \cdot | \breve{\mathbf{y}},\mathbf{v}^{s-1}) \quad \forall t
\end{align}
where $\breve{\mathbf{y}} := (\breve{\mathbf{y}}_1,\ldots,\breve{\mathbf{y}}_n)$.
Fig.~\ref{fig:fba} depicts the FBA structure by a factor graph. SDD has $S=1$; see~\cite[Fig.~4]{plabst2022achievable}.

We increase the memory $\widetilde{K}$ so that $\widetilde{K}/S$ is an integer to simplify the analysis. Now partition the channel state (the past $\widetilde{K}$ input symbols) into disjoint strings for $j=1,\dots,S$: 
\begin{align}
    \St_{j,t} &:=  \left( X_{\kappa(j,t)-\ell} \right)_{\ell = 0}^{\widetilde{K}-1}  \setminus  \mathbf{V}^{s-1}  
    \label{eq:state_partition_unknown}
    \\
    \St_{j,t}^c &:=  \left( X_{\kappa(j,t)-\ell} \right)_{\ell = 0}^{\widetilde{K}-1}  \setminus \St_{j,t}
    \label{eq:state_partition_known}
\end{align}
where $\St_{j,t}$ has $\widetilde{K}_s = (S-s+1) \cdot \widetilde{K}/S$ unknown symbols and $\St_{j,t}^c$ has $\widetilde{K}_s^c = (s-1) \cdot \widetilde{K}/S$ known symbols. For example, 
consider $\widetilde{K}=3$, $S=3$, and $s=2$, for which $\widetilde{K}_s=2$, $\widetilde{K}_s^c=1$, the known symbols are $\mathbf{V}^{s-1}=X_1,X_4,X_7,\dots$, and thus 
\begin{align}
    \St_{j,t}, \St_{j,t}^c = 
    \begin{cases}
        (X_2,X_3), (X_1) & j = 3, t = 1\\
        (X_2,X_3), (X_4) & j = 1, t = 2\\
        (X_3,X_5), (X_4) & j = 2, t = 2\\
        (X_5,X_6), (X_4) & j = 3, t = 2\\
        \quad \vdots 
    \end{cases}.
    \label{eq:example_partition}
\end{align}
For $s>1$, one may simplify by marginalizing only over the states $H_{j,t}$. 
Note that the states $H_{S,t},H_{1,t+1},\ldots, H_{s-1,t+1}$ are the same for all $t$, as the channel inputs from time steps $\kappa(1,t)$ to $\kappa(s-1,t+1)$ are known; see~\eqref{eq:example_partition}. 
The factor graph in Fig.~\ref{fig:fba} thus uses only $H_{S,t}$.
The FBA complexity decreases with increasing $s$ because the effective memory $\widetilde{K}_s$ decreases with $s$. 

For each stage $s$, the FBA computes APPs by processing on Fig.~\ref{fig:fba} over two paths. The first path runs forward in time and recursively calculates forward state metrics, while the second path runs backward in time and computes backward state metrics. Both paths use decoded symbols $\mathbf{V}^{s-1} = \mathbf{v}^{s-1}$ from previous SIC stages. Define the respective forward and backward state metrics for $j=s,s+1,\dots,S$ as 
\begin{align}
    &\mufwsub{j,t}{\st_{j,t}} := p\big(\breve{\mathbf{y}}_{1}^{\kappa(j,t)}, (v_{u,\ell})_{1\leq u < s, 1 \leq \ell \leq t}, \st_{j,t}\big)
    \label{eq:fw_sm}\\
    &\mubwsub{j,t}{\st_{j,t}} := p\big(\breve{\mathbf{y}}_{\kappa(j,t)+1}^{n},  (v_{u,\ell} )_{1\leq u < s, t < \ell \leq N} \mid \st_{j,t}, \st_{j,t}^c \big) .
    \label{eq:bw_sm}
\end{align}
The APPs for stage $s$, for all $t$, and $a \in \mathcal{A}$ are obtained by
\begin{align}
    P_{V_{s,t}|\breve{\mathbf{Y}}, \mathbf{V}^{s-1}}( a | \breve{\mathbf{y}},\mathbf{v}^{s-1}) \propto \sum_{\st_{s,t} : v_{s,t} = a } \! \mufwsub{s,t}{\st_{s,t}} \cdot \mubwsub{s,t}{\st_{s,t}}
    .
    \label{eq:app_s}
\end{align}

\begin{figure*}[!b]
\begin{align}
    \textbf{\text{Forward recursion:}} \quad 
    \mufwsub{{j,t}}{\st_{j,t}}  
    = 
    \begin{cases}
    \mathlarger{\sum\limits_{\st_{j-1,t}, v_{j,t}}} \mufwsub{j-1,t}{\st_{j-1,t}}
    \cdot \fI_{j,t}(\st_{j,t}, \st_{j-1,t},  v_{j,t})  \cdot \fII_t(\st_{j-1,t}), 
    & j = s  
    \\[1.0em]
    \mathlarger{\sum\limits_{\st_{j-1,t}, v_{j,t}}} \mufwsub{j-1,t}{\st_{j-1,t}}
    \cdot \fI_{j,t}(\st_{j,t}, \st_{j-1,t},  v_{j,t}),
    & j = s+1,\dots,S 
    \label{eq:f_fw} 
    \end{cases}
\end{align}
\begin{align}
    \textbf{\text{Backward recursion:}} \quad 
    \mubwsub{{j-1,t}}{\st_{j-1,t}}
    = 
    \begin{cases}
     \mathlarger{\sum\limits_{\st_{j,t},  v_{j,t}}} \mubwsub{{j,t}}{\st_{j,t}}   \cdot \fI_{j,t}(\st_{j,t}, \st_{j-1,t},  v_{j,t}) \cdot  \fII_t(\st_{j-1,t}),  
    & j = s  
    \\[1.0em]
     \mathlarger{\sum\limits_{\st_{j,t},  v_{j,t}} \mubwsub{{j,t}}{\st_{j,t}}}   \cdot \fI_{j,t}(\st_{j,t}, \st_{j-1,t},  v_{j,t}), 
    & j = s+1,\dots,S
    \label{eq:f_bw}
    \end{cases}
\end{align} 
\end{figure*}
Finally, the forward and backward state metrics are calculated recursively via~\eqref{eq:f_fw}-\eqref{eq:f_bw} on the bottom of the next page, where
\begin{align}
    & \fI_{j,t}(\st_{j,t}, \st_{j-1,t},  v_{j,t}) = p
    (\breve{\mathbf{y}}_{\kappa(j,t)}, \st_{j,t},   v_{j,t} \mid \st_{j-1,t}, \st_{j-1,t}^c )
    \label{eq:fI}
\end{align}
and the expression $\fII_t(a)$ is
\begin{align}
     \prod_{j =1}^{s-1} p_{\breve{\mathbf{Y}}_{\kappa(j,t)} \mid  \St_{j-1,t}, \St_{j-1,t}^c, V_{j,t}}(\breve{\mathbf{y}}_{\kappa(j,t)} \mid a, \st_{j-1,t}^c, v_{j,t}).
    \label{eq:fII}
\end{align}
where index pairs are mapped as
\begin{align}
   \begin{rcases}
    (0,t)  \\
    (s-1,t) 
    \end{rcases} 
    \mapsto (S,t-1)
    \label{eq:paridx_map}
\end{align}
because of the parallel indexing; see Fig.~\ref{fig:SP_conversion} and the factor graph in Fig.~\ref{fig:fba}.
We compare the recursions~\eqref{eq:f_fw}-\eqref{eq:f_bw} to the classic FBA in~\cite[Sec.~III-IV]{plabst2022achievable} for different $s$: 
\begin{itemize}
    \item For $s=1$ we have $\fII_t(\cdot) = 1$ and~\eqref{eq:fI} is independent of the index $j \in \{1,\ldots,S\}$. Thus, the state recursions are also independent of $j \in \{1,\ldots,S\}$. 
    This corresponds to the classic FBA~\cite[Sec.~III-IV]{plabst2022achievable} that performs time-invariant state recursions. 
    \item For $1 < s < S$, the state recursions vary with $j \in \{s,\ldots, S\}$ due to the $j$-dependency of~\eqref{eq:fI} and the additional factor~\eqref{eq:fII}. The FBA uses periodically time-varying state recursions. 
    \item For $s = S$, the FBA applies only the first case of~\eqref{eq:f_fw} and~\eqref{eq:f_bw}, respectively, which results in time-invariant recursions
\end{itemize}
Note that the first cases of~\eqref{eq:f_fw} and~\eqref{eq:f_bw} include the factor~\eqref{eq:fII} that depends on known symbols $(v_{j,t})_{j=1}^{s-1}$; cf. the example~\eqref{eq:example_partition}. The first forward and last backward state metrics are initialized with symbols drawn from a uniform distribution.

\section{NN Equalizer}
\label{sec:app_nn}
The FBA is too complex to implement even for relatively small memory $\widetilde{K}$ or alphabets $\mathcal{A}$, so one must use mismatched models. In every SIC stage $s$, we use NN-equalizers with an FBA structure that are bidirectional, recurrent, and periodically time-varying.

\subsection{NN Inputs and Outputs}%
Fig.~\ref{fig:tvrnn_architecture} shows an NN equalizer with multiple layers. Each layer is bidirectional with a forward and backward path and internal states similar to the FBA~\cite{jiang2019learn,kim2020physical,haykin2000adaptive}. The layers imitate message passing on the factor graph of Fig.~\ref{fig:fba}. 
We found that processing the inputs~\eqref{eq:single_input} successively, as in the FBA, requires many RNN states that store soft information about previous inputs. 
To reduce the number of states, the first NN layer effectively acts as a (nonlinear) channel shortening filter~\cite{rusek2012optimal} that processes overlapping blocks of channel outputs and SIC symbols.
The NN layers may be interpreted as turbo detection and decoding modules exchanging extrinsic information~\cite{ten2004design}. We next discuss the NN for SIC stage $s$.

\subsubsection{Inputs} 
Suppose the NN inputs are real-valued. (One may also use complex-valued inputs; the choice is not crucial to the performance.) In SIC stage $s$, we process inputs for the remaining stages $j = s,\ldots,S$; cf. the FBA recursions~\eqref{eq:f_fw} and~\eqref{eq:f_bw}. Define
\begin{align}
    \mathbf{r}^1_{j,t} := (\overline{\mathbf{y}}_{j,t},\overline{\mathbf{v}}_{j,t}) \;\dimR{L_\mathrm{Y} + L_\mathrm{IC}}
    \label{eq:r_vec}
\end{align}
with $L_\mathrm{IC}$ symbols $\overline{\mathbf{v}}_{j,t}$ and $L_\mathrm{Y}$ channel outputs
\begin{align}
    \overline{\mathbf{y}}_{j,t} := ( y_{N_\text{os}  \cdot \kappa(j,t) +\, u } )_{u=-\Delta}^{\nabla} \;\dimR{L_\mathrm{Y}}
    \label{eq:chunk_obs}
\end{align}
where $\Delta:=\lfloor (L_\mathrm{Y}-1)/2\rfloor$ and $\nabla:=\lceil (L_\mathrm{Y}-1)/2\rceil$ correspond to symbols before and after transmission of the symbol $v_{j,t} = x_{\kappa(j,t)}$, respectively.
Next, collect the $L_\text{IC}$ symbols among $\mathbf{v}^{s-1}$ that are closest to $v_{j,t} = x_{\kappa(j,t)}$ in the vector
\begin{align}
   \overline{\mathbf{v}}_{j,t} :=  ( x_\kappa \mid \kappa \in  \mathcal{V}_{j,t} ) \; \dimR{L_\text{IC}} 
   \label{eq:chunk_ic}
\end{align}
where by ``closest" we mean
\begin{align}
    \mathcal{V}_{j,t} = \argmin_{\lvert \mathcal{U}\rvert = L_\text{IC}}   \sum\limits_{\substack{a \in \mathcal{U}\; \\ x_a \in \mathbf{v}^{s-1}}} |a - \kappa(j,t)| .
    \label{eq:set_tkappa}
\end{align}
Selecting the IC symbols~\eqref{eq:chunk_ic} is useful for symmetric impulse responses where $|g_\osidx|^2$ is nearly maximum at $\osidx=0$ and decreases with $|\osidx|$ beyond some threshold; see Fig.~\ref{fig:combined_filter_taps}. We use zero-padding to extend the vectors~\eqref{eq:chunk_obs} where necessary.

\subsubsection{Outputs}
After processing the $(\mathbf{r}^1_{j,t})_{j=s}^S$ in~\eqref{eq:r_vec} for all $t$, the RNN outputs APP estimates for the current stage $s$: 
\begin{align}
    Q_{V_{s,t}|\mathbf{Y},\mathbf{V}^{s-1}}(\,\cdot \mid \mathbf{y},\mathbf{v}^{s-1}) \quad  \forall t.
    \label{eq:app_estimate}
\end{align}

\subsection{RNN Structure}
\label{sec:rnnstructure}
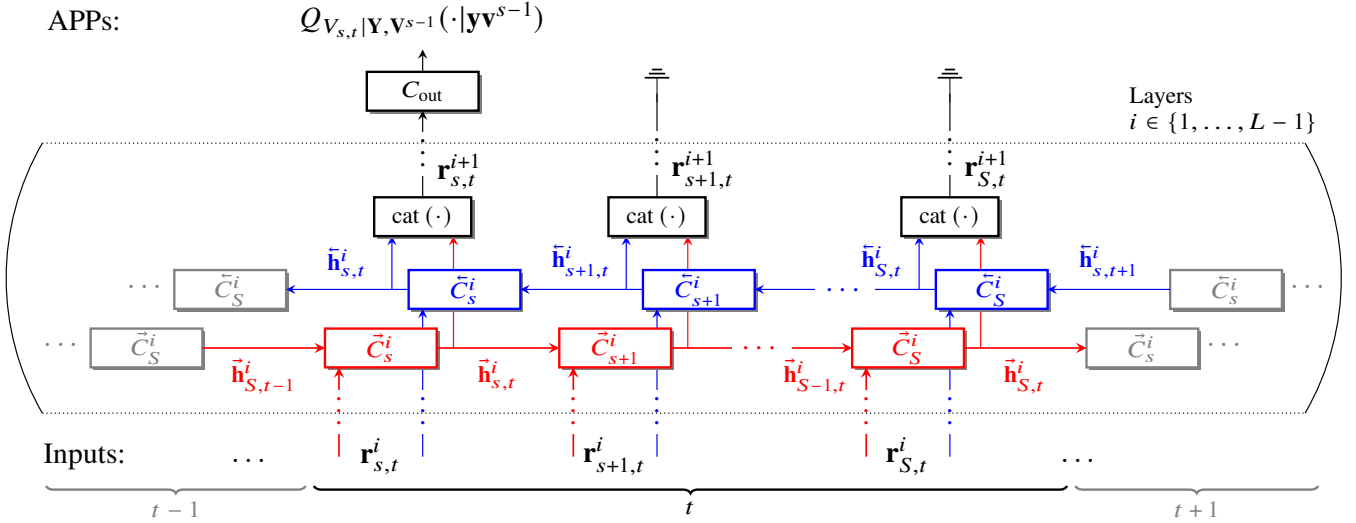
\begin{figure*}
    \centering
    \hspace*{-20pt}
    \resizebox{2.15\columnwidth}{!}
    {
    \tikzset{
block/.style = {drop shadow={shadow xshift=0.1em,shadow yshift=-0.1em,}, draw, thick, fill=white, fill opacity=1,rectangle, minimum height=1.32em, minimum width=3.8em,font=\footnotesize,outer sep=0,inner sep=0},
conblock/.style = {black,draw=black,drop shadow={shadow xshift=0.1em,shadow yshift=-0.1em,}, draw, thick, fill=white, fill opacity=1,rectangle, minimum height=1.32em, minimum width=3.3em,font=\footnotesize,outer sep=0,inner sep=0},
sum/.style= {draw, fill=white, circle, node distance=1cm,inner sep=0,outer sep=0},
input/.style = {coordinate},
output/.style= {coordinate},
pinstyle/.style = {pin edge={to-,thin,black}},
mydash/.style={anchor=center,fill=white,rectangle},
graystyle/.style={draw=gray,opacity=1,text=gray}
}

\pgfdeclarelayer{background}
\pgfdeclarelayer{foreground}
\pgfsetlayers{background,main,foreground}   %

\begin{tikzpicture}[auto, red, draw=red, node distance=3.0cm,>=stealth,line width=0.2mm]

    \node[input](input){}; 
    \node[block,right of=input,node distance=2cm](C1){$\Cfw{s}{i}$}; 
    \node[block,right of=C1,node distance=2.8cm](C2){$\Cfw{s+1}{i}$}; 
    \node[block,right of=C2,node distance=3.5cm](C3){$\Cfw{S}{i}$}; 
    
    \node[output,below of=C1,node distance=1.3cm](out1){}; 
    \node[output,below of=C2,node distance=1.3cm](out2){}; 
    \node[output,below of=C3,node distance=1.3cm](out3){}; 
    
    \def\stackshift{0.5cm} 
    \node[conblock,above of=C1,node distance=1.0cm,xshift=\stackshift,yshift=0.57cm](conblock1){$\vstack{(\cdot)}$}; 
    \node[conblock,above of=C2,node distance=1.0cm,xshift=\stackshift,yshift=0.57cm](conblock2){$\vstack{(\cdot)}$}; 
    \node[conblock,above of=C3,node distance=1.0cm,xshift=\stackshift,yshift=0.57cm](conblock3){$\vstack{(\cdot)}$}; 
    
    \node[output,above of=conblock1,node distance=0.7cm](output1){}; 
    \node[output,above of=conblock2,node distance=0.7cm](output2){}; 
    \node[output,above of=conblock3,node distance=0.7cm](output3){}; 
    
    \node[draw=black,black,block,above of=output1,node distance=0.8cm](app1){$C_\text{out}$}; 
    \node[black,above of=output2,node distance=0.9cm](app2){}; 
    \node[black,above of=output3,node distance=0.9cm](app3){}; 
    
    \node[block,left of=C1,node distance=2.8cm,graystyle](C0){$\Cfw{S}{i}$}; 
    \node[block,right of=C3,node distance=2.8cm,graystyle](C4){$\Cfw{s}{i}$}; 
    
    \node[left of=C0,node distance=1.0cm,graystyle,draw=none](){$\cdots$}; 
    \node[right of=C4,node distance=1.0cm,graystyle,draw=none](){$\cdots$}; 
    
    \draw[->] (out1)node[black]{$\mathbf{r}^i_{s,t}$}  
    ++(-\stackshift,0) node(inter1){} --node[mydash]{$\rvdots$} (C1.south -| inter1); 
    \draw[->] (out2)node[black]{$\mathbf{r}^i_{s+1,t}$} ++(-\stackshift,0) node(inter2){} --node[mydash]{$\rvdots$} (C2.south -| inter2);
    \draw[->](out3)node[black]{$\mathbf{r}^i_{S,t}$} ++(-\stackshift,0) node(inter3){} --node[mydash]{$\rvdots$} (C3.south -| inter3);
    
    \draw[->] (C0) -- node[below,font=\footnotesize]{$\hfw{S,t-1}{i}$} (C1);
    \draw[->] (C1) -- node[below,font=\footnotesize]{$\hfw{s,t}{i}$} (C2);
    \draw[->] (C2) --node[mydash,pos=.5] {$\dots$}  (C3) node[below ,xshift=-1.1cm,font=\footnotesize]{$\hfw{S-1,t}{i}$}; 
    \draw[->] (C3) --node[below,font=\footnotesize]{$\hfw{S,t}{i}$} (C4);
    
    \begin{pgfonlayer}{background}
        \draw[red,->] (C1.east) ++(0.2,0) node[](inter1){} -- (conblock1.south -| inter1);
        \draw[red,->] (C2.east) ++(0.2,0) node[](inter2){} -- (conblock2.south -| inter2);
        \draw[red,->] (C3.east) ++(0.2,0) node[](inter3){} -- (conblock3.south -| inter3);
    \end{pgfonlayer}

    \draw[black,->] (conblock1) node[above right,xshift=0.05cm,yshift=0.20cm]{$\mathbf{r}^{i+1}_{s,t}$} -- node[mydash,pos=0.5]{$\rvdots$} (app1); 
    
    \draw[black,-] (conblock2) node[above right,xshift=0.05cm,yshift=0.20cm]{$\mathbf{r}^{i+1}_{s+1,t}$} --node[mydash,pos=.425]{$\rvdots$} (app2); 
    \draw[black,-] (conblock3) node[above right,xshift=0.05cm,yshift=0.20cm]{$\mathbf{r}^{i+1}_{S,t}$} --node[mydash,pos=.425]{$\rvdots$} (app3);
    
    \draw[->,black] (app1) -- ++(0,0.5) node[above](appdesc){$Q_{V_{s,t}|\mathbf{Y},\mathbf{V}^{s-1}}(\cdot|\mathbf{y}\mathbf{v}^{s-1})$};
    
    \node[black,ground,line width=0.3pt,rotate around={180:(app2)},scale=0.7,yshift=0.15cm] at (app2.north){};
    \node[black,ground,line width=0.3pt,rotate around={180:(app3)},scale=0.7,yshift=0.15cm] at (app3.north){};

    \node[below left,text=black,yshift=0.1cm,xshift=-1.2cm] at(out1){$\cdots$};
    \node[below right,text=black,yshift=0.1cm,xshift=+1.7cm] at(out3){$\cdots$};

    \begin{pgfonlayer}{background}
    \begin{scope}[draw=blue,blue,yshift=0.70cm,xshift=2*\stackshift]
        \node[input](input){}; 
        \node[block,right of=input,node distance=2cm](C1){$\Cbw{s}{i}$}; 
        \node[block,right of=C1,node distance=2.8cm](C2){$\Cbw{s+1}{i}$}; 
        \node[block,right of=C2,node distance=3.5cm](C3){$\Cbw{S}{i}$}; 
        
        \node[block,left of=C1,node distance=2.8cm,graystyle](C0){$\Cbw{S}{i}$}; 
        \node[block,right of=C3,node distance=2.8cm,graystyle](C4){$\Cbw{s}{i}$}; 
        
        \node[left of=C0,node distance=1.0cm,graystyle,draw=none](){$\cdots$}; 
        \node[right of=C4,node distance=1.0cm,graystyle,draw=none](){$\cdots$};

        \draw[<-] (C0) --node[above,font=\footnotesize]{$\hbw{s,t}{i}$} (C1);
        \draw[<-] (C1) --  node[above,font=\footnotesize]{$\hbw{s+1,t}{i}$}
        (C2);
        \draw[<-] (C2) --node[mydash,pos=.5] {$\dots$} (C3) node[above ,xshift=-1.3cm,font=\footnotesize]{$\hbw{S,t}{i}$} ; 
        \draw[->] (C1.west) ++(-0.2,0) node[](inter1){} -- (conblock1.south -| inter1);
        \draw[->] (C2.west) ++(-0.2,0) node[](inter2){} -- (conblock2.south -| inter2);
        \draw[->] (C3.west) ++(-0.2,0) node[](inter3){} -- (conblock3.south -| inter3);

        \draw[<-] (C3) --node[above,font=\footnotesize] {$\hbw{s,t+1}{i}$} (C4);
        
        \draw[->] (out1) ++(\stackshift,0) node(inter1){} --node[pos=.3,mydash]{$\rvdots$} (C1.south -| inter1); 
        \draw[->] (out2) ++(\stackshift,0) node(inter2){} --node[pos=.3,mydash]{$\rvdots$} (C2.south -| inter2); 
        \draw[->] (out3) ++(\stackshift,0) node(inter3){} --node[pos=.3,mydash]{$\rvdots$} (C3.south -| inter3); 
    \end{scope}
    \end{pgfonlayer}

\draw [
    black,
    thick,
    decoration={
        brace,
        mirror,
        raise=0.5cm
    },
    decorate,
    font=\footnotesize
] ($ (out1) + (-0.8,+0.15) $) -- ($ (out3) + (1.9,+0.15) $)
node [pos=0.5,anchor=north,yshift=-0.55cm] {$t$}; 

\draw [
    gray,
    thick,
    decoration={
        brace,
        mirror,
        raise=0.5cm
    },
    decorate,
    font=\footnotesize
] ($ (out1) + (8.3,+0.15) $) -- ($ (out3) + (4.9,+0.15)$) node [pos=0.5,anchor=north,yshift=-0.55cm] {$t+1$}; 

\draw [
    gray,
    thick,
    decoration={
        brace,
        mirror,
        raise=0.5cm
    },
    decorate,
    font=\footnotesize
] ($ (out1) + (-4.0,+0.15) $) -- ($ (out3) + (-7.2,+0.15) $)
node [pos=0.5,anchor=north,yshift=-0.55cm] {$t-1$};

\draw [black] ($(C0 -| out1) + (-1.1,0.05)$) ++(150:3.4) node(arclstart){} arc (150:210:3.23) node(arclend){};
\draw [black] ($(C0 -| out1) + (8.1,0.05)$) ++(30:3.4) node(arcrstart){}  arc (30:-30:3.23) node(arcrend){};
\node[black,yshift=-0.33cm,xshift=-0.85cm,text width=2.5cm,font=\footnotesize] at (arcrstart |- app3){Layers \\ $i \in \{1,\ldots,\maxlay-1\}$};

\draw[black,densely dotted] (arclstart.center) -- (arcrstart.center); 
\draw[black,densely dotted] (arclend.center) -- (arcrend.center); 
   
\node [black,xshift=0.5cm] at(arclstart |- out3){Inputs:};
\node [black,xshift=0.5cm] at(arclstart |- appdesc){APPs:};

\end{tikzpicture}
    }
    \caption{Bidirectional time-varying RNN for stage $s$.}
    \label{fig:tvrnn_architecture}
\end{figure*}

The RNN processes its inputs in forward and backward chronological orders:
\begin{align}
\begin{array}{llllll}
    \ldots & \mathbf{r}^1_{s,t}, & \mathbf{r}^1_{s+1,t}, & \ldots, & \mathbf{r}^1_{S,t}, & \\
    & \mathbf{r}^1_{s,t+1},& \mathbf{r}^1_{s+1,t+1},& \ldots,& \mathbf{r}^1_{S,t+1}& \ldots
    \end{array}
    .
    \label{eq:unrolled_input}
\end{align}
Stage $s$ processes a total of $N (S-s+1)$ inputs; see Fig.~\ref{fig:tvrnn_architecture}.

The process producing~\eqref{eq:unrolled_input} is stationary. This means that for $s>1$, the sub-process for stages $s$ to $S$  in~\eqref{eq:unrolled_input} is cyclostationary with period {$\Gamma = S-s+1$} due to the SIC partitioning of Sec.~\ref{sec:sic}. We now encounter the following issue. Classic RNNs~\cite[Chap.~10.3]{goodfellow2016deep} use time-invariant input and state processing, but such RNNs can perform poorly with cyclostationary inputs; see~\cite[p.~390]{goodfellow2016deep}. Motivated by the FBA state recursions~\eqref{eq:f_fw}-\eqref{eq:f_bw}, which are periodically time-varying for $s>1$ with period $\Gamma$, we allow periodically time-varying input processing and state recursions in the RNN. 
For $s=1$, the process~\eqref{eq:unrolled_input} is stationary and the FBA recursions~\eqref{eq:f_fw}-\eqref{eq:f_bw} are time-invariant; in this case we may use the classic RNNs~\cite[Chap.~10.3]{goodfellow2016deep} for which $\Gamma = 1$.
Fig.~\ref{fig:tvrnn_architecture} shows the structure of such a bidirectional RNN with $\maxlay$ layers: $\maxlay-1$ layers are recurrent with forward and backward paths, and the last layer is feedforward. 

Consider the forward path in layer $i$. This path has cells $\Cfw{s}{i}, \Cfw{s+1}{i}, \ldots, \Cfw{S}{i}$ that repeat periodically for $t=1,\ldots,N$, just like the node in the factor graph in Fig.~\ref{fig:fba}. Compared to classic RNNs, the number of RNN parameters increases by a factor of $\Gamma$, but the computational complexity remains the same~\cite[Chap.~10.3]{goodfellow2016deep}. 
RNNs have internal states $(\fw{\mathbf{h}}^i_{j,t})_{j=s}^S$ in the forward and backward paths, where $t= 1,\ldots, N$.

To illustrate further, let the input dimensions of the recurrent layers and the output layer be $(\lay_1, \ldots, \lay_{\maxlay-1}, \lay_\maxlay)$, where $\lay_1 = L_\mathrm{Y} + L_\text{IC}$. We convert the pairs~\eqref{eq:r_vec} to column vectors. The state recursion via the cell $\Cfw{j}{i}$, $j = s,\ldots,S$, is
\begin{align}
    \hfw{j,t}{i} = f\big(
    \Wfw{\text{in},{j}}{i}( \mathbf{r}_{j,t}^i)  + \Wfw{j-1}{i}( \hfw{j-1,t}{i} )\big)
    \; \dimR{\lay_{i+1}/2}
\end{align}
where $f(\cdot)$ is the per-entry rectified linear unit (ReLU)~\cite{fukushima1969relu} and we use the mapping~\eqref{eq:paridx_map} and $s-1 \mapsto S$ for the parallel indexing. The input and state maps for $j = s,\ldots,S$ are
\begin{align*}
    &\Wfw{\text{in},j}{i}: \mathbb{R}^{\lay_{i}} \mapsto \mathbb{R}^{\lay_{i+1}/2}, && 
    \Wfw{\text{in},j}{i}(\mathbf{r}_{j,t}^i) = \fw{\mathbf{W}}_{\text{in},j}^i \mathbf{r}_{j,t}^i +  \fw{\mathbf{b}}^i_{\text{in},j}
    \\
    &\Wfw{j}{i}: \mathbb{R}^{\lay_{i+1}/2} \mapsto \mathbb{R}^{\lay_{i+1}/2}, && \Wfw{j}{i}(\hfw{j,t}{i}) = \fw{\mathbf{W}}^i_j \hfw{j,t}{i} +  \fw{\mathbf{b}}^i_j
\end{align*}
where $\fw{\mathbf{W}}_{\text{in},j}^i$, $\fw{\mathbf{W}}_j^i$ are input and state recursion matrices, and $\fw{\mathbf{b}}_{\text{in},j}^i$, $\fw{\mathbf{b}}^i_j$ are input and state bias vectors, respectively. The backward path works analogously; see Fig.~\ref{fig:tvrnn_architecture}. 

To complete the recurrent layer processing, the outputs of the two paths are concatenated for $j = s,\ldots,S$ and all $t$:  
\begin{align}
    \mathbf{r}_{j,t}^{i+1}
    = \vstack( \hfw{j,t}{i}, \hbw{j,t}{i}) \;\;\dimR{\lay_{i+1}}
    \label{eq:rnn_output_concat}
\end{align}
and \eqref{eq:rnn_output_concat} is passed to the next recurrent layer if $i+1 < \maxlay$. If $i + 1 = \maxlay$, the last cell $C_\text{out}$ performs final processing, i.e., for all $t$ and fixing $j=s$ we have
\begin{align}
    Q_{V_{s,t}|\mathbf{Y}, \mathbf{V}^{s-1}}(\cdot|\mathbf{y},\mathbf{v}^{s-1}) = 
    \phi( 
    \mathbf{W}_\text{out} \mathbf{r}_{s,t}^{\maxlay} + \mathbf{b}_\text{out}) \;\;\dimR{\lvert \mathcal{A} \rvert}
    \label{eq:app_sic_s_rnn}
\end{align}
where $\mathbf{W}_\text{out} \dimR{|\mathcal{A}| \times \lay_\maxlay}$, $\mathbf{b}_\text{out} \dimR{|\mathcal{A}|}$,
and the ``softmax'' function $\phi(\cdot)$~\cite{bishop2006pattern} generates a PMF interpreted as symbol-wise APPs. To initialize the RNN, we set the first forward and last backward states in all recurrent layers to zero.

\subsection{Achievable Rates and NN Optimization}
The NN-equalizers approximate APPs via~\eqref{eq:app_sic_s_rnn}, which gives a lower bound on the SIC rate (see~\cite[Sec.~VI~D.]{prinz2023successive},~\cite{arnoldsimulationmi}):
\begin{align}
    I_{n,\text{SIC}} \geq
    \frac{1}{S} \sum_{s=1}^S \! \underbrace{\frac{1}{\siclength}  \sum_{t=1}^N \! H(V_{s,t}) + \mathbb{E}[\log_2 Q(V_{s,t} | \bm{Y},\bm{V}^{s-1})]}_{\textstyle := I^{s}_{q,N\!,\text{SIC}}}
    \label{eq:lb_sic_rate}
\end{align}
where the expectation is over the actual $p_{V_{s,t}, \mathbf{Y}, \mathbf{V}^{s-1}}$. The expression $I^{s}_{q,N\!,\text{SIC}}$ is the  mismatched rate of SIC stage $s$ and we define the limiting rate as $I^{s}_{q,\text{SIC}} := \lim_{N\rightarrow \infty} I^{s}_{q,N\!,\text{SIC}}$.

We wish to maximize $I^{s}_{q,N\!,\text{SIC}}$, or equivalently minimize the expectation in~\eqref{eq:lb_sic_rate} (a cross entropy) for each SIC stage $s$. To simplify notation, collect the NN parameters (the NN weights and biases) in a string $\mathbf{\Theta}$. 
We estimate $I^{s}_{q,N\!,\text{SIC}}$ via simulation and formulate the optimization problem~\cite[Sec.~4.1]{bocherer2022mlcomm} as
\begin{align}
    &\argmin_{(Q_{V_{s,t} | \mathbf{Y}, \mathbf{V}^{s-1}})_{t=1}^N}
    - 
    \frac{1}{\siclength} \sum_{t=1}^\siclength 
    \left\langle\,  \log_2 Q_{V_{s,t}|\mathbf{Y}, \mathbf{V}^{s-1}}(v_{s,t} \mid \mathbf{y},  {\mathbf{v}^{s-1}}) 
    \,\right\rangle 
    \nonumber
    \\
    &\quad \textrm{subject to }
    (Q_{V_{s,t}|\mathbf{Y}, \mathbf{V}^{s-1}})_{t=1}^N = f_\text{NN}(\mathbf{y} , {\mathbf{v}^{s-1}}  ; \mathbf{\Theta})
    \label{eq:nn_optim}
\end{align}
where $\langle \cdot \rangle$ denotes Monte-Carlo averaging over $N_\text{blk}$ transmit symbols and pairs $(\mathbf{v}^{(w)},\,\mathbf{y}^{(w)})_{w=1}^{N_\text{blk}}$, and the APPs are a function of the NN $f_\text{NN}(\mathbf{y}, \mathbf{v}^{s-1}; \mathbf{\Theta})$.
The solution to~\eqref{eq:nn_optim} is usually approximated by batch stochastic gradient descent (SGD).

\subsection{Receiver Complexity}
Table~\ref{tab:complexity} compares the number of multiplications per symbol APP calculation for the FBA, bit-wise GS~\cite{prinz2023successive}, and the NN. 
The FBA complexity is exponential in $\widetilde{K}$~\eqref{eq:total_memory}. In contrast, the GS complexity is quadratic in $\widetilde{K}$ and linear in the number of bits per source symbol $m$, iterations $N_\text{iter}$, and parallel samplers $N_\text{par}$. Matrix-vector multiplications dominate the NN complexity, which depends on the number of NN layers and their sizes $(\lay_i)_{i=1}^\maxlay$.
\begin{table}
\caption{Algorithmic complexity per APP estimate.}
\centering
{\renewcommand{\arraystretch}{1.0}
\begin{tabular}{ll} 
\toprule
\textbf{Algorithm} & \textbf{Multiplications}  \\ \midrule
FBA~\cite{plabst2022achievable} & $\mathcal{O}(S \cdot \lvert \mathcal{A} \rvert^{\widetilde{K} + 1})$\\ Bit-wise GS~\cite{prinz2023successive} & $\mathcal{O}(S \cdot \widetilde{K}^2 \cdot m \cdot N_\text{iter} \cdot N_\text{par})$\\
NN & $\mathcal{O}\big(S \cdot \big(\sum_{i=1}^{\maxlay-1} \lay_{i}\lay_{i+1} + \lay_{i+1}^2/2  + \lay_\maxlay \cdot |\mathcal{A}|  \big)\big)$\\
\bottomrule
\end{tabular}}
\label{tab:complexity}
\end{table}%
Note that we use real-valued NN inputs. For complex-valued inputs~\eqref{eq:r_vec}, composite real representations must be used, doubling the dimensions $L_\mathrm{Y}$ and $L_\text{IC}$. 
The FBA and GS also require complex multiplications when the modulation format and filters are complex-valued. We do not consider these cases in  Table~\ref{tab:complexity}.

\section{Numerical Results}
\label{sec:numerical}
We study short-haul optical fiber links with a SLD, as described in~Sec.~\ref{sec:fiber_optic_application}. We compare the mismatched FBA, GS, and NN equalizers for bipolar and complex modulation alphabets; see~\cite{prinz2023successive,plabst2022achievable,tasbihi2021direct,mecozzi_capacity_amdd_2018}. The program code is available at~\cite{nnsiccode2024plabst}.

\subsection{System Parameters}
The system parameters are listed in Table~\ref{tab:simparams}. The DAC performs sinc pulse shaping at the symbol rate \SI{35}{\giga Bd}. We consider $L_\text{fib} = 0$ (back-to-back) and $L_\text{fib} = \SI{30}{\kilo\meter}$ of SSMF without fiber attenuation. The $L_\text{fib} = \SI{0}{}$ model is reasonable for several kilometers of fiber in the O-band where chromatic dispersion is low; see~\cite[Chap.~2.6.2]{schneider2004nonlinear}. 
The pulse shape and dispersion introduce long memory in the combined channel $g(t)$; see Fig.~\ref{fig:combined_filter_taps}.
The SLD doubles the bandwidth of the signal $X(t)$ to $2B$. The receive filter $h(t)$ is a sinc pulse with bandwidth $2B$, and oversampling $Y(t)$ with $N_\text{os} = 2$ provides sufficient statistics. 

We approximate $g(t)$ by a discrete filter $g_k$ with $K_g =  151 \cdot N_\text{sim} +1$ taps, resulting in a memory of $\widetilde{K}_g = 151$ symbols. This choice has $g_k$ containing $99.9\%$ of the energy of $g(t)$. The receive filter $h(t)$ does not influence the transmit signal component, and the discrete-time noise $N_k$ is AWGN with variance $\sigma^2$. We thus have $\widetilde{K}_h = 0$ and the total system memory~\eqref{eq:total_memory} is $\widetilde{K} = \widetilde{K}_g$.
We simulate blocks with $n \geq \SI{20e3}{}$ symbols.
The average transmit power is %
\begin{align}
    P_\text{tx} = \frac{\mathbb{E}\left[ \lVert X(t) \rVert^2 \right]}{n \cdot  T_\mathrm{s}}
\end{align} 
and the filtered noise variance is $\sigma^2 = 1$ so $\text{SNR}=P_\text{tx}$.
We use the $M$-ary modulations: 
\begin{itemize}
    \item unipolar $M$-PAM with $\mathcal{A}=\{0,1,\ldots,2^m-1\}$;
    \item bipolar $M$-ASK with $\mathcal{A}=\{\pm1,\pm3,\ldots \pm (2^m-1)\}$;
    \item $M$-SQAM~\cite{plabst2022achievable} with $\mathcal{A} = \left\{ \pm a, \pm \mathrm{j}a \mid a=1,2,\ldots,M/4  \right\}$
\end{itemize}
where SQAM refers to star-QAM.
We use differential phase encoding before the DAC to help resolve phase ambiguities; see~\cite[Appendix]{prinz2023successive}. 
The spectral efficiency in $\SI{}{bit\per\second\per\hertz}$ corresponds to the information rate in bpcu because the time-bandwidth product of the sinc pulse is 1.
\begin{table}
\caption{Short-reach fiber-optic system~\cite[Sec. II A]{plabst2022achievable}.}
\centering
{\renewcommand{\arraystretch}{0.85}
\begin{tabular}{ll} 
\toprule
\textbf{Parameters} &  \textbf{}   \\\midrule
Fiber length & $L_\text{fib} \in \{0, \SI{30}{\kilo\meter}\}$ \\
Attenuation factor & \SI{0}{dB \per\kilo\meter}  \\
Carrier wavelength & \SI{1550}{\nano\meter} (C band transmission)\\
Group velocity dispersion & $\beta_2=\SI{-2.168e-23}{\second^2\per\kilo\meter}$\\
Symbol rate & $B = \SI{35}{\giga Bd}$ \\
\midrule
DAC and SSMF  &$g(t) = B\cdot \mathrm{sinc}(Bt) * g_\text{SSMF}(t)$ \\
Frequency response of SSMF &$G_\text{SSMF}(f) = \exp{(\mathrm{j}\, \beta_2/2 (2\pi f)^2 L_\text{fib})}$\\
Receive photo-diode (SLD) &$\nonlp{x} = \lvert x \rvert^2$ \\
Real-valued post-SLD AWGN & $N'(t)$, $\text{PSD} = N_0/2$\\
Receive filter &$h(t) = 2B \,\mathrm{sinc}(2Bt)$\\
Oversampling factors & $N_\text{os} = N_\text{sim} = 2$\\
Filtered  and sampled noise & $N_k$, AWGN with variance $\sigma^2 =  N_0 B$ \\
\bottomrule
\end{tabular}
}
\label{tab:simparams}
\end{table}
\subsection{NN Optimization}
\label{sec:nnoptim}

We optimize one NN per SIC stage and SNR. The NNs are initialized with optimized parameters from a lower SNR, if available\footnote{The latest program version~\cite[v1.1]{nnsiccode2024plabst} applies learning rate scheduling to train NNs faster and without initialization from a previous SNR.}.
We approximate~\eqref{eq:nn_optim} via batch SGD with moment extension, namely ADAM~\cite{kingma2017adam}. RNNs may experience numerical instabilities during SGD when the number $N (S-s+1)$ of sequential inputs~\eqref{eq:unrolled_input} is large~\cite{pascanu2013difficulty}. During training, we limit the inputs to $T_\text{RNN}$, which results in $N = T_\text{RNN}/(S-s+1)$ symbols per SIC stage. The number of inputs $T_\text{RNN}$ and size $L_\text{Y}$ indicate the maximum symbol memory that the RNN can capture and is approximately $\widetilde{N}_\text{RNN} \approx \lfloor L_\text{Y}/N_\text{os} \rfloor + (T_\text{RNN}-1)$.

We perform batch SGD with $N_\text{batch}$ inputs and the strings $(\mathbf{v}^{(w)}, \mathbf{y}^{(w)})_{w=1}^{N_\text{batch}}$ where $N = T_\text{RNN}/(S-s+1)$. We take $N_\text{iter}$ gradient steps with step size\footnote{The decay rates for the gradient moment estimates are chosen as in~\cite{kingma2017adam}.} $\beta_\text{lr}$. To improve numerical stability at high SNR, the NN inputs~\eqref{eq:r_vec} are normalized to unit variance.  
We use Monte-Carlo simulations to evaluate NN-SIC rates with $N_\text{blk}$ frames, each having $n$ symbols; see~\eqref{eq:lb_sic_rate}-\eqref{eq:nn_optim}. The NN parameters are found empirically; see Table~\ref{tab:nnparams}. The symbols used for the NNs are summarized in Tab.~\ref{tab:listofsymbols}. Table~\ref{tab:complexity_application} and Fig.~\ref{fig:complexity} show the algorithmic complexities for the examples of Sec.~\ref{subsec:SIC-Rates} by listing the number of multiplications for the FBA, GS, and NNs for stage $s=1$.

\begin{table*}
\centering
\caption{NN Parameters.}
\label{tab:nnparams}
\renewcommand{\arraystretch}{0.4} 

\pgfplotstableset{
    columns={modulation,specify,layer,nrnn,Ls,Tr,B,V,n,lr,I},
    font={\footnotesize},
    col sep = comma,
    every head row/.style={before row=\toprule,after row=\midrule},
    every last row/.style={after row=\bottomrule},
    columns/modulation/.append style={column type = {l},string type,column name={\textbf{Modulation}}},
    columns/specify/.append style={column type = {l},string type,column name={\textbf{}}},
    columns/layer/.append style={column type = {l},string type,string replace*={|}{, },string replace*={[}{}, string replace*={]}{},  string replace*={, A}{}},
    columns/nrnn/.append style={column name={$\widetilde{N}_\text{RNN}$},string replace*={nrnn=}{}},
    columns/lr/.append style={column name={$\beta_\text{lr}$},string replace*={lr=}{}},
    columns/Tr/.append style={column name={$T_\text{RNN}$},string replace*={Tr=}{}},
    columns/B/.append style={column name={$N_\text{batch}$},string replace*={B=}{}},
    columns/S/.append style={column name={$S$},string replace*={S=}{}},
    columns/Ls/.append style={string replace*={Ls=}{},column name={$L_\text{IC}$}},
    columns/I/.append style={sci,column name={$N_\text{iter}$},string replace*={I=}{}},
    columns/V/.append style={sci,column name={$N_\text{blk}$},string replace*={V=}{}},
    columns/n/.append style={sci,column name={$n$},string replace*={n=}{}},
    columns/L/.append style={column name={$L_\text{fib}\,\mathrm{[km]}$},string replace*={L=}{},multiply by=1e-3},
    columns/Rs/.append style={column name={$B\,\mathrm{[GBd]}$},string replace*={Rs=}{},multiply by=1e-9},
    columns/C/.append style={sci,column name={$C_\text{mul}$},string replace*={C=}{},string replace*={.txt}{}},
}

\begin{filecontents*}{params_all.csv}
ver,modulation,specify,layer,nrnn,lr,Tr,B,S,Ls,I,V,n,P,a,L,Rs,C
v1.5,$M=4$,|,[32|64|A],nrnn=47,lr=1E-03,Tr=32,B=128,S=2,Ls=16,I=1.0E+04,V=1.0E+03,n=6.0E+04,P=151,a=0,L=0.0E+00,Rs=3.5E+10,C=5.4E+03.txt
v1.5,$M=8$,|,[64|128|64|A],nrnn=95,lr=5E-04,Tr=64,B=64,S=2,Ls=32,I=6.0E+04,V=3.0E+03,n=6.0E+04,P=151,a=0,L=0.0E+00,Rs=3.5E+10,C=3.1E+04.txt
v1.5,$M=16$,|,[84|128|128|A],nrnn=105,lr=2E-04,Tr=64,B=64,S=2,Ls=64,I=6.0E+04,V=3.0E+03,n=6.0E+04,P=151,a=0,L=0.0E+00,Rs=3.5E+10,C=5.4E+04.txt
v1.5,$M=32$,|,[84|128|128|A],nrnn=105,lr=1E-04,Tr=64,B=64,S=2,Ls=64,I=6.0E+04,V=3.0E+03,n=6.0E+04,P=151,a=0,L=0.0E+00,Rs=3.5E+10,C=5.6E+04.txt
v1.5,$M=64$,|,[84|128|128|128|64|A],nrnn=125,lr=5E-05,Tr=84,B=84,S=2,Ls=100,I=7.5E+04,V=3.0E+03,n=6.0E+04,P=151,a=0,L=0.0E+00,Rs=3.5E+10,C=9.5E+04.txt
v1.5,$M=128$,|,[84|128|128|128|128|A],nrnn=125,lr=5E-05,Tr=84,B=84,S=2,Ls=100,I=7.5E+04,V=3.0E+03,n=6.0E+04,P=151,a=0,L=0.0E+00,Rs=3.5E+10,C=1.2E+05.txt
v1.5,$M=4$,PAM/ASK,[64|128|64|A],nrnn=95,lr=5E-04,Tr=64,B=128,S=2,Ls=32,I=2.0E+04,V=1.0E+03,n=6.0E+04,P=151,a=0,L=3.0E+04,Rs=3.5E+10,C=3.1E+04.txt
v1.5,$M=4$,SQAM,[64|256|128|128|A],nrnn=95,lr=5E-04,Tr=64,B=128,S=2,Ls=32,I=2.0E+04,V=1.0E+03,n=6.0E+04,P=151,a=0,L=3.0E+04,Rs=3.5E+10,C=1.3E+05.txt
v1.5,$M=8$,PAM/ASK,[64|128|128|A],nrnn=115,lr=3E-04,Tr=84,B=128,S=2,Ls=32,I=5.0E+04,V=1.0E+03,n=6.0E+04,P=151,a=0,L=3.0E+04,Rs=3.5E+10,C=4.6E+04.txt
v1.5,$M=8$,SQAM,[64|256|128|128|A],nrnn=115,lr=3E-04,Tr=84,B=128,S=2,Ls=32,I=5.0E+04,V=1.0E+03,n=6.0E+04,P=151,a=0,L=3.0E+04,Rs=3.5E+10,C=1.3E+05.txt
v1.5,$M=16$,|,[84|200|128|128|A],nrnn=161,lr=5E-05,Tr=120,B=64,S=6,Ls=64,I=8.0E+04,V=3.0E+03,n=8.0E+04,P=151,a=0,L=3.0E+04,Rs=3.5E+10,C=1.1E+05.txt
v1.5,$M=32$,|,[100|200|200|200|168|A],nrnn=169,lr=4E-05,Tr=120,B=64,S=6,Ls=64,I=1.0E+05,V=7.0E+03,n=8.0E+04,P=151,a=0,L=3.0E+04,Rs=3.5E+10,C=2.3E+05.txt
v1.5,$M=64$,|,[100|300|300|300|240|A],nrnn=169,lr=4E-05,Tr=120,B=64,S=6,Ls=100,I=1.0E+05,V=7.0E+03,n=8.0E+04,P=151,a=0,L=3.0E+04,Rs=3.5E+10,C=4.9E+05.txt
\end{filecontents*}
\pgfplotstabletypeset[
every row no 6/.style={before row=\midrule},  %
columns={modulation,specify,layer,nrnn,Ls,Tr,B,V,n,lr,I,L},
columns/layer/.append style={column name={$L_\mathrm{Y},\ell_2,\ell_3,\ell_4,\ell_5$}}]{params_all.csv}
\end{table*}

\begin{table}
\caption{List of NN Symbols.}
\centering
{\renewcommand{\arraystretch}{0.95}
\setlength{\tabcolsep}{0.5em} 
\begin{tabular}{lll} 
\toprule
\textbf{Description} &  \textbf{Symbol} &  Occurrence  \\\midrule
Number of layers & $L$  & Fig.~\ref{fig:tvrnn_architecture}\\
Number of channel outputs & $L_\mathrm{Y}$   & \eqref{eq:chunk_obs}  \\
Number of IC symbols & $L_\mathrm{IC}$  & \eqref{eq:chunk_ic}  \\
Input size of first layer & $\ell_1 = L_\mathrm{Y} + L_\mathrm{IC}$  & \eqref{eq:r_vec}, Sec.~\ref{sec:rnnstructure}    \\
Input size of layer $i$, $1 \leq i \leq L$  & $\ell_i$  &  Sec.~\ref{sec:rnnstructure}\\
Batch size for SGD & $N_\text{batch}$ & Sec.~\ref{sec:nnoptim} \\
Step size for SGD & $\beta_\text{lr}$ & Sec.~\ref{sec:nnoptim} \\
Number of SGD iterations & $N_\text{iter}$ & Sec.~\ref{sec:nnoptim} \\
Number of sequential training inputs & $T_\text{RNN}$ &  Sec.~\ref{sec:nnoptim} \\
Time-varying period of NN & $\Gamma$ & Sec.~\ref{sec:rnnstructure}\\
Approximate NN memory & $\widetilde{N}_\text{RNN}$ & Sec.~\ref{sec:nnoptim} \\
\midrule
Number of frames for validation & $N_\text{blk}$ & Sec.~\ref{sec:nnoptim} \\
Number of symbols per frame & $n$ &  Sec.~\ref{sec:nnoptim}\\
\bottomrule
\end{tabular}
}
\label{tab:listofsymbols}
\end{table}
\subsection{SIC Rates}
\label{subsec:SIC-Rates}
We plot the SIC rates in bits per channel use (bpcu) for the mismatched
FBA~\cite{plabst2022achievable}, GS~\cite{prinz2023successive}, and NN detectors. In general, the FBA must use a mismatched channel memory $\widetilde{N} \ll \widetilde{K}$ because its complexity grows exponentially in $\widetilde{K}$; see Tab.~\ref{tab:complexity}. Similarly, we use a mismatched memory $\widetilde{N} < \widetilde{K}$ for GS.

\begin{figure}[!t]
    \centering
    \input{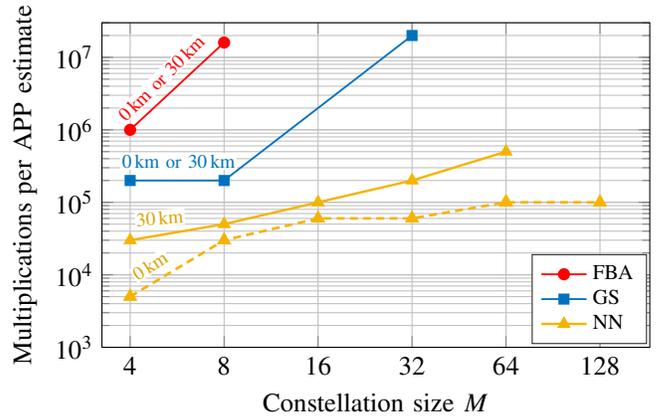}

\begin{tikzpicture}
	\begin{axis}[
    scale only axis,
    width=1.55*\gridwidth,
    height=1*\gridheight,
    xmode=log,
    ymode=log,
    xmajorgrids,
    xminorgrids,
    ymajorgrids,
    yminorgrids,
    yminorticks=true,
    grid=both,
    xmin=2^1.7,
    ymin=1E3,
    ymax=3E7,
    ytick={1E3,1E4,1E5,1E6,1E7,1E8},
    xtick={4,8,16,32,64,128},
    xticklabels={4,8,16,32,64,128},
    xlabel={Constellation size $M$},
    ylabel={Multiplications per APP estimate},
    legend style={legend cell align=left,  draw=white!15!black, font=\footnotesize,  
    at={(0.995,0.01)},anchor=south east,row sep=-1.5pt},
    ]

    \addplot[FBA,ASK,mark=*,solid,  mark size=1.8pt, mark options={line width=1pt,solid},line width=0.7pt] coordinates { 
    (2^2,1E6) %
    (2^3,16E6) %
    } node [pos=0.4, above=4pt,rotate=45,opacitylabel,font=\scriptsize,] {$\SI{0}{\kilo\meter}$ or $\SI{30}{\kilo\meter}$}; 
    \addlegendentry{FBA};

    \addplot+[GS,PAM,mark=square*, solid, mark size=1.6pt, mark options={line width=1pt,solid},line width=0.7pt] coordinates { 
    (2^2,2E5) %
    (2^3,2E5) %
    (2^5,2E7) %
    } node [pos=0.065, above=4pt,rotate=0,opacitylabel,font=\scriptsize,] {$\SI{0}{\kilo\meter}$ or $\SI{30}{\kilo\meter}$}; 
    \addlegendentry{GS};
    
    \addplot+[RNN,forget plot, mark=triangle*,densely dashed, mark size=2pt, mark options={line width=1.2pt,solid},line width=1pt] coordinates { 
    (2^2,5E3) %
    (2^3,3E4) %
    (2^4,6E4) %
    (2^5,6E4) %
    (2^6,1E5) %
    (2^6,1E5) %
    (2^7,1E5) %
    } node [pos=0.1, above=3pt,rotate=33,opacitylabel,font=\scriptsize,text=mycolorrnn!90!black] {$\SI{0}{\kilo\meter}$};

    \addplot+[RNN,mark=triangle*,solid, mark size=2pt, mark options={line width=1pt,solid},line width=0.9pt] coordinates { 
    (2^2,3E4) %
    (2^3,5E4) %
    (2^4,1E5) %
    (2^5,2E5) %
    (2^6,5E5) %
    } node [pos=0.07, above=3pt,rotate=4,opacitylabel,font=\scriptsize,text=mycolorrnn!90!black] {$\SI{30}{\kilo\meter}$}; 
    \addlegendentry{NN};

	\end{axis}
\end{tikzpicture}
    \caption{Algorithmic complexity versus constellation size for SSMF lengths $L_\text{fib} = \SI{0}{\kilo\meter}$ and $L_\text{fib} = \SI{30}{\kilo\meter}$ .}
    \label{fig:complexity}
\end{figure}

Consider 4-PAM/ASK/SQAM with FBA-SIC and NN-SIC. We compare rates with $L_\text{fib} = \SI{0}{}$ and $L_\text{fib} = \SI{30}{\kilo\meter}$ SSMF in Fig.~\ref{fig:Q4_L=0km} and Fig.~\ref{fig:Q4_L=30km}, respectively. We compute the JDD upper bounds (UBs) of~\cite[Eq.~(45)]{arnoldsimulationmi} by using the mismatched FBA with $\widetilde{N} = 9$, and where the auxiliary channel is optimized according to~\cite[Sec.~III~C]{plabst2022achievable}.
The rates for bit-wise GS~\cite[Fig.~8]{prinz2023successive} with $\widetilde{N} = 9$, $N_\text{par} = 20$, and $N_\text{iter} = 60$ are similar to the FBA-SIC rates. 
Observe from Table~\ref{tab:complexity_application} and Fig.~\ref{fig:complexity} that the algorithmic complexity remains large for the NN-based approach, which motivates looking for simplifications or alternative algorithms.

{
\setlength{\tabcolsep}{3.5pt}
\begin{table}
\caption{Number of multiplications per APP estimate.}
\newcommand*{\formatnum}[1]{%
    \pgfmathprintnumber[
        sci,
        precision=0,
        sci zerofill=true,
        ]{#1}}%
\centering
{
\newcommand{\upbar}{\tikz[overlay] \draw (0,0.6em)--(0,0em);}
\newcommand{\downbar}{\tikz[overlay] \draw (0,0.5em)--(0,-0.65em);}
\renewcommand{\arraystretch}{0.65}
\begin{tabular}{lcl lll} 
\toprule
\multicolumn{3}{l}{\textbf{Parameters}} & \textbf{FBA} & \textbf{GS} & \textbf{NN} (real/complex)\\\midrule
$M=4$ & $L_\text{fib}=\SI{0}{\kilo\meter}$ & Fig.~\ref{fig:Q4_L=0km}  & \ok{\formatnum{1048576}} & \ok{\formatnum{2E5}} & \ok{\formatnum{5.4E+03}} \\
$M=8$ &  \downbar &  Fig.~\ref{fig:Q8-Q128_L=0km} & $-$  & $-$  &
\ok{\formatnum{3.1E+04}} \\
$M=16$ &  \downbar &  Fig.~\ref{fig:Q8-Q128_L=0km} & $-$  & $-$  &  \ok{\formatnum{6.2E+04}}  \\
$M=32$ &  \downbar &  Fig.~\ref{fig:rnn_vs_gibbs_q32},\ref{fig:Q32_L=0km_SIC_stage_rate} & $-$  & \ok{\formatnum{1.8E7}} & \ok{\formatnum{5.6E+04}} \\
$M=64$ &  \downbar &  Fig.~\ref{fig:Q8-Q128_L=0km} & $-$  & $-$  & \formatnum{9.5E+04} \\
$M=128$ &  \upbar &  Fig.~\ref{fig:Q8-Q128_L=0km} & $-$  & $-$  & \formatnum{1.3E5} \\
\midrule
$M=4$ &  $L_\text{fib}=\SI{30}{\kilo\meter}$ &   Fig.~\ref{fig:Q4_L=30km} & \ok{\formatnum{1048576}} & \ok{\formatnum{2E5}} & 
\ok{\formatnum{3.1E+04}} / \ok{\formatnum{1.1E+05}} 
\\
$M=8$ & \downbar &  Fig.~\ref{fig:Q8_L=30km} & \ok{\formatnum{16777216}} & \ok{\formatnum{2.4E5}} &
\ok{\formatnum{4.6E+04}} /  \ok{\formatnum{1.1E+05}} 
\\ 
$M=16$ &  \downbar &  Fig.~\ref{fig:Q8-Q128_L=30km} & $-$  & $-$  & \ok{\formatnum{1.1E+05}}  \\
$M=32$ &  \downbar &  Fig.~\ref{fig:rnn_vs_gibbs_q32_L30},\ref{fig:Q32_L=30km_SIC_stage_rate} & $-$  & \ok{\formatnum{1.8E7}} & \ok{\formatnum{2.4E5}}  \\
$M=64$ &  \upbar &  Fig.~\ref{fig:Q8-Q128_L=30km} & $-$  & $-$  & \ok{\formatnum{4.9E5}} \\
\bottomrule
\end{tabular}}
\label{tab:complexity_application}
\end{table}
}

Fig.~\ref{fig:Q4_L=0km} and Fig.~\ref{fig:Q4_L=30km} show that the rates increase with $S$. NN-SIC is at least as good as the mismatched FBA-SIC for all modulation formats. The NN shows large gains over the mismatched FBA for $S>2$ and medium to high SNRs because of the FBA channel mismatch, i.e., the FBA memory is much smaller than the NN memory; see the third column in Tab.~\ref{tab:nnparams}.
For 4-PAM/ASK and $L_\text{fib} =0$, the best NN curves reduce the gap to the UBs to \SI{0.8}{dB} at around $80\%$ of the maximum rate. The rates for 4-SQAM and $L_\text{fib}=0$ saturate early. This is likely due to phase symmetries in 4-SQAM; see~\cite[Appendix]{prinz2023successive}. 
For $L_\text{fib} = \SI{30}{\kilo\meter}$, the gaps to the UB reduce to $\approx \SI{1}{dB}$ for all modulations. For 4-SQAM, we had to use a larger NN than for 4-PAM/ASK; see Fig.~\ref{fig:Q4_L=30km}. For $L_\text{fib}=0$ and $L_\text{fib}=\SI{30}{\kilo\meter}$, two and four SIC levels are needed to approach the JDD rate, respectively.

The NN outperforms GS and the mismatched FBA with much smaller complexity; see Table~\ref{tab:complexity} and Fig.~\ref{fig:complexity}. For $M=4$ and $L_\text{fib}=0$, the NN requires 40 times fewer multiplications than GS and 200 times fewer than the FBA. For real-valued modulation and $L_\text{fib} = \SI{30}{\kilo\meter}$, the NN is roughly ten times less complex than GS and 30 times less complex than the mismatched FBA. For 4-SQAM, the complexity of GS and the NN is similar. Fig.~\ref{fig:Q8_L=30km} shows similar results for 8-PAM/ASK/SQAM for $L_\text{fib} = \SI{30}{\kilo\meter}$, where we chose SIC with up to $S=6$ stages.

\begin{figure*}[!htb]
    \centering
    \begin{minipage}[t]{0.33\linewidth}
         \pgfplotstableread{
    X Y
  -13.0000    0.0036
  -12.0000    0.0054
  -11.0000    0.0085
  -10.0000    0.0134
   -9.0000    0.0241
   -8.0000    0.0361
   -7.0000    0.0590
   -6.0000    0.0854
   -5.0000    0.1321
   -4.0000    0.2019
   -3.0000    0.2939
   -2.0000    0.4250
   -1.0000    0.5778
         0    0.7590
    1.0000    0.9479
    2.0000    1.1474
    3.0000    1.3367
    4.0000    1.5044
    5.0000    1.6353
    6.0000    1.7396
    7.0000    1.8053
    8.0000    1.8610
    9.0000    1.8986
   10.0000    1.9427
   11.0000    1.9666
   12.0000    1.9873
   13.0000    1.9975
   14.0000    1.9994
   15.0000    2.0000
   16.0000    2.0001
   17.0000    2.0001
   18.0000    2.0000
}{\PAMrcosspannine}

\pgfplotstableread{
    X Y
  -13.0000    0.0038
  -12.0000    0.0060
  -11.0000    0.0095
  -10.0000    0.0150
   -9.0000    0.0236
   -8.0000    0.0370
   -7.0000    0.0578
   -6.0000    0.0897
   -5.0000    0.1377
   -4.0000    0.2081
   -3.0000    0.3081
   -2.0000    0.4448
   -1.0000    0.6236
         0    0.8472
    1.0000    1.1160
    2.0000    1.4288
    3.0000    1.7846
    4.0000    2.1827
}{\PAMCovUpperBound}

\pgfdeclarelayer{background}
\pgfdeclarelayer{foreground}
\pgfsetlayers{background,main,foreground}
\input{plots/plot_settings}
\begin{tikzpicture}[]

\begin{axis}[%
yminorticks=true,
xmajorgrids,
ymajorgrids,
yminorgrids,
minor x tick num=1,
minor y tick num=4,
grid=both,
legend style={legend cell align=left,  draw=white!15!black, font=\footnotesize,  
},
legend style={at={(1.00,0.00)},anchor=south east},
xlabel style={font=\color{white!15!black}},
ylabel style={font=\color{white!15!black}},
axis background/.style={fill=white},
scale only axis,
width=\gridwidth,
height=\gridheight,
xmin=-5,
xmax=15,
xlabel={$\mathrm{P}_{\mathrm{tx}}$ in [dB]},
ymin=0,
ymax=2.001,
ylabel={Rate},
title={4-PAM},
ylabel shift = -4pt,
]

\pgfplotstableread{
X Y
 -5.0000    0.1495
-4.0000    0.2168
-3.0000    0.3096
-2.0000    0.4372
-1.0000    0.6097
     0    0.7985
1.0000    1.0075
2.0000    1.2449
3.0000    1.4871
4.0000    1.7318
5.0000    1.9667
6.0000    2.2078
7.0000    2.4685
}{\ASKLoeligerUpperBoundIX}

\addplot [PAM,UB,opacity=0.8,name path global=UB] table {\ASKLoeligerUpperBoundIX};
\addlegendentry{UB, $\widetilde{N} = 9$};

\addplot[PAM,FBA,] table [x=power, y=rate, col sep=space]{plots/raw/Q4/TOBI/4-PAM_M=1_alpha=0.00_L=0km_mem=9_n=30000_daniel.txt};\addlegendentry{FBA, $\widetilde{N} = 9$}; %

\addplot[PAM,FBA,forget plot] table [x=power, y=rate, col sep=space]{plots/raw/Q4/TOBI/4-PAM_M=2_alpha=0.00_L=0km_mem=9_n=30000_daniel.txt};%

\addplot[PAM,FBA,forget plot] table [x=power, y=rate, col sep=space]{plots/raw/Q4/TOBI/4-PAM_M=3_alpha=0.00_L=0km_mem=9_n=20000_daniel.txt};%

\addplot[PAM,FBA,forget plot] table [x=power, y=rate, col sep=space]{plots/raw/Q4/TOBI/4-PAM_M=4_alpha=0.00_L=0km_mem=9_n=20000_daniel.txt};%

\addplot[PAM,RNN] table [x=Ptxdb, y=SIC1, col sep=comma]{plots/V1.5/Q4/v1.5,4-PAM,32_64_4,lr=1E-03,Tr=32,B=128,S=2,Ls=16,I=1.0E+04,V=1.0E+03,n=6.0E+04,P=151,a=0,L=0.0E+00,Rs=3.5E+10,C=5.4E+03.txt};\addlegendentry{NN, $\widetilde{N}_\text{RNN}=47$}; %

\addplot[PAM,RNN] table [x=Ptxdb, y=IqXY, col sep=comma,forget plot]{plots/V1.5/Q4/v1.5,4-PAM,32_64_4,lr=1E-03,Tr=32,B=128,S=2,Ls=16,I=1.0E+04,V=1.0E+03,n=6.0E+04,P=151,a=0,L=0.0E+00,Rs=3.5E+10,C=5.4E+03.txt}; 

\addplot[PAM,RNN] table [x=Ptxdb, y=IqXY, col sep=comma,forget plot]{plots/V1.5/Q4/v1.5,4-PAM,32_64_4,lr=1E-03,Tr=32,B=128,S=3,Ls=16,I=1.0E+04,V=1.0E+03,n=6.0E+04,P=151,a=0,L=0.0E+00,Rs=3.5E+10,C=5.4E+03.txt}; 

\addplot[PAM,RNN,name path global=s4pam] table [x=Ptxdb, y=IqXY, col sep=comma,forget plot]{plots/V1.5/Q4/v1.5,4-PAM,32_64_4,lr=1E-03,Tr=36,B=128,S=4,Ls=16,I=1.0E+04,V=1.0E+03,n=6.0E+04,P=151,a=0,L=0.0E+00,Rs=3.5E+10,C=5.4E+03.txt};

\path[name path global=line] (axis cs:\pgfkeysvalueof{/pgfplots/xmin},1.6) -- (axis cs: \pgfkeysvalueof{/pgfplots/xmax},1.6);
\path[name intersections={of=line and UB, name=p1}, name intersections={of=line and s4pam, name=p2}];
\draw[arr] let \p1=(p1-1), \p2=(p2-1) in (p1-1) -- ([xshift=-0.0cm]p2-1) node [left,midway,opacitylabel,font=\footnotesize,yshift=0.0cm,xshift=-0.3cm] {%
	\pgfplotsconvertunittocoordinate{x}{\x1}%
	\pgfplotscoordmath{x}{datascaletrafo inverse to fixed}{\pgfmathresult}%
	\edef\valueA{\pgfmathresult}%
	\pgfplotsconvertunittocoordinate{x}{\x2}%
	\pgfplotscoordmath{x}{datascaletrafo inverse to fixed}{\pgfmathresult}%
	\pgfmathparse{\pgfmathresult - \valueA}%
	\pgfmathprintnumber{\pgfmathresult} dB
};

\end{axis}
\end{tikzpicture}%
    \end{minipage}%
    \begin{minipage}[t]{0.33\linewidth}
        \pgfplotstableread{
    X Y
  -13.0000    0.0039
  -12.0000    0.0062
  -11.0000    0.0098
  -10.0000    0.0155
   -9.0000    0.0245
   -8.0000    0.0386
   -7.0000    0.0605
   -6.0000    0.0943
   -5.0000    0.1456
   -4.0000    0.2222
   -3.0000    0.3332
   -2.0000    0.4890
   -1.0000    0.6992
         0    0.9708
    1.0000    1.3072
    2.0000    1.7071
    3.0000    2.1654
    4.0000       Inf
    5.0000       Inf
    6.0000       Inf
    7.0000       Inf
    8.0000       Inf
    9.0000       Inf
   10.0000       Inf
   11.0000       Inf
   12.0000       Inf
   13.0000       Inf
   14.0000       Inf
   15.0000       Inf
   16.0000       Inf
   17.0000       Inf
   18.0000       Inf
   }{\ASKCovUpperBound}

\pgfdeclarelayer{background}
\pgfdeclarelayer{foreground}
\pgfsetlayers{background,main,foreground}

\input{plots/plot_settings}
\begin{tikzpicture}[]

\begin{axis}[%
yminorticks=true,
xmajorgrids,
ymajorgrids,
yminorgrids,
minor x tick num=1,
minor y tick num=4,
grid=both,
legend style={legend cell align=left,  draw=white!15!black, font=\scriptsize,  
},
legend style={at={(1.00,0.00)},anchor=south east},
xlabel style={font=\color{white!15!black}},
ylabel style={font=\color{white!15!black}},
axis background/.style={fill=white},
scale only axis,
width=\gridwidth,
height=\gridheight,
xmin=-5,
xmax=15,
xlabel={$\mathrm{P}_{\mathrm{tx}}$ in [dB]},
ymin=0,
ymax=2.001,
ylabel={Rate},
title={4-ASK},
ylabel shift = -4pt
]

\pgfplotstableread{
    X Y
   -5.0000    0.1454
   -4.0000    0.2167
   -3.0000    0.3213
   -2.0000    0.4744
   -1.0000    0.6503
         0    0.8757
    1.0000    1.1392
    2.0000    1.4137
    3.0000    1.6998
    4.0000    1.9674
    5.0000    2.2157
    6.0000    2.4732
    7.0000    2.7491
}{\ASKLoeligerUpperBoundIX}

\addplot [ASK,UB,opacity=0.8,name path global=UB] table {\ASKLoeligerUpperBoundIX};
\addlegendentry{UB, $\widetilde{N} = 9$};

\addplot[ASK,FBA] table [x=power, y=rate, col sep=space]{plots/raw/Q4/TOBI/4-ASK_M=1_alpha=0.00_L=0km_mem=9_n=20000_daniel.txt};\addlegendentry{FBA, $\widetilde{N} = 9$}; %

\addplot[ASK,FBA,forget plot] table [x=power, y=rate, col sep=space]{plots/raw/Q4/TOBI/4-ASK_M=2_alpha=0.00_L=0km_mem=9_n=20000_daniel.txt};%

\addplot[ASK,FBA,forget plot] table [x=power, y=rate, col sep=space]{plots/raw/Q4/TOBI/4-ASK_M=3_alpha=0.00_L=0km_mem=9_n=20000_daniel.txt}; %

\addplot[ASK,FBA,forget plot] table [x=power, y=rate, col sep=space]{plots/raw/Q4/TOBI/4-ASK_M=4_alpha=0.00_L=0km_mem=9_n=20000_daniel.txt};%

\addplot[ASK,RNN] table [x=Ptxdb, y=SIC1, col sep=comma]{plots/V1.5/Q4/v1.5,4-ASK,32_64_4,lr=1E-03,Tr=32,B=128,S=2,Ls=16,I=1.0E+04,V=1.0E+03,n=6.0E+04,P=151,a=0,L=0.0E+00,Rs=3.5E+10,C=5.4E+03.txt};\addlegendentry{NN, $\widetilde{N}_\text{RNN}=47$}; %

\addplot[ASK,RNN] table [x=Ptxdb, y=IqXY, col sep=comma,forget plot]{plots/V1.5/Q4/v1.5,4-ASK,32_64_4,lr=1E-03,Tr=32,B=128,S=2,Ls=16,I=1.0E+04,V=1.0E+03,n=6.0E+04,P=151,a=0,L=0.0E+00,Rs=3.5E+10,C=5.4E+03.txt}; 

\addplot[ASK,RNN] table [x=Ptxdb, y=IqXY, col sep=comma,forget plot]{plots/V1.5/Q4/v1.5,4-ASK,32_64_4,lr=1E-03,Tr=32,B=128,S=3,Ls=16,I=1.0E+04,V=1.0E+03,n=6.0E+04,P=151,a=0,L=0.0E+00,Rs=3.5E+10,C=5.4E+03.txt}; 

\addplot[ASK,RNN,name path global=s4ask] table [x=Ptxdb, y=IqXY, col sep=comma,forget plot]{plots/V1.5/Q4/v1.5,4-ASK,32_64_4,lr=1E-03,Tr=36,B=128,S=4,Ls=16,I=1.0E+04,V=1.0E+03,n=6.0E+04,P=151,a=0,L=0.0E+00,Rs=3.5E+10,C=5.4E+03.txt};

\begin{pgfonlayer}{foreground}
\path[name path global=line] (axis cs:\pgfkeysvalueof{/pgfplots/xmin},1.6) -- (axis cs: \pgfkeysvalueof{/pgfplots/xmax},1.6);
\path[name intersections={of=line and UB, name=p1}, name intersections={of=line and s4ask, name=p2}];
\draw[arr] let \p1=(p1-1), \p2=(p2-1) in (p1-1) -- ([xshift=-0.0cm]p2-1) node [left,midway,opacitylabel,font=\footnotesize,yshift=0.0cm,xshift=-0.3cm] {%
	\pgfplotsconvertunittocoordinate{x}{\x1}%
	\pgfplotscoordmath{x}{datascaletrafo inverse to fixed}{\pgfmathresult}%
	\edef\valueA{\pgfmathresult}%
	\pgfplotsconvertunittocoordinate{x}{\x2}%
	\pgfplotscoordmath{x}{datascaletrafo inverse to fixed}{\pgfmathresult}%
	\pgfmathparse{\pgfmathresult - \valueA}%
	\pgfmathprintnumber{\pgfmathresult} dB
};
\end{pgfonlayer}
\end{axis}

\end{tikzpicture}%
    \end{minipage}%
    \begin{minipage}[t]{0.33\linewidth}
        \pgfplotstableread{
    X Y
 -13.000000000000000   0.001188044177414
 -12.000000000000000   0.001881779741266
 -11.000000000000000   0.002979554474699
 -10.000000000000000   0.004715104241672
  -9.000000000000000   0.007455026637460
  -8.000000000000000   0.011770840308635
  -7.000000000000000   0.018545130013262
  -6.000000000000000   0.029120862706446
  -5.000000000000000   0.045495373664651
  -4.000000000000000   0.070537302019467
  -3.000000000000000   0.108156026314185
  -2.000000000000000   0.163282177130346
  -1.000000000000000   0.241471835345610
                   0   0.348030643529216
   1.000000000000000   0.486835194396268
   2.000000000000000   0.659349026306563
   2.999999999999999   0.864339381613759
   4.000000000000000   1.098396646549094
   5.000000000000000   1.356908474012931
   6.000000000000000   1.635022472478486
   7.000000000000000   1.928313973403565
   8.000000000000000   2.233107295218103
   }{\QAMCovUpperBound}

\input{plots/plot_settings}
\begin{tikzpicture}[]
\begin{axis}[%
yminorticks=true,
xmajorgrids,
ymajorgrids,
yminorgrids,
minor x tick num=1,
minor y tick num=4,
grid=both,
legend style={legend cell align=left,  draw=white!15!black, font=\scriptsize,  
},
legend style={at={(1.00,0.00)},anchor=south east},
xlabel style={font=\color{white!15!black}},
ylabel style={font=\color{white!15!black}},
axis background/.style={fill=white},
scale only axis,
width=\gridwidth,
height=\gridheight,
xmin=-5,
xmax=15,
xlabel={$\mathrm{P}_{\mathrm{tx}}$ in [dB]},
ymin=0,
ymax=2.001,
ylabel={Rate},
title={QPSK},
ylabel shift = -4pt
]

\pgfplotstableread{
    X Y
   -5.0000    0.0561
   -4.0000    0.0763
   -3.0000    0.1100
   -2.0000    0.1607
   -1.0000    0.2479
         0    0.3485
    1.0000    0.4761
    2.0000    0.6434
    3.0000    0.8352
    4.0000    1.0493
    5.0000    1.2843
    6.0000    1.5463
    7.0000    1.8325
    8.0000    2.1331
    9.0000    2.4608
}{\QPSKLoeligerUpperBoundIX}

\addplot [QAM,UB,opacity=0.8,name path=UB,postaction={decorate,decoration={raise=0.6ex,text along path,text color={black},text align={right, right indent=1.1cm},text={}}}] table {\QPSKLoeligerUpperBoundIX};
\addlegendentry{UB, $\widetilde{N} = 9$};

\addplot[QAM,FBA] table [x=power, y=rate, col sep=space]{plots/raw/Q4/TOBI/4-QPSK_M=1_alpha=0.00_L=0km_mem=9_n=20000_daniel.txt};\addlegendentry{FBA, $\widetilde{N} = 9$}; %

\addplot[QAM,FBA,forget plot] table [x=power, y=rate, col sep=space]{plots/raw/Q4/TOBI/4-QPSK_M=2_alpha=0.00_L=0km_mem=9_n=20000_daniel.txt};%

\addplot[QAM,FBA,forget plot] table [x=power, y=rate, col sep=space]{plots/raw/Q4/TOBI/4-QPSK_M=3_alpha=0.00_L=0km_mem=9_n=20000_daniel.txt};%

\addplot[QAM,FBA,forget plot] table [x=power, y=rate, col sep=space]{plots/raw/Q4/TOBI/4-QPSK_M=4_alpha=0.00_L=0km_mem=9_n=20000_daniel.txt};%

\addplot[QAM,RNN] table [x=Ptxdb, y=SIC1, col sep=comma]{plots/V1.5/Q4/v1.5,4-SQAM,32_64_4,lr=1E-03,Tr=32,B=128,S=2,Ls=16,I=1.0E+04,V=1.0E+03,n=6.0E+04,P=151,a=0,L=0.0E+00,Rs=3.5E+10,C=6.4E+03.txt};
\addlegendentry{NN, $\widetilde{N}_\text{RNN}=47$}

\addplot[QAM,RNN] table [x=Ptxdb, y=IqXY, forget plot, col sep=comma]{plots/V1.5/Q4/v1.5,4-SQAM,32_64_4,lr=1E-03,Tr=32,B=128,S=2,Ls=16,I=1.0E+04,V=1.0E+03,n=6.0E+04,P=151,a=0,L=0.0E+00,Rs=3.5E+10,C=6.4E+03.txt};

\addplot[QAM,RNN,forget plot] table [x=Ptxdb, y=IqXY, col sep=comma]{plots/V1.5/Q4/v1.5,4-SQAM,32_64_4,lr=1E-03,Tr=32,B=128,S=3,Ls=16,I=1.0E+04,V=1.0E+03,n=6.0E+04,P=151,a=0,L=0.0E+00,Rs=3.5E+10,C=6.4E+03.txt};

\addplot[QAM,RNN,forget plot] table [x=Ptxdb, y=IqXY, col sep=comma]{plots/V1.5/Q4/v1.5,4-SQAM,32_64_4,lr=1E-03,Tr=36,B=128,S=4,Ls=16,I=1.0E+04,V=1.0E+03,n=6.0E+04,P=151,a=0,L=0.0E+00,Rs=3.5E+10,C=6.4E+03.txt};

\end{axis}
\end{tikzpicture}%
    \end{minipage}%
    \caption{SIC rates for $L_\text{fib}=0$ and $S=1,2,3,4$ stages. Mismatched FBA and UB memory $\widetilde{N} = 9$, RNN memory $\widetilde{N}_\text{RNN} = 47$. The lower curves of a particular style show the SDD ($S=1$) rates, while the upper curves show the $S=4$ rates.} 
    \label{fig:Q4_L=0km}
    \bigskip\bigskip
    \centering
    \begin{minipage}[t]{0.33\linewidth}
         \pgfplotstableread{
    X Y
   -5.9944    0.0604
   -4.9944    0.0930
   -3.9944    0.1414
   -2.9944    0.2111
   -1.9944    0.3081
   -0.9944    0.4381
    0.0056    0.6053
    1.0056    0.8125
    2.0056    1.0611
    3.0056    1.3516
    4.0056    1.6844
    5.0056    2.0597
    6.0056    2.4771
    7.0056    2.9357
    8.0056    3.4329
    9.0056    3.9653
   10.0056    4.5283
   11.0056    5.1166
   12.0056    5.7251
  }{\PAMCovUpperBound}

\pgfdeclarelayer{background}
\pgfdeclarelayer{foreground}
\pgfsetlayers{background,main,foreground}

\input{plots/plot_settings}
\begin{tikzpicture}[]

\begin{axis}[%
yminorticks=true,
xmajorgrids,
ymajorgrids,
yminorgrids,
minor x tick num=1,
minor y tick num=4,
grid=both,
legend style={legend cell align=left,  draw=white!15!black, font=\scriptsize,  
},
legend style={at={(1.00,0.00)},anchor=south east},
xlabel style={font=\color{white!15!black}},
ylabel style={font=\color{white!15!black}},
axis background/.style={fill=white},
scale only axis,
width=\gridwidth,
height=\gridheight,
xmin=-5,
xmax=10,
xlabel={$\mathrm{P}_{\mathrm{tx}}$ in [dB]},
ymin=0,
ymax=2.001,
ylabel={Rate},
title={4-PAM},
ylabel shift = -4pt
]

\pgfplotstableread{
X Y
-4.9944    0.0908
-3.9944    0.1327
-2.9944    0.2247
-1.9944    0.3123
-0.9944    0.4344
0.0056    0.5938
1.0056    0.7909
2.0056    0.9979
3.0056    1.2385
4.0056    1.4974
5.0056    1.7944
6.0056    2.0962
7.0056    2.4102
}{\ASKLoeligerUpperBoundIX}

\addplot [PAM,UB,opacity=0.8,name path global=UB] table {\ASKLoeligerUpperBoundIX};
\addlegendentry{UB, $\widetilde{N} = 9$};

\addplot[PAM,FBA] table [x=power, y=rate, col sep=space]{plots/raw/Q4/TOBI/4-PAM_M=1_alpha=0.00_L=30km_mem=9_n=30000.txt};\addlegendentry{FBA, $\widetilde{N} = 9$}; %

\addplot[PAM,FBA,forget plot] table [x=power, y=rate, col sep=space]{plots/raw/Q4/TOBI/4-PAM_M=2_alpha=0.00_L=30km_mem=9_n=30000.txt};%

\addplot[PAM,FBA,forget plot] table [x=power, y=rate, col sep=space]{plots/raw/Q4/TOBI/4-PAM_M=3_alpha=0.00_L=30km_mem=9_n=30000.txt}; %

\addplot[PAM,FBA,forget plot] table [x=power, y=rate, col sep=space]{plots/raw/Q4/TOBI/4-PAM_M=4_alpha=0.00_L=30km_mem=9_n=30000.txt};%

\addplot[PAM,RNN] table [x=Ptxdb, y=SIC1, col sep=comma]{plots/V1.5/Q4_L30/v1.5,4-PAM,64_128_64_4,lr=5E-04,Tr=64,B=128,S=2,Ls=32,I=2.0E+04,V=1.0E+03,n=6.0E+04,P=151,a=0,L=3.0E+04,Rs=3.5E+10,C=3.1E+04.txt};
\addlegendentry{NN, $\widetilde{N}_\text{RNN}=95$}

\addplot[PAM,RNN,forget plot] table [x=Ptxdb, y=IqXY, col sep=comma]{plots/V1.5/Q4_L30/v1.5,4-PAM,64_128_64_4,lr=5E-04,Tr=64,B=128,S=2,Ls=32,I=2.0E+04,V=1.0E+03,n=6.0E+04,P=151,a=0,L=3.0E+04,Rs=3.5E+10,C=3.1E+04.txt};

\addplot[PAM,RNN,forget plot] table [x=Ptxdb, y=IqXY, col sep=comma]{plots/V1.5/Q4_L30/v1.5,4-PAM,64_128_64_4,lr=5E-04,Tr=64,B=128,S=3,Ls=32,I=2.0E+04,V=1.0E+03,n=6.0E+04,P=151,a=0,L=3.0E+04,Rs=3.5E+10,C=3.1E+04.txt};

\addplot[PAM,RNN,forget plot,name path global=s4pam] table [x=Ptxdb, y=IqXY, col sep=comma]{plots/V1.5/Q4_L30/v1.5,4-PAM,64_128_64_4,lr=5E-04,Tr=66,B=128,S=4,Ls=32,I=2.0E+04,V=1.0E+03,n=6.0E+04,P=151,a=0,L=3.0E+04,Rs=3.5E+10,C=3.1E+04.txt};

\begin{pgfonlayer}{foreground}
\path[name path global=line] (axis cs:\pgfkeysvalueof{/pgfplots/xmin},1.6) -- (axis cs: \pgfkeysvalueof{/pgfplots/xmax},1.6);
\path[name intersections={of=line and UB, name=p1}, name intersections={of=line and s4pam, name=p2}];
\draw[arr] let \p1=(p1-1), \p2=(p2-1) in (p1-1) -- ([xshift=-0.0cm]p2-1) node [left,midway,opacitylabel,font=\footnotesize,yshift=0.0cm,xshift=-0.3cm] {%
	\pgfplotsconvertunittocoordinate{x}{\x1}%
	\pgfplotscoordmath{x}{datascaletrafo inverse to fixed}{\pgfmathresult}%
	\edef\valueA{\pgfmathresult}%
	\pgfplotsconvertunittocoordinate{x}{\x2}%
	\pgfplotscoordmath{x}{datascaletrafo inverse to fixed}{\pgfmathresult}%
	\pgfmathparse{\pgfmathresult - \valueA}%
	\pgfmathprintnumber{\pgfmathresult} dB
};
\end{pgfonlayer}
\end{axis}

\end{tikzpicture}%
    \end{minipage}%
    \begin{minipage}[t]{0.33\linewidth}
        \pgfplotstableread{
    X Y
   -6.0000    0.0754
   -5.0000    0.1169
   -4.0000    0.1792
   -3.0000    0.2705
   -2.0000    0.4004
   -1.0000    0.5785
         0    0.8125
    1.0000    1.1075
    2.0000    1.4643
    3.0000    1.8800
    4.0000    2.3491
    5.0000    2.8640
    6.0000    3.4169
    7.0000    3.9997
    8.0000    4.6053
    9.0000    5.2278
   10.0000    5.8625
   11.0000    6.5056
   12.0000    7.1547
  }{\ASKCovUpperBound}

\pgfdeclarelayer{background}
\pgfdeclarelayer{foreground}
\pgfsetlayers{background,main,foreground}

\input{plots/plot_settings}
\begin{tikzpicture}[]

\begin{axis}[%
yminorticks=true,
xmajorgrids,
ymajorgrids,
yminorgrids,
minor x tick num=1,
minor y tick num=4,
grid=both,
legend style={legend cell align=left,  draw=white!15!black, font=\scriptsize,  
},
legend style={at={(1.00,0.00)},anchor=south east},
xlabel style={font=\color{white!15!black}},
ylabel style={font=\color{white!15!black}},
axis background/.style={fill=white},
scale only axis,
width=\gridwidth,
height=\gridheight,
xmin=-5,
xmax=10,
xlabel={$\mathrm{P}_{\mathrm{tx}}$ in [dB]},
ymin=0,
ymax=2.001,
ylabel={Rate},
title={4-ASK},
ylabel shift = -4pt
]

\pgfplotstableread{
X Y
   -5.0000    0.1111
   -4.0000    0.1817
   -3.0000    0.2756
   -2.0000    0.3972
   -1.0000    0.5686
         0    0.7795
    1.0000    1.0320
    2.0000    1.3157
    3.0000    1.6302
    4.0000    1.9724
    5.0000    2.3280
    6.0000    2.6987
    7.0000    3.0672
}{\ASKLoeligerUpperBoundIX}

\addplot [ASK,UB,opacity=0.8,name path global=UB] table {\ASKLoeligerUpperBoundIX};
\addlegendentry{UB, $\widetilde{N} = 9$};

\addplot[ASK,FBA] table [x=power, y=rate, col sep=space]{plots/raw/Q4/TOBI/4-ASK_M=1_alpha=0.00_L=30km_mem=9_n=30000.txt};\addlegendentry{FBA, $\widetilde{N} = 9$}; %

\addplot[ASK,FBA,forget plot] table [x=power, y=rate, col sep=space]{plots/raw/Q4/TOBI/4-ASK_M=2_alpha=0.00_L=30km_mem=9_n=30000.txt};%

\addplot[ASK,FBA,forget plot] table [x=power, y=rate, col sep=space]{plots/raw/Q4/TOBI/4-ASK_M=3_alpha=0.00_L=30km_mem=9_n=30000.txt}; %

\addplot[ASK,FBA,forget plot] table [x=power, y=rate, col sep=space]{plots/raw/Q4/TOBI/4-ASK_M=4_alpha=0.00_L=30km_mem=9_n=30000.txt};%

\addplot[ASK,RNN] table [x=Ptxdb, y=SIC1, col sep=comma]{plots/V1.5/Q4_L30/v1.5,4-ASK,64_128_64_4,lr=5E-04,Tr=64,B=128,S=2,Ls=32,I=2.0E+04,V=1.0E+03,n=6.0E+04,P=151,a=0,L=3.0E+04,Rs=3.5E+10,C=3.1E+04.txt};
\addlegendentry{NN, $\widetilde{N}_\text{RNN}=95$}

\addplot[ASK,RNN,forget plot] table [x=Ptxdb, y=IqXY, col sep=comma]{plots/V1.5/Q4_L30/v1.5,4-ASK,64_128_64_4,lr=5E-04,Tr=64,B=128,S=2,Ls=32,I=2.0E+04,V=1.0E+03,n=6.0E+04,P=151,a=0,L=3.0E+04,Rs=3.5E+10,C=3.1E+04.txt};

\addplot[ASK,RNN,forget plot] table [x=Ptxdb, y=IqXY, col sep=comma]{plots/V1.5/Q4_L30/v1.5,4-ASK,64_128_64_4,lr=5E-04,Tr=64,B=128,S=3,Ls=32,I=2.0E+04,V=1.0E+03,n=6.0E+04,P=151,a=0,L=3.0E+04,Rs=3.5E+10,C=3.1E+04.txt};

\addplot[ASK,RNN,forget plot,name path global=s4ask] table [x=Ptxdb, y=IqXY, col sep=comma]{plots/V1.5/Q4_L30/v1.5,4-ASK,64_128_64_4,lr=5E-04,Tr=66,B=128,S=4,Ls=32,I=2.0E+04,V=1.0E+03,n=6.0E+04,P=151,a=0,L=3.0E+04,Rs=3.5E+10,C=3.1E+04.txt};

\begin{pgfonlayer}{foreground}
\path[name path global=line] (axis cs:\pgfkeysvalueof{/pgfplots/xmin},1.6) -- (axis cs: \pgfkeysvalueof{/pgfplots/xmax},1.6);
\path[name intersections={of=line and UB, name=p1}, name intersections={of=line and s4ask, name=p2}];
\draw[arr] let \p1=(p1-1), \p2=(p2-1) in (p1-1) -- ([xshift=-0.0cm]p2-1) node [left,midway,opacitylabel,font=\footnotesize,yshift=0.0cm,xshift=-0.3cm] {%
	\pgfplotsconvertunittocoordinate{x}{\x1}%
	\pgfplotscoordmath{x}{datascaletrafo inverse to fixed}{\pgfmathresult}%
	\edef\valueA{\pgfmathresult}%
	\pgfplotsconvertunittocoordinate{x}{\x2}%
	\pgfplotscoordmath{x}{datascaletrafo inverse to fixed}{\pgfmathresult}%
	\pgfmathparse{\pgfmathresult - \valueA}%
	\pgfmathprintnumber{\pgfmathresult} dB
};
\end{pgfonlayer}
\end{axis}

\end{tikzpicture}%
    \end{minipage}%
    \begin{minipage}[t]{0.33\linewidth}
         \pgfplotstableread{
    X Y
   -6.0000    0.0704
   -5.0000    0.1092
   -4.0000    0.1674
   -3.0000    0.2530
   -2.0000    0.3749
   -1.0000    0.5424
         0    0.7633
    1.0000    1.0428
    2.0000    1.3823
    3.0000    1.7796
    4.0000    2.2296
    5.0000    2.7254
    6.0000    3.2596
    7.0000    3.8248
    8.0000    4.4143
    9.0000    5.0227
   10.0000    5.6454
   11.0000    6.2787
   12.0000    6.9200
  }{\QAMCovUpperBound}

\pgfdeclarelayer{background}
\pgfdeclarelayer{foreground}
\pgfsetlayers{background,main,foreground}

\input{plots/plot_settings}
\begin{tikzpicture}[]

\begin{axis}[%
yminorticks=true,
xmajorgrids,
ymajorgrids,
yminorgrids,
minor x tick num=1,
minor y tick num=4,
grid=both,
legend style={legend cell align=left,  draw=white!15!black, font=\scriptsize,  
},
legend style={at={(1.00,0.00)},anchor=south east},
xlabel style={font=\color{white!15!black}},
ylabel style={font=\color{white!15!black}},
axis background/.style={fill=white},
scale only axis,
width=\gridwidth,
height=\gridheight,
xmin=-5,
xmax=10,
xlabel={$\mathrm{P}_{\mathrm{tx}}$ in [dB]},
ymin=0,
ymax=2.001,
ylabel={Rate},
title={QPSK},
ylabel shift = -4pt
]

\pgfplotstableread{
    X Y
   -5.0000    0.1199
   -4.0000    0.1605
   -3.0000    0.2521
   -2.0000    0.3807
   -1.0000    0.5377
         0    0.7489
    1.0000    0.9915
    2.0000    1.2662
    3.0000    1.5852
    4.0000    1.9283
    5.0000    2.2769
    6.0000    2.6119
    7.0000    2.9602
}{\QAMLoeligerUpperBoundXII}

\addplot [QAM,UB,opacity=0.8,name path global=UB] table {\QAMLoeligerUpperBoundXII};
\addlegendentry{UB, $\widetilde{N} = 9$};

\addplot[QAM,FBA] table [x=power, y=rate, col sep=space]{plots/raw/Q4/TOBI/4-QPSK_M=1_alpha=0.00_L=30km_mem=9_n=30000_.txt};\addlegendentry{FBA, $\widetilde{N} = 9$}; %

\addplot[QAM,FBA,forget plot] table [x=power, y=rate, col sep=space]{plots/raw/Q4/TOBI/4-QPSK_M=2_alpha=0.00_L=30km_mem=9_n=30000_.txt};%

\addplot[QAM,FBA,forget plot] table [x=power, y=rate, col sep=space]{plots/raw/Q4/TOBI/4-QPSK_M=3_alpha=0.00_L=30km_mem=9_n=30000_.txt}; %

\addplot[QAM,FBA,forget plot] table [x=power, y=rate, col sep=space]{plots/raw/Q4/TOBI/4-QPSK_M=4_alpha=0.00_L=30km_mem=9_n=30000_.txt};%

\addplot[QAM,RNN] table [x=Ptxdb, y=SIC1, col sep=comma]{plots/V1.5/Q4_L30/v1.5,4-SQAM,64_256_128_128_4,lr=5E-04,Tr=64,B=128,S=2,Ls=32,I=2.0E+04,V=1.0E+03,n=6.0E+04,P=151,a=0,L=3.0E+04,Rs=3.5E+10,C=1.3E+05.txt};
\addlegendentry{NN, $\widetilde{N}_\text{RNN}=95$};

\addplot[QAM,RNN,forget plot] table [x=Ptxdb, y=IqXY, col sep=comma]{plots/V1.5/Q4_L30/v1.5,4-SQAM,64_256_128_128_4,lr=5E-04,Tr=64,B=128,S=2,Ls=32,I=2.0E+04,V=1.0E+03,n=6.0E+04,P=151,a=0,L=3.0E+04,Rs=3.5E+10,C=1.3E+05.txt};

\addplot[QAM,RNN,forget plot] table [x=Ptxdb, y=IqXY, col sep=comma]{plots/V1.5/Q4_L30/v1.5,4-SQAM,64_256_128_128_4,lr=5E-04,Tr=64,B=128,S=3,Ls=32,I=2.0E+04,V=1.0E+03,n=6.0E+04,P=151,a=0,L=3.0E+04,Rs=3.5E+10,C=1.3E+05.txt};

\addplot[QAM,RNN,forget plot,name path global=s4sqam] table [x=Ptxdb, y=IqXY, col sep=comma]{plots/V1.5/Q4_L30/v1.5,4-SQAM,64_256_128_128_4,lr=5E-04,Tr=66,B=128,S=4,Ls=32,I=2.0E+04,V=1.0E+03,n=6.0E+04,P=151,a=0,L=3.0E+04,Rs=3.5E+10,C=1.3E+05.txt};

\begin{pgfonlayer}{foreground}
\path[name path global=line] (axis cs:\pgfkeysvalueof{/pgfplots/xmin},1.6) -- (axis cs: \pgfkeysvalueof{/pgfplots/xmax},1.6);
\path[name intersections={of=line and UB, name=p1}, name intersections={of=line and s4sqam, name=p2}];
\draw[arr] let \p1=(p1-1), \p2=(p2-1) in (p1-1) -- ([xshift=-0.0cm]p2-1) node [left,midway,opacitylabel,font=\footnotesize,yshift=0.0cm,xshift=-0.3cm] {%
	\pgfplotsconvertunittocoordinate{x}{\x1}%
	\pgfplotscoordmath{x}{datascaletrafo inverse to fixed}{\pgfmathresult}%
	\edef\valueA{\pgfmathresult}%
	\pgfplotsconvertunittocoordinate{x}{\x2}%
	\pgfplotscoordmath{x}{datascaletrafo inverse to fixed}{\pgfmathresult}%
	\pgfmathparse{\pgfmathresult - \valueA}%
	\pgfmathprintnumber{\pgfmathresult} dB
};
\end{pgfonlayer}
\end{axis}

\end{tikzpicture}%
    \end{minipage} %
    \caption{SIC rates for $L_\text{fib}=\SI{30}{\kilo\meter}$ and $S=1,2,3,4$ stages. Mismatched FBA and UB memory $\widetilde{N} = 9$, RNN memory $\widetilde{N}_\text{RNN} = 95$. The lower curves of a particular style show the SDD ($S=1$) rates, while the upper curves show the $S=4$ rates. }
    \label{fig:Q4_L=30km}
    \bigskip\bigskip
\centering
    \begin{minipage}[t]{0.33\linewidth}
         \pgfplotstableread{
    X Y
  -12.9939    0.0022
  -11.9939    0.0035
  -10.9939    0.0056
   -9.9939    0.0089
   -8.9939    0.0140
   -7.9939    0.0220
   -6.9939    0.0345
   -5.9939    0.0537
   -4.9939    0.0829
   -3.9939    0.1264
   -2.9939    0.1894
   -1.9939    0.2777
   -0.9939    0.3969
    0.0061    0.5511
    1.0061    0.7433
    2.0061    0.9745
    3.0061    1.2452
    4.0061    1.5557
    5.0061    1.9062
    6.0061    2.2973
    7.0061    2.7288
    8.0061    3.1999
    9.0061    3.7081
   10.0061    4.2498
   11.0061    4.8205
   12.0061    5.4149
   13.0061    6.0280
   14.0061    6.6553
   15.0061    7.2929
   16.0061    7.9379
   17.0061    8.5881
   18.0061    9.2420
}{\PAMCovUpperBound}

\pgfdeclarelayer{background}
\pgfdeclarelayer{foreground}
\pgfsetlayers{background,main,foreground}

\input{plots/plot_settings}
\begin{tikzpicture}[]

\begin{axis}[%
yminorticks=true,
xmajorgrids,
ymajorgrids,
yminorgrids,
minor x tick num=1,
minor y tick num=4,
grid=both,
legend style={legend cell align=left,  draw=white!15!black, font=\scriptsize,  
},
legend style={at={(1.00,0.00)},anchor=south east},
xlabel style={font=\color{white!15!black}},
ylabel style={font=\color{white!15!black}},
axis background/.style={fill=white},
scale only axis,
width=\gridwidth,
height=\gridheight,
xmin=-5,
xmax=15,
xlabel={$\mathrm{P}_{\mathrm{tx}}$ in [dB]},
ymin=0,
ymax=3.001,
ylabel={Rate},
title={8-PAM},
ylabel shift = -4pt
]

\pgfplotstableread{
    X Y
-4.993932989836233   0.083734298263019
  -3.993932989836232   0.131084706110043
  -2.993932989836233   0.186934627217982
  -1.993932989836232   0.285393567396054
  -0.993932989836232   0.395210381694465
   0.006067010163768   0.560541682685218
   1.006067010163768   0.744815461949774
   2.006067010163768   0.946315488502890
   3.006067010163767   1.205051371855740
   4.006067010163767   1.485013008072389
   5.006067010163767   1.815428635069021
   6.006067010163767   2.167207242960034
   7.006067010163767   2.590162159226839
   8.006067010163768   3.021924403063920
   9.006067010163768   3.520562694409054
  10.006067010163768   4.048585711483219
  11.006067010163768   4.621492262829195
}{\PAMLoeligerUpperBoundXII}
 
\addplot [ASK,UB,black,opacity=0.8,name path global=UB] table {\PAMLoeligerUpperBoundXII};
\addlegendentry{UB, $\widetilde{N} = 7$};

\addplot[PAM,FBA,] table [x=power, y=rate, col sep=space]{plots/raw/Q8/TOBI/8-PAM_M=1_alpha=0.00_L=30km_mem=7.txt};\addlegendentry{FBA, $\widetilde{N} = 7$};

\addplot[PAM,FBA,forget plot] table [x=power, y=rate, col sep=space]{plots/raw/Q8/TOBI/8-PAM_M=2_alpha=0.00_L=30km_mem=7.txt};

\addplot[PAM,FBA,forget plot] table [x=power, y=rate, col sep=space]{plots/raw/Q8/TOBI/8-PAM_M=3_alpha=0.00_L=30km_mem=7.txt};

\addplot[PAM,FBA,forget plot] table [x=power, y=rate, col sep=space]{plots/raw/Q8/TOBI/8-PAM_M=4_alpha=0.00_L=30km_mem=7.txt};

\addplot[PAM,RNN] table [x=Ptxdb, y=SIC1, col sep=comma]{plots/V1.5/Q8_L30/v1.5,8-PAM,64_128_128_8,lr=3E-04,Tr=84,B=128,S=2,Ls=32,I=5.0E+04,V=1.0E+03,n=6.0E+04,P=151,a=0,L=3.0E+04,Rs=3.5E+10,C=4.6E+04.txt};
\addlegendentry{NN, $\widetilde{N}_\text{RNN}=115$}

\addplot[PAM,RNN,forget plot] table [x=Ptxdb, y=IqXY, col sep=comma]{plots/V1.5/Q8_L30/v1.5,8-PAM,64_128_128_8,lr=3E-04,Tr=84,B=128,S=2,Ls=32,I=5.0E+04,V=1.0E+03,n=6.0E+04,P=151,a=0,L=3.0E+04,Rs=3.5E+10,C=4.6E+04.txt};

\addplot[PAM,RNN,forget plot] table [x=Ptxdb, y=IqXY, col sep=comma]{plots//V1.5/Q8_L30/v1.5,8-PAM,64_128_128_8,lr=3E-04,Tr=84,B=128,S=3,Ls=32,I=5.0E+04,V=1.0E+03,n=6.0E+04,P=151,a=0,L=3.0E+04,Rs=3.5E+10,C=4.6E+04.txt};

\addplot[PAM,RNN,forget plot,name path global=s4pam] table [x=Ptxdb, y=IqXY, col sep=comma]{plots/V1.5/Q8_L30/v1.5,8-PAM,64_128_128_8,lr=3E-04,Tr=84,B=128,S=4,Ls=32,I=5.0E+04,V=1.0E+03,n=6.0E+04,P=151,a=0,L=3.0E+04,Rs=3.5E+10,C=4.6E+04.txt};

\addplot[PAM,RNN,forget plot,name path global=s6pam] table [x=Ptxdb, y=IqXY, col sep=comma]{plots/V1.5/Q8_L30/v1.5,8-PAM,64_128_128_8,lr=3E-04,Tr=120,B=128,S=6,Ls=32,I=5.0E+04,V=1.0E+03,n=6.0E+04,P=151,a=0,L=3.0E+04,Rs=3.5E+10,C=4.6E+04.txt};

\begin{pgfonlayer}{foreground}
\path[name path global=line] (axis cs:\pgfkeysvalueof{/pgfplots/xmin},2.4) -- (axis cs: \pgfkeysvalueof{/pgfplots/xmax},2.4);
\path[name intersections={of=line and UB, name=p1}, name intersections={of=line and s6pam, name=p2}];
\draw[arr] let \p1=(p1-1), \p2=(p2-1) in (p1-1) -- ([xshift=-0.0cm]p2-1) node [left,midway,opacitylabel,font=\footnotesize,yshift=0.0cm,xshift=-0.3cm] {%
	\pgfplotsconvertunittocoordinate{x}{\x1}%
	\pgfplotscoordmath{x}{datascaletrafo inverse to fixed}{\pgfmathresult}%
	\edef\valueA{\pgfmathresult}%
	\pgfplotsconvertunittocoordinate{x}{\x2}%
	\pgfplotscoordmath{x}{datascaletrafo inverse to fixed}{\pgfmathresult}%
	\pgfmathparse{\pgfmathresult - \valueA}%
	\pgfmathprintnumber{\pgfmathresult} dB
};
\end{pgfonlayer}

\end{axis}

\end{tikzpicture}%
    \end{minipage}%
    \begin{minipage}[t]{0.33\linewidth}
        \pgfplotstableread{
    X Y
 -13.0000    0.0032
  -12.0000    0.0051
  -11.0000    0.0081
  -10.0000    0.0128
   -9.0000    0.0203
   -8.0000    0.0319
   -7.0000    0.0502
   -6.0000    0.0784
   -5.0000    0.1215
   -4.0000    0.1864
   -3.0000    0.2817
   -2.0000    0.4172
   -1.0000    0.6029
         0    0.8469
    1.0000    1.1538
    2.0000    1.5239
    3.0000    1.9535
    4.0000    2.4360
    5.0000    2.9634
    6.0000    3.5270
    7.0000    4.1188
    8.0000    4.7317
    9.0000    5.3600
   10.0000    5.9993
   11.0000    6.6464
   12.0000    7.2990
   13.0000    7.9554
   14.0000    8.6144
   15.0000    9.2752
   16.0000    9.9372
   17.0000   10.6001
   18.0000   11.2635
   }{\ASKCovUpperBound}

\pgfdeclarelayer{background}
\pgfdeclarelayer{foreground}
\pgfsetlayers{background,main,foreground}
\input{plots/plot_settings}

\begin{tikzpicture}[]

\begin{axis}[%
yminorticks=true,
xmajorgrids,
ymajorgrids,
yminorgrids,
minor y tick num=4,
minor x tick num=1,
minor y tick num=4,
grid=both,
legend style={legend cell align=left,  draw=white!15!black, font=\scriptsize,  
},
legend style={at={(1.00,0.00)},anchor=south east},
xlabel style={font=\color{white!15!black}},
ylabel style={font=\color{white!15!black}},
axis background/.style={fill=white},
scale only axis,
width=\gridwidth,
height=\gridheight,
xmin=-5,
xmax=15,
xlabel={$\mathrm{P}_{\mathrm{tx}}$ in [dB]},
ymin=0,
ymax=3.001,
ylabel={Rate},
title={8-ASK},
ylabel shift = -4pt
]

\pgfplotstableread{
    X Y
  -5.000000000000000   0.127659464033783
  -4.000000000000000   0.183545551465475
  -3.000000000000000   0.283199792957735
  -2.000000000000000   0.396136952053306
  -1.000000000000000   0.570516691930347
                   0   0.790497707047351
   1.000000000000000   1.064669330409921
   2.000000000000000   1.355159123505398
   2.999999999999999   1.697875540692198
   4.000000000000000   2.089244923545045
   5.000000000000000   2.519118956528239
   6.000000000000000   2.996732195270778
   7.000000000000000   3.503883175120827
}{\ASKLoeligerUpperBoundXII}
 
\addplot [ASK,UB,black,opacity=0.8,name path global=UB] table {\ASKLoeligerUpperBoundXII};
\addlegendentry{UB, $\widetilde{N} = 7$};

\addplot[ASK,FBA,] table [x=power, y=rate, col sep=space]{plots/raw/Q8/TOBI/8-ASK_M=1_alpha=0.00_L=30km_mem=7.txt}; %
\addlegendentry{FBA, $\widetilde{N} = 7$}; %

\addplot[ASK,FBA,forget plot] table [x=power, y=rate, col sep=space]{plots/raw/Q8/TOBI/8-ASK_M=2_alpha=0.00_L=30km_mem=7.txt};%

\addplot[ASK,FBA,forget plot] table [x=power, y=rate, col sep=space]{plots/raw/Q8/TOBI/8-ASK_M=3_alpha=0.00_L=30km_mem=7.txt};%

\addplot[ASK,FBA,forget plot] table [x=power, y=rate, col sep=space]{plots/raw/Q8/TOBI/8-ASK_M=4_alpha=0.00_L=30km_mem=7.txt};%

\addplot[ASK,RNN] table [x=Ptxdb, y=SIC1, col sep=comma]{plots/V1.5/Q8_L30/v1.5,8-ASK,64_128_128_8,lr=3E-04,Tr=84,B=128,S=2,Ls=32,I=5.0E+04,V=1.0E+03,n=6.0E+04,P=151,a=0,L=3.0E+04,Rs=3.5E+10,C=4.6E+04.txt};
\addlegendentry{NN, $\widetilde{N}_\text{RNN}=115$}

\addplot[ASK,RNN,forget plot] table [x=Ptxdb, y=IqXY, col sep=comma]{plots/V1.5/Q8_L30/v1.5,8-ASK,64_128_128_8,lr=3E-04,Tr=84,B=128,S=2,Ls=32,I=5.0E+04,V=1.0E+03,n=6.0E+04,P=151,a=0,L=3.0E+04,Rs=3.5E+10,C=4.6E+04.txt};

\addplot[ASK,RNN,forget plot] table [x=Ptxdb, y=IqXY, col sep=comma]{plots/V1.5/Q8_L30/v1.5,8-ASK,64_128_128_8,lr=3E-04,Tr=84,B=128,S=3,Ls=32,I=5.0E+04,V=1.0E+03,n=6.0E+04,P=151,a=0,L=3.0E+04,Rs=3.5E+10,C=4.6E+04.txt};

\addplot[ASK,RNN,forget plot,name path global=s4ask] table [x=Ptxdb, y=IqXY, col sep=comma]{plots/V1.5/Q8_L30/v1.5,8-ASK,64_128_128_8,lr=3E-04,Tr=84,B=128,S=4,Ls=32,I=5.0E+04,V=1.0E+03,n=6.0E+04,P=151,a=0,L=3.0E+04,Rs=3.5E+10,C=4.6E+04.txt};

\addplot[ASK,RNN,forget plot,name path global=s6ask] table [x=Ptxdb, y=IqXY, col sep=comma]{plots/V1.5/Q8_L30/v1.5,8-ASK,64_128_128_8,lr=3E-04,Tr=120,B=128,S=6,Ls=32,I=5.0E+04,V=1.0E+03,n=6.0E+04,P=151,a=0,L=3.0E+04,Rs=3.5E+10,C=4.6E+04.txt};

\begin{pgfonlayer}{foreground}
\path[name path global=line] (axis cs:\pgfkeysvalueof{/pgfplots/xmin},2.4) -- (axis cs: \pgfkeysvalueof{/pgfplots/xmax},2.4);
\path[name intersections={of=line and UB, name=p1}, name intersections={of=line and s6ask, name=p2}];
\draw[arr] let \p1=(p1-1), \p2=(p2-1) in (p1-1) -- ([xshift=-0.0cm]p2-1) node [left,midway,opacitylabel,font=\footnotesize,yshift=0.0cm,xshift=-0.3cm] {%
	\pgfplotsconvertunittocoordinate{x}{\x1}%
	\pgfplotscoordmath{x}{datascaletrafo inverse to fixed}{\pgfmathresult}%
	\edef\valueA{\pgfmathresult}%
	\pgfplotsconvertunittocoordinate{x}{\x2}%
	\pgfplotscoordmath{x}{datascaletrafo inverse to fixed}{\pgfmathresult}%
	\pgfmathparse{\pgfmathresult - \valueA}%
	\pgfmathprintnumber{\pgfmathresult} dB
};
\end{pgfonlayer}

\end{axis}

\end{tikzpicture}%
    \end{minipage}%
    \begin{minipage}[t]{0.33\linewidth}
        \pgfdeclarelayer{background}
\pgfdeclarelayer{foreground}
\pgfsetlayers{background,main,foreground}

\input{plots/plot_settings}
\begin{tikzpicture}[]

\begin{axis}[%
yminorticks=true,
xmajorgrids,
ymajorgrids,
yminorgrids,
minor x tick num=1,
minor y tick num=4,
grid=both,
legend style={legend cell align=left,  draw=white!15!black, font=\scriptsize,  
},
legend style={at={(1.00,0.00)},anchor=south east},
xlabel style={font=\color{white!15!black}},
ylabel style={font=\color{white!15!black}},
axis background/.style={fill=white},
scale only axis,
width=\gridwidth,
height=\gridheight,
xmin=-5,
xmax=15,
xlabel={$\mathrm{P}_{\mathrm{tx}}$ in [dB]},
ymin=0,
ymax=3.001,
ylabel={Rate},
title={8-SQAM},
ylabel shift = -4pt
]

\pgfplotstableread{
    X Y
    -5.000000000000000   0.128162784543899
  -4.000000000000000   0.180217365128368
  -3.000000000000000   0.272820358365454
  -2.000000000000000   0.398325581834539
  -1.000000000000000   0.562751304907378
                   0   0.781331124302505
   1.000000000000000   1.045587119743641
   2.000000000000000   1.352991959882176
   2.999999999999999   1.710197075225446
   4.000000000000000   2.112611338739478
   5.000000000000000   2.555633349849050
   6.000000000000000   3.046513184341896
   7.000000000000000   3.563983953281022
}{\QAMLoeligerUpperBoundXII}

\addplot [QAM,UB,opacity=0.8,name path global=UB] table {\QAMLoeligerUpperBoundXII};
\addlegendentry{UB, $\widetilde{N} = 7$};

\addplot[QAM,FBA,] table [x=power, y=rate, col sep=space]{plots/raw/Q8/TOBI/8-SQAM_M=1_alpha=0.00_L=30km_mem=7.txt};%
\addlegendentry{FBA, $\widetilde{N} = 7$}; %

\addplot[QAM,FBA,forget plot] table [x=power, y=rate, col sep=space]{plots/raw/Q8/TOBI/8-SQAM_M=2_alpha=0.00_L=30km_mem=7.txt};%

\addplot[QAM,FBA,forget plot] table [x=power, y=rate, col sep=space]{plots/raw/Q8/TOBI/8-SQAM_M=3_alpha=0.00_L=30km_mem=7.txt};%

\addplot[QAM,FBA,forget plot] table [x=power, y=rate, col sep=space]{plots/raw/Q8/TOBI/8-SQAM_M=4_alpha=0.00_L=30km_mem=7.txt};%

\addplot[QAM,RNN] table [x=Ptxdb, y=SIC1, col sep=comma]{plots/V1.5/Q8_L30/v1.5,8-SQAM,64_256_128_128_8,lr=3E-04,Tr=84,B=128,S=2,Ls=32,I=5.0E+04,V=1.0E+03,n=6.0E+04,P=151,a=0,L=3.0E+04,Rs=3.5E+10,C=1.3E+05.txt}; \addlegendentry{NN, $\widetilde{N}_\text{RNN}=115$}

\addplot[QAM,RNN,forget plot] table [x=Ptxdb, y=IqXY, col sep=comma]{plots/V1.5/Q8_L30/v1.5,8-SQAM,64_256_128_128_8,lr=3E-04,Tr=84,B=128,S=2,Ls=32,I=5.0E+04,V=1.0E+03,n=6.0E+04,P=151,a=0,L=3.0E+04,Rs=3.5E+10,C=1.3E+05.txt}; 

\addplot[QAM,RNN,forget plot] table [x=Ptxdb, y=IqXY, col sep=comma]{plots/V1.5/Q8_L30/v1.5,8-SQAM,64_256_128_128_8,lr=3E-04,Tr=84,B=128,S=3,Ls=32,I=5.0E+04,V=1.0E+03,n=6.0E+04,P=151,a=0,L=3.0E+04,Rs=3.5E+10,C=1.3E+05.txt}; 

\addplot[QAM,RNN,forget plot,name path global=s4qam] table [x=Ptxdb, y=IqXY, col sep=comma]{plots/V1.5/Q8_L30/v1.5,8-SQAM,64_256_128_128_8,lr=3E-04,Tr=84,B=128,S=4,Ls=32,I=5.0E+04,V=1.0E+03,n=6.0E+04,P=151,a=0,L=3.0E+04,Rs=3.5E+10,C=1.3E+05.txt}; 

\addplot[QAM,RNN,forget plot,name path global=s6qam] table [x=Ptxdb, y=IqXY, col sep=comma]{plots/V1.5/Q8_L30/v1.5,8-SQAM,64_256_128_128_8,lr=3E-04,Tr=120,B=128,S=6,Ls=32,I=5.0E+04,V=1.0E+03,n=6.0E+04,P=151,a=0,L=3.0E+04,Rs=3.5E+10,C=1.3E+05.txt};

\begin{pgfonlayer}{foreground}
\path[name path global=line] (axis cs:\pgfkeysvalueof{/pgfplots/xmin},2.4) -- (axis cs: \pgfkeysvalueof{/pgfplots/xmax},2.4);
\path[name intersections={of=line and UB, name=p1}, name intersections={of=line and s6qam, name=p2}];
\draw[arr] let \p1=(p1-1), \p2=(p2-1) in (p1-1) -- ([xshift=-0.0cm]p2-1) node [left,midway,opacitylabel,font=\footnotesize,yshift=0.0cm,xshift=-0.3cm] {%
	\pgfplotsconvertunittocoordinate{x}{\x1}%
	\pgfplotscoordmath{x}{datascaletrafo inverse to fixed}{\pgfmathresult}%
	\edef\valueA{\pgfmathresult}%
	\pgfplotsconvertunittocoordinate{x}{\x2}%
	\pgfplotscoordmath{x}{datascaletrafo inverse to fixed}{\pgfmathresult}%
	\pgfmathparse{\pgfmathresult - \valueA}%
	\pgfmathprintnumber{\pgfmathresult} dB
};
\end{pgfonlayer}

\end{axis}

\end{tikzpicture}%
    \end{minipage}%
\caption{SIC rates for $L_\text{fib}=\SI{30}{\kilo\meter}$, $S=1,2,3,4$ stages and an additional plot with $S=6$ stages for the NN. Mismatched FBA and UB memory $\widetilde{N} = 7$, RNN memory $\widetilde{N}_\text{RNN} = 115$. The lower curves of a particular style show the SDD ($S=1$) rates, while the upper curves show the largest $S$ rates.}
\label{fig:Q8_L=30km}
\end{figure*}

Fig.~\ref{fig:rnn_vs_gibbs_q32} and Fig.~\ref{fig:rnn_vs_gibbs_q32_L30} compare the performance of NNs and bit-wise GS for 32-PAM/ASK/SQAM with $L_\text{fib} = 0$ and $S=2$, and $L_\text{fib} =\SI{30}{\kilo\meter}$ and $S=6$, respectively. The FBA is infeasible for large modulation alphabets. GS runs with $N_\text{par} = 64$ samplers, $\widetilde{N} = 21$, and $N_\text{iter} = 125$. The first $25$ iterations constitute the ``burn-in'' period and are discarded. The performance of the NN and GS match at low SNRs, except for 32-SQAM and $L_\text{fib} = \SI{30}{\kilo\meter}$ where GS is slightly better. At high SNRs, GS stalls; see~\cite{senst2011rao} and~\cite[Fig.~8]{prinz2023successive}. The NN achieves the maximum rates for 32-PAM/ASK and saturates slightly earlier for 32-SQAM. The NN outperforms GS and is over 300 times less complex for $L_\text{fib} =\SI{0}{}$ and roughly 100 times less complex for $L_\text{fib} =\SI{30}{\kilo\meter}$.

\begin{figure}
         \centering
         \input{plots/plot_settings}
\begin{tikzpicture}[]

\begin{axis}[%
yminorticks=true,
xmajorgrids,
ymajorgrids,
yminorgrids,
minor x tick num=1,
minor y tick num=5,
grid=both,
legend style={legend cell align=left,  draw=white!15!black, font=\footnotesize,  
legend pos=south east
},
xlabel style={font=\color{white!15!black}},
ylabel style={font=\color{white!15!black}},
axis background/.style={fill=white},
scale only axis,
width=1.50*\gridwidth,
height=1.0*\gridheight,
xmin=-4,
xmax=24,
xlabel={$\mathrm{P}_{\mathrm{tx}}$ in [dB]},
ymin=0,
ymax=5.01,
ytick={0,1,2,3,4,5},
ylabel={Rate},
title={GS versus NN},
ylabel shift = -4pt
]

\pgfplotstableread[col sep=space]{plots/V1.3/GS/paper__Gibbs__32-PAM__S=2__alpha=0.00__L=0km__mem=21__it=100__par=64__n=40000__burn=25.txt}\tmpPAM

\pgfplotstablecreatecol[col sep=space, create col/copy column from table={plots/V1.3/GS/paper__Gibbs__32-PAM__S=2__alpha=0.00__L=0km__mem=21__it=100__par=64__n=40000__burn=25_2.txt}{rate}]{rate2}{\tmpPAM}

\pgfplotstablecreatecol[col sep=space, create col/copy column from table={plots/V1.3/GS/paper__Gibbs__32-PAM__S=2__alpha=0.00__L=0km__mem=21__it=100__par=64__n=40000__burn=25_3.txt}{rate}]{rate3}{\tmpPAM}

\pgfplotstablecreatecol[col sep=space, create col/copy column from table={plots/V1.3/GS/paper__Gibbs__32-PAM__S=2__alpha=0.00__L=0km__mem=21__it=100__par=64__n=40000__burn=25_4.txt}{rate}]{rate4}{\tmpPAM}

\pgfplotstablecreatecol[col sep=space, create col/copy column from table={plots/V1.3/GS/paper__Gibbs__32-PAM__S=2__alpha=0.00__L=0km__mem=21__it=100__par=64__n=40000__burn=25_5.txt}{rate}]{rate5}{\tmpPAM}

\pgfplotstablecreatecol[col sep=space, create col/copy column from table={plots/V1.3/GS/paper__Gibbs__32-PAM__S=2__alpha=0.00__L=0km__mem=21__it=100__par=64__n=40000__burn=25_6.txt}{rate}]{rate6}{\tmpPAM}

\addplot[GS,PAM,forget plot] table [col sep=space,x=power, y expr={ 1/6*( \thisrow{rate} +  \thisrow{rate2} + \thisrow{rate3} + \thisrow{rate4} + \thisrow{rate5}  + \thisrow{rate6}) }] {\tmpPAM};

\pgfplotstableread[col sep=space]{plots/V1.3/GS/paper__Gibbs__32-ASK__S=2__alpha=0.00__L=0km__mem=21__it=100__par=64__n=40000__burn=25.txt}\tmpASK

\pgfplotstablecreatecol[col sep=space, create col/copy column from table={plots/V1.3/GS/paper__Gibbs__32-ASK__S=2__alpha=0.00__L=0km__mem=21__it=100__par=64__n=40000__burn=25_2.txt}{rate}]{rate2}{\tmpASK}

\pgfplotstablecreatecol[col sep=space, create col/copy column from table={plots/V1.3/GS/paper__Gibbs__32-ASK__S=2__alpha=0.00__L=0km__mem=21__it=100__par=64__n=40000__burn=25_3.txt}{rate}]{rate3}{\tmpASK}

\pgfplotstablecreatecol[col sep=space, create col/copy column from table={plots/V1.3/GS/paper__Gibbs__32-ASK__S=2__alpha=0.00__L=0km__mem=21__it=100__par=64__n=40000__burn=25_4.txt}{rate}]{rate4}{\tmpASK}

\addplot[GS,ASK,forget plot] table [col sep=space,x=power, y expr={1/4*( \thisrow{rate} + \thisrow{rate2}  +  \thisrow{rate3}  +  \thisrow{rate4} )
}] {\tmpASK};

\pgfplotstableread[col sep=space]{plots/V1.3/GS/paper__Gibbs__32-ring-qam__S=2__alpha=0.00__L=0km__mem=21__it=100__par=64__n=40000__burn=25__m=3-2__phase-offset=0__jlt_spacing=1.txt}\tmpSQAM

\pgfplotstablecreatecol[col sep=space, create col/copy column from table={plots/V1.3/GS/paper__Gibbs__32-ring-qam__S=2__alpha=0.00__L=0km__mem=21__it=100__par=64__n=40000__burn=25__m=3-2__phase-offset=0__jlt_spacing=1_2.txt}{rate}]{rate2}{\tmpSQAM}

\addplot[GS,QAM,forget plot] table [col sep=space,x=power, y expr={1/2*( \thisrow{rate} +  \thisrow{rate2} )
}] {\tmpSQAM};

\addplot[PAM] table [x=Ptxdb, y=IqXY, col sep=comma]{plots/V1.5/Q32/v1.5,32-PAM,84_128_128_32,lr=1E-04,Tr=64,B=64,S=2,Ls=64,I=6.0E+04,V=3.0E+03,n=6.0E+04,P=151,a=0,L=0.0E+00,Rs=3.5E+10,C=5.6E+04.txt};
\addlegendentry{32-PAM}

\addplot[ASK] table [x=Ptxdb, y=IqXY, col sep=comma]{plots/V1.5/Q32/v1.5,32-ASK,84_128_128_32,lr=1E-04,Tr=64,B=64,S=2,Ls=64,I=6.0E+04,V=3.0E+03,n=6.0E+04,P=151,a=0,L=0.0E+00,Rs=3.5E+10,C=5.6E+04.txt};
\addlegendentry{32-ASK}

\addplot[QAM] table [x=Ptxdb, y=IqXY, col sep=comma]{plots/V1.5/Q32/v1.5,32-SQAM,84_128_128_32,lr=1E-04,Tr=64,B=64,S=2,Ls=64,I=6.0E+04,V=3.0E+03,n=6.0E+04,P=151,a=0,L=0.0E+00,Rs=3.5E+10,C=6.4E+04.txt};
\addlegendentry{32-SQAM}

\draw (axis cs:16.8,3.9) ellipse (0.12cm and 0.35cm) node[below,yshift=-0.47cm,align=center,text width=1.2cm,font=\footnotesize,opacitylabel]{Gibbs\\ Sampling};
\draw (axis cs:12,4.2) ellipse (0.5cm and 0.12cm) node[left,xshift=-0.68cm,font=\footnotesize,opacitylabel]{NN};

\end{axis}

\end{tikzpicture}%
         \caption{SIC rates for $L_\text{fib}=0$ and $S=2$. RNN: $\widetilde{N}_\text{RNN}=105$. Bit-wise GS~\cite{prinz2023successive}: $\widetilde{N}=21$, $N_\text{iter}=125$ and $N_\text{par}=64$. }
         \label{fig:rnn_vs_gibbs_q32}
\end{figure}
\begin{figure}
         \centering
         \input{plots/plot_settings}
\begin{tikzpicture}[]

\begin{axis}[%
yminorticks=true,
xmajorgrids,
ymajorgrids,
yminorgrids,
minor x tick num=1,
minor y tick num=5,
grid=both,
legend style={legend cell align=left,  draw=white!15!black, font=\footnotesize,  
legend pos=south east
},
xlabel style={font=\color{white!15!black}},
ylabel style={font=\color{white!15!black}},
axis background/.style={fill=white},
scale only axis,
width=1.50*\gridwidth,
height=1.0*\gridheight,
xmin=-4,
xmax=24,
xlabel={$\mathrm{P}_{\mathrm{tx}}$ in [dB]},
ymin=0,
ymax=5.01,
ytick={0,1,2,3,4,5},
ylabel={Rate},
title={GS versus NN},
ylabel shift = -4pt
]

\pgfplotstableread[col sep=space]{plots/V1.3/GS/paper__Gibbs__32-PAM__S=6__alpha=0.00__L=30km__mem=21__it=100__par=64__n=40000__burn=25.txt}\tmpPAM
\addplot[GS,PAM,forget plot] table [col sep=space,x=power, y expr={ \thisrow{rate} }] {\tmpPAM};

\pgfplotstableread[col sep=space]{plots/V1.3/GS/paper__Gibbs__32-ASK__S=6__alpha=0.00__L=30km__mem=21__it=100__par=64__n=40000__burn=25.txt}\tmpASK

\addplot[GS,ASK,forget plot] table [col sep=space,x=power, y expr={\thisrow{rate}}] {\tmpASK};

\pgfplotstableread[col sep=space]{plots/V1.3/GS/paper__Gibbs__32-ring-qam__S=6__alpha=0.00__L=30km__mem=21__it=100__par=64__n=40000__burn=25__m=3-2__phase-offset=0__jlt_spacing=1.txt}\tmpSQAM

\addplot[GS,QAM,forget plot] table [col sep=space,x=power, y expr={\thisrow{rate} }] {\tmpSQAM};

\addplot[PAM] table [x=Ptxdb, y=IqXY, col sep=comma]{plots/V1.5/Q32_L30/v1.5,32-PAM,100_200_200_200_168_32,lr=4E-05,Tr=120,B=64,S=6,Ls=64,I=1.0E+05,V=7.0E+03,n=8.0E+04,P=151,a=0,L=3.0E+04,Rs=3.5E+10,C=2.3E+05.txt}; 
\addlegendentry{32-PAM}; 

\addplot[ASK] table [x=Ptxdb, y=IqXY, col sep=comma]{plots/V1.5/Q32_L30/v1.5,32-ASK,100_200_200_200_168_32,lr=4E-05,Tr=120,B=64,S=6,Ls=64,I=1.0E+05,V=7.0E+03,n=8.0E+04,P=151,a=0,L=3.0E+04,Rs=3.5E+10,C=2.3E+05.txt}; 
\addlegendentry{32-ASK}; 

\addplot[QAM] table [x=Ptxdb, y=IqXY, col sep=comma]{plots/V1.5/Q32_L30/v1.5,32-SQAM,100_200_200_200_168_32,lr=4E-05,Tr=120,B=64,S=6,Ls=64,I=1.0E+05,V=7.0E+03,n=8.0E+04,P=151,a=0,L=3.0E+04,Rs=3.5E+10,C=2.4E+05.txt};
\addlegendentry{32-SQAM};

\draw (axis cs:16.7,3.2) ellipse (0.12cm and 0.5cm) node[below,yshift=-0.55cm,align=center,text width=1.2cm,font=\footnotesize,opacitylabel]{Gibbs\\ Sampling};
\draw (axis cs:11.0,3.5) ellipse (0.5cm and 0.12cm) node[left,xshift=-0.68cm,font=\footnotesize,opacitylabel]{NN};

\end{axis}

\end{tikzpicture}%
         \caption{SIC rates for $L_\text{fib}=\SI{30}{\kilo\meter}$ and $S=6$. RNN: $\widetilde{N}_\text{RNN}=169$. Bit-wise GS~\cite{prinz2023successive}: $\widetilde{N}=21$, $N_\text{iter}=125$ and $N_\text{par}=64$.}
         \label{fig:rnn_vs_gibbs_q32_L30}
\end{figure}

\begin{figure*}
    \centering
    \begin{minipage}[t]{0.32\textwidth}
    \centering
        \input{plots/plot_settings}
\begin{tikzpicture}[]

\begin{axis}[%
yminorticks=true,
xmajorgrids,
ymajorgrids,
yminorgrids,
minor x tick num=1,
minor y tick num=4,
grid=both,
legend style={legend cell align=left,  draw=white!15!black, font=\footnotesize,  %
legend pos=south east
},
xlabel style={font=\color{white!15!black}},
ylabel style={font=\color{white!15!black}},
axis background/.style={fill=white},
scale only axis,
width=1*\gridwidth,
height=1*\gridheight,
xmin=-4,
xmax=24,
xlabel={$\mathrm{P}_{\mathrm{tx}}$ in [dB]},
ymin=0,
ymax=5.01,
ytick={0,1,2,3,4,5},
ylabel={Rate},
title={32-PAM},
ylabel shift = -4pt
]

\addplot[PAM,black,forget plot,postaction={decorate,decoration={raise=-1.5ex,text along path,text color={black},text align={left, left indent=4.0cm},text={|\footnotesize|SDD}}}] table [x=Ptxdb, y=SIC1, col sep=comma]{plots/V1.5/Q32/v1.5,32-PAM,84_128_128_32,lr=1E-04,Tr=64,B=64,S=2,Ls=64,I=6.0E+04,V=3.0E+03,n=6.0E+04,P=151,a=0,L=0.0E+00,Rs=3.5E+10,C=5.6E+04.txt}; %

\addplot[PAM,forget plot] table [x=Ptxdb, y=IqXY, col sep=comma]{plots/V1.5/Q32/v1.5,32-PAM,84_128_128_32,lr=1E-04,Tr=64,B=64,S=2,Ls=64,I=6.0E+04,V=3.0E+03,n=6.0E+04,P=151,a=0,L=0.0E+00,Rs=3.5E+10,C=5.6E+04.txt};

\addplot[PAM,forget plot] table [x=Ptxdb, y=IqXY, col sep=comma]{plots/V1.5/Q32/v1.5,32-PAM,84_128_128_32,lr=1E-04,Tr=64,B=64,S=3,Ls=64,I=6.0E+04,V=3.0E+03,n=6.0E+04,P=151,a=0,L=0.0E+00,Rs=3.5E+10,C=5.6E+04.txt};

\addplot[PAM,forget plot] table [x=Ptxdb, y=IqXY, col sep=comma]{plots/V1.5/Q32/v1.5,32-PAM,84_128_128_32,lr=1E-04,Tr=66,B=64,S=4,Ls=64,I=6.0E+04,V=3.0E+03,n=6.0E+04,P=151,a=0,L=0.0E+00,Rs=3.5E+10,C=5.6E+04.txt};

\addplot[PAM,forget plot,dashed,black] table [x=Ptxdb, y=IqXY, col sep=comma]{plots/V1.5/Q32/v1.5,32-PAM,84_128_128_32,lr=1E-04,Tr=120,B=64,S=6,Ls=64,I=6.0E+04,V=3.0E+03,n=6.0E+04,P=151,a=0,L=0.0E+00,Rs=3.5E+10,C=5.6E+04.txt};

\end{axis}

\end{tikzpicture}%
    \end{minipage}
    \begin{minipage}[t]{0.32\textwidth}
    \centering
        \input{plots/plot_settings}
\begin{tikzpicture}[]

\begin{axis}[%
yminorticks=true,
xmajorgrids,
ymajorgrids,
yminorgrids,
minor x tick num=1,
minor y tick num=4,
grid=both,
legend style={legend cell align=left,  draw=white!15!black, font=\footnotesize,  %
legend pos=south east
},
xlabel style={font=\color{white!15!black}},
ylabel style={font=\color{white!15!black}},
axis background/.style={fill=white},
scale only axis,
width=1*\gridwidth,
height=1*\gridheight,
xmin=-4,
xmax=24,
xlabel={$\mathrm{P}_{\mathrm{tx}}$ in [dB]},
ymin=0,
ymax=5.01,
ytick={0,1,2,3,4,5},
ylabel={Rate},
title={32-ASK},
ylabel shift = -4pt,
restrict y to domain=0:5
]

\addplot[ASK,black,forget plot,postaction={decorate,decoration={raise=-1.5ex,text along path,text color={black},text align={left, left indent=3.6cm},text={|\footnotesize|SDD}}}] table [x=Ptxdb, y=SIC1, col sep=comma]{plots/V1.5/Q32/v1.5,32-ASK,84_128_128_32,lr=1E-04,Tr=64,B=64,S=2,Ls=64,I=6.0E+04,V=3.0E+03,n=6.0E+04,P=151,a=0,L=0.0E+00,Rs=3.5E+10,C=5.6E+04.txt}; %

\addplot[ASK,forget plot] table [x=Ptxdb, y=IqXY, col sep=comma]{plots/V1.5/Q32/v1.5,32-ASK,84_128_128_32,lr=1E-04,Tr=64,B=64,S=2,Ls=64,I=6.0E+04,V=3.0E+03,n=6.0E+04,P=151,a=0,L=0.0E+00,Rs=3.5E+10,C=5.6E+04.txt};

\addplot[ASK,forget plot] table [x=Ptxdb, y=IqXY, col sep=comma]{plots/V1.5/Q32/v1.5,32-ASK,84_128_128_32,lr=1E-04,Tr=64,B=64,S=3,Ls=64,I=6.0E+04,V=3.0E+03,n=6.0E+04,P=151,a=0,L=0.0E+00,Rs=3.5E+10,C=5.6E+04.txt};

\addplot[ASK,forget plot] table [x=Ptxdb, y=IqXY, col sep=comma]{plots/V1.5/Q32/v1.5,32-ASK,84_128_128_32,lr=1E-04,Tr=66,B=64,S=4,Ls=64,I=6.0E+04,V=3.0E+03,n=6.0E+04,P=151,a=0,L=0.0E+00,Rs=3.5E+10,C=5.6E+04.txt};

\addplot[ASK,forget plot,dashed, black] table [x=Ptxdb, y=IqXY, col sep=comma]{plots/V1.5/Q32/v1.5,32-ASK,84_128_128_32,lr=1E-04,Tr=120,B=64,S=6,Ls=64,I=6.0E+04,V=3.0E+03,n=6.0E+04,P=151,a=0,L=0.0E+00,Rs=3.5E+10,C=5.6E+04.txt};

\end{axis}

\end{tikzpicture}%
    \end{minipage}
    \begin{minipage}[t]{0.32\textwidth}
    \centering
        \input{plots/plot_settings}
\begin{tikzpicture}[]

\begin{axis}[%
yminorticks=true,
xmajorgrids,
ymajorgrids,
yminorgrids,
minor x tick num=1,
minor y tick num=4,
grid=both,
legend style={legend cell align=left,  draw=white!15!black, font=\footnotesize,  %
legend pos=south east
},
xlabel style={font=\color{white!15!black}},
ylabel style={font=\color{white!15!black}},
axis background/.style={fill=white},
scale only axis,
width=1*\gridwidth,
height=1*\gridheight,
xmin=-4,
xmax=24,
xlabel={$\mathrm{P}_{\mathrm{tx}}$ in [dB]},
ymin=0,
ymax=5.01,
ytick={0,1,2,3,4,5},
ylabel={Rate},
title={32-SQAM},
ylabel shift = -4pt
]

\addplot[QAM,black,forget plot,postaction={decorate,decoration={raise=-1.5ex,text along path,text color={black},text align={left, left indent=4.0cm},text={|\footnotesize|SDD}}}] table [x=Ptxdb, y=SIC1, col sep=comma]{plots/V1.5/Q32/v1.5,32-SQAM,84_128_128_32,lr=1E-04,Tr=64,B=64,S=2,Ls=64,I=6.0E+04,V=3.0E+03,n=6.0E+04,P=151,a=0,L=0.0E+00,Rs=3.5E+10,C=6.4E+04.txt}; %

\addplot[QAM,forget plot] table [x=Ptxdb, y=IqXY, col sep=comma]{plots/V1.5/Q32/v1.5,32-SQAM,84_128_128_32,lr=1E-04,Tr=64,B=64,S=2,Ls=64,I=6.0E+04,V=3.0E+03,n=6.0E+04,P=151,a=0,L=0.0E+00,Rs=3.5E+10,C=6.4E+04.txt};

\addplot[QAM,forget plot] table [x=Ptxdb, y=IqXY, col sep=comma]{plots/V1.5/Q32/v1.5,32-SQAM,84_128_128_32,lr=1E-04,Tr=64,B=64,S=3,Ls=64,I=6.0E+04,V=3.0E+03,n=6.0E+04,P=151,a=0,L=0.0E+00,Rs=3.5E+10,C=6.4E+04.txt};

\addplot[QAM,forget plot] table [x=Ptxdb, y=IqXY, col sep=comma]{plots/V1.5/Q32/v1.5,32-SQAM,84_128_128_32,lr=1E-04,Tr=66,B=64,S=4,Ls=64,I=6.0E+04,V=3.0E+03,n=6.0E+04,P=151,a=0,L=0.0E+00,Rs=3.5E+10,C=6.4E+04.txt};

\addplot[QAM,forget plot,black,dashed] table [x=Ptxdb, y=IqXY, col sep=comma]{plots/V1.5/Q32/v1.5,32-SQAM,84_128_128_32,lr=1E-04,Tr=120,B=64,S=6,Ls=64,I=6.0E+04,V=3.0E+03,n=6.0E+04,P=151,a=0,L=0.0E+00,Rs=3.5E+10,C=6.4E+04.txt};

\end{axis}

\end{tikzpicture}%
    \end{minipage}
    \caption{SIC rates with $S=1,2,3,4,6$, $L_\text{fib}=0$ and $\widetilde{N}_\text{RNN}=105$. The lower curves show the SDD ($S=1$) rates, while the upper curves show the $S=6$ rates.}
    \label{fig:Q32_L=0km_SIC_stage_rate}
 \end{figure*}

The results in Fig.~\ref{fig:Q32_L=0km_SIC_stage_rate}-\ref{fig:Q8-Q128} consider NN-SIC only because GS ``stalls'' and the FBA is infeasible. The rates for 32-PAM/ASK/SQAM, $L_\text{fib}=0$ and $L_\text{fib}=\SI{30}{\kilo\meter}$ are shown in Fig.~\ref{fig:Q32_L=0km_SIC_stage_rate} and Fig.~\ref{fig:Q32_L=30km_SIC_stage_rate}, respectively. We use SIC with up to six stages; the dashed black curves show the rates for stage six. For $L_\text{fib}=0$, 32-ASK and 32-SQAM require two SIC stages to approach JDD performance, while 32-PAM performs well even with SDD.
For $L_\text{fib}=\SI{30}{\kilo\meter}$, 32-PAM and 32-ASK require at least four stages, while 32-SQAM may need more than six SIC stages. 

Fig.~\ref{fig:Q8-Q128_L=0km} compares NN-SIC rates for $L_\text{fib}=\SI{0}{}$, $S=2$ and modulations up to $M=128$. ASK/SQAM with NN-SIC achieves large energy gains over classic PAM. The gains increase with $M$ and reach $\approx \SI{3}{dB}$ for $M=128$.
Fig.~\ref{fig:Q8-Q128_L=30km} plots rates for $L_\text{fib}=\SI{30}{\kilo\meter}$, $S=6$ and modulations up to $M=64$. The gains over PAM are $\approx \SI{2.1}{dB}$. The SQAM rates saturate because SQAM has phase symmetries; see~\cite[Appendix]{prinz2023successive}.
The NN complexity scales roughly linearly with the modulation alphabet size $M$.
\begin{figure*}
    \centering
    \begin{minipage}[t]{0.32\textwidth}
    \centering
     \input{plots/plot_settings}
\begin{tikzpicture}[]

\begin{axis}[%
yminorticks=true,
xmajorgrids,
ymajorgrids,
yminorgrids,
minor x tick num=1,
minor y tick num=4,
grid=both,
legend style={legend cell align=left,  draw=white!15!black, font=\footnotesize,  legend pos=south east
},
xlabel style={font=\color{white!15!black}},
ylabel style={font=\color{white!15!black}},
axis background/.style={fill=white},
scale only axis,
width=1*\gridwidth,
height=1*\gridheight,
xmin=-4,
xmax=24,
xlabel={$\mathrm{P}_{\mathrm{tx}}$ in [dB]},
ymin=0,
ymax=5.01,
ytick={0,1,2,3,4,5},
ylabel={Rate},
title={32-PAM},
ylabel shift = -4pt
]

\addplot[PAM,black,forget plot,postaction={decorate,decoration={raise=-1.5ex,text along path,text color={black},text align={left, left indent=4.0cm},text={|\footnotesize|SDD}}}] table [x=Ptxdb, y=SIC1, col sep=comma]{plots/V1.5/Q32_L30/v1.5,32-PAM,100_200_200_200_168_32,lr=4E-05,Tr=120,B=64,S=2,Ls=64,I=1.0E+05,V=7.0E+03,n=8.0E+04,P=151,a=0,L=3.0E+04,Rs=3.5E+10,C=2.3E+05.txt}; %

\addplot[PAM,forget plot] table [x=Ptxdb, y=IqXY, col sep=comma]{plots/V1.5/Q32_L30/v1.5,32-PAM,100_200_200_200_168_32,lr=4E-05,Tr=120,B=64,S=2,Ls=64,I=1.0E+05,V=7.0E+03,n=8.0E+04,P=151,a=0,L=3.0E+04,Rs=3.5E+10,C=2.3E+05.txt};

\addplot[PAM,forget plot] table [x=Ptxdb, y=IqXY, col sep=comma]{plots/V1.5/Q32_L30/v1.5,32-PAM,100_200_200_200_168_32,lr=4E-05,Tr=120,B=64,S=3,Ls=64,I=1.0E+05,V=7.0E+03,n=8.0E+04,P=151,a=0,L=3.0E+04,Rs=3.5E+10,C=2.3E+05.txt};

\addplot[PAM,forget plot] table [x=Ptxdb, y=IqXY, col sep=comma]{plots/V1.5/Q32_L30/v1.5,32-PAM,100_200_200_200_168_32,lr=4E-05,Tr=120,B=64,S=4,Ls=64,I=1.0E+05,V=7.0E+03,n=8.0E+04,P=151,a=0,L=3.0E+04,Rs=3.5E+10,C=2.3E+05.txt};

\addplot[PAM,forget plot,dashed,black] table [x=Ptxdb, y=IqXY, col sep=comma]{plots/V1.5/Q32_L30/v1.5,32-PAM,100_200_200_200_168_32,lr=4E-05,Tr=120,B=64,S=6,Ls=64,I=1.0E+05,V=7.0E+03,n=8.0E+04,P=151,a=0,L=3.0E+04,Rs=3.5E+10,C=2.3E+05.txt};

\end{axis}

\end{tikzpicture}%
    \end{minipage}
    \begin{minipage}[t]{0.32\textwidth}
    \centering
     \input{plots/plot_settings}
\begin{tikzpicture}[]

\begin{axis}[%
yminorticks=true,
xmajorgrids,
ymajorgrids,
yminorgrids,
minor x tick num=1,
minor y tick num=4,
grid=both,
legend style={legend cell align=left,  draw=white!15!black, font=\footnotesize,  legend pos=south east
},
xlabel style={font=\color{white!15!black}},
ylabel style={font=\color{white!15!black}},
axis background/.style={fill=white},
scale only axis,
width=1*\gridwidth,
height=1*\gridheight,
xmin=-4,
xmax=24,
xlabel={$\mathrm{P}_{\mathrm{tx}}$ in [dB]},
ymin=0,
ymax=5.0,
ytick={0,1,2,3,4,5},
ylabel={Rate},
title={32-ASK},
ylabel shift = -4pt
]

\addplot[black,forget plot,postaction={decorate,decoration={raise=-1.5ex,text along path,text color={black},text align={left, left indent=4.0cm},text={|\footnotesize|SDD}}}] table [x=Ptxdb, y=SIC1, col sep=comma]{plots/V1.5/Q32_L30/v1.5,32-ASK,100_200_200_200_168_32,lr=4E-05,Tr=120,B=64,S=2,Ls=64,I=1.0E+05,V=7.0E+03,n=8.0E+04,P=151,a=0,L=3.0E+04,Rs=3.5E+10,C=2.3E+05.txt}; %

\addplot[ASK,forget plot] table [x=Ptxdb, y=IqXY, col sep=comma]{plots/V1.5/Q32_L30/v1.5,32-ASK,100_200_200_200_168_32,lr=4E-05,Tr=120,B=64,S=2,Ls=64,I=1.0E+05,V=7.0E+03,n=8.0E+04,P=151,a=0,L=3.0E+04,Rs=3.5E+10,C=2.3E+05.txt}; 

\addplot[ASK,forget plot] table [x=Ptxdb, y=IqXY, col sep=comma]{plots/V1.5/Q32_L30/v1.5,32-ASK,100_200_200_200_168_32,lr=4E-05,Tr=120,B=64,S=3,Ls=64,I=1.0E+05,V=7.0E+03,n=8.0E+04,P=151,a=0,L=3.0E+04,Rs=3.5E+10,C=2.3E+05.txt}; 

\addplot[ASK,forget plot] table [x=Ptxdb, y=IqXY, col sep=comma]{plots/V1.5/Q32_L30/v1.5,32-ASK,100_200_200_200_168_32,lr=4E-05,Tr=120,B=64,S=4,Ls=64,I=1.0E+05,V=7.0E+03,n=8.0E+04,P=151,a=0,L=3.0E+04,Rs=3.5E+10,C=2.3E+05.txt}; 

\addplot[ASK,forget plot,densely dashed,black] table [x=Ptxdb, y=IqXY, col sep=comma]{plots/V1.5/Q32_L30/v1.5,32-ASK,100_200_200_200_168_32,lr=4E-05,Tr=120,B=64,S=6,Ls=64,I=1.0E+05,V=7.0E+03,n=8.0E+04,P=151,a=0,L=3.0E+04,Rs=3.5E+10,C=2.3E+05.txt};

\end{axis}

\end{tikzpicture}%
    \end{minipage}
    \begin{minipage}[t]{0.32\textwidth}
    \centering
     \input{plots/plot_settings}
\begin{tikzpicture}[]

\begin{axis}[%
yminorticks=true,
xmajorgrids,
ymajorgrids,
yminorgrids,
minor x tick num=1,
minor y tick num=4,
grid=both,
legend style={legend cell align=left,  draw=white!15!black, font=\footnotesize,  at={(0.02,0.98)},anchor=north west
},
xlabel style={font=\color{white!15!black}},
ylabel style={font=\color{white!15!black}},
axis background/.style={fill=white},
scale only axis,
width=1*\gridwidth,
height=1*\gridheight,
xmin=-4,
xmax=24,
xlabel={$\mathrm{P}_{\mathrm{tx}}$ in [dB]},
ymin=0,
ymax=5.01,
ytick={0,1,2,3,4,5},
ylabel={Rate},
title={32-SQAM},
ylabel shift = -4pt
]

\addplot[QAM,black,forget plot,postaction={decorate,decoration={raise=-1.5ex,text along path,text color={black},text align={left, left indent=4.0cm},text={|\footnotesize|SDD}}}] table [x=Ptxdb, y=SIC1, col sep=comma]{plots/V1.5/Q32_L30/v1.5,32-SQAM,100_200_200_200_168_32,lr=4E-05,Tr=120,B=64,S=2,Ls=64,I=1.0E+05,V=7.0E+03,n=8.0E+04,P=151,a=0,L=3.0E+04,Rs=3.5E+10,C=2.4E+05.txt}; %

\addplot[QAM,forget plot] table [x=Ptxdb, y=IqXY, col sep=comma]{plots/V1.5/Q32_L30/v1.5,32-SQAM,100_200_200_200_168_32,lr=4E-05,Tr=120,B=64,S=2,Ls=64,I=1.0E+05,V=7.0E+03,n=8.0E+04,P=151,a=0,L=3.0E+04,Rs=3.5E+10,C=2.4E+05.txt};

\addplot[QAM,forget plot] table [x=Ptxdb, y=IqXY, col sep=comma]{plots/V1.5/Q32_L30/v1.5,32-SQAM,100_200_200_200_168_32,lr=4E-05,Tr=120,B=64,S=3,Ls=64,I=1.0E+05,V=7.0E+03,n=8.0E+04,P=151,a=0,L=3.0E+04,Rs=3.5E+10,C=2.4E+05.txt}; 

\addplot[QAM,forget plot] table [x=Ptxdb, y=IqXY, col sep=comma]{plots/V1.5/Q32_L30/v1.5,32-SQAM,100_200_200_200_168_32,lr=4E-05,Tr=120,B=64,S=4,Ls=64,I=1.0E+05,V=7.0E+03,n=8.0E+04,P=151,a=0,L=3.0E+04,Rs=3.5E+10,C=2.4E+05.txt}; 

\addplot[QAM,forget plot,densely dashed, black] table [x=Ptxdb, y=IqXY, col sep=comma]{plots/V1.5/Q32_L30/v1.5,32-SQAM,100_200_200_200_168_32,lr=4E-05,Tr=120,B=64,S=6,Ls=64,I=1.0E+05,V=7.0E+03,n=8.0E+04,P=151,a=0,L=3.0E+04,Rs=3.5E+10,C=2.4E+05.txt};

\end{axis}

\end{tikzpicture}%
 \end{minipage}
    \caption{SIC rates with $S=1,2,3,4,6$, $L_\text{fib}=\SI{30}{\kilo\meter}$ and $\widetilde{N}_\text{RNN}=169$. The lower curves show the SDD ($S=1$) rates, while the upper curves show the $S=6$ rates.}
 \label{fig:Q32_L=30km_SIC_stage_rate}
\end{figure*}
\begin{figure*}
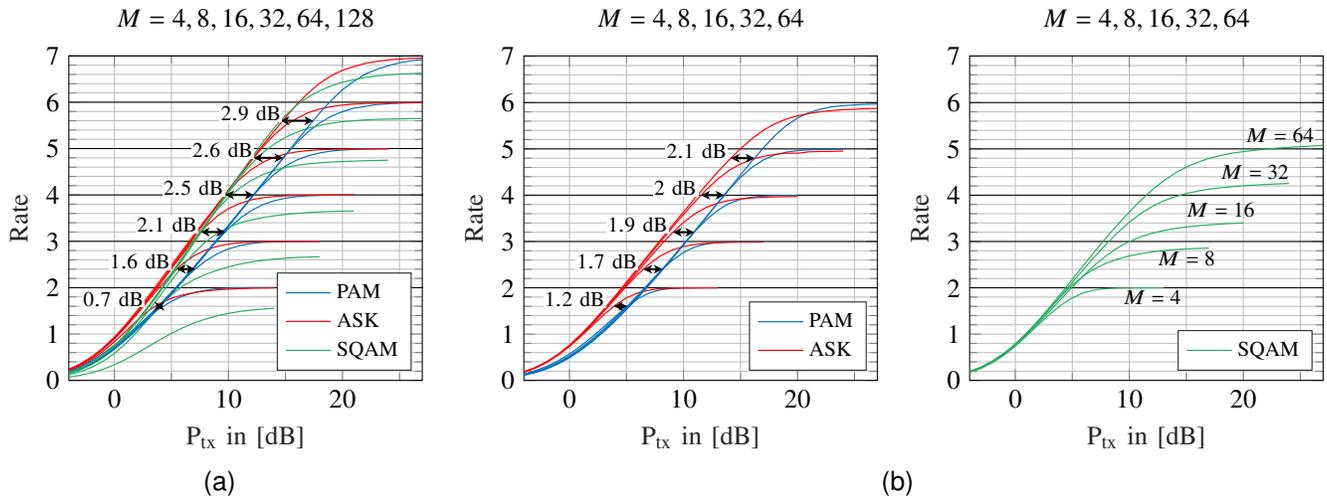

\centering
\centering
\subfloat[\label{fig:Q8-Q128_L=0km}]%
{
\begin{minipage}[t]{0.32\textwidth}
\input{plots/Q8-Q128_L=0km.tex}
\end{minipage}
}%
\subfloat[\label{fig:Q8-Q128_L=30km}]
{
\begin{minipage}[t]{0.32\textwidth}
\input{plots/plot_settings}
\begin{tikzpicture}[]

\pgfdeclarelayer{background}
\pgfdeclarelayer{foreground}
\pgfsetlayers{background,main,foreground}

\begin{axis}[%
yminorticks=true,
xmajorgrids,
ymajorgrids,
yminorgrids,
minor x tick num=1,
minor y tick num=4,
grid=both,
legend style={legend cell align=left,  draw=white!15!black, font=\footnotesize,  %
legend pos=south east
},
xlabel style={font=\color{white!15!black}},
ylabel style={font=\color{white!15!black}},
axis background/.style={fill=white},
scale only axis,
width=1*\gridwidth,
height=1*\gridheight,
xmin=-4,
xmax=27,
xlabel={$\mathrm{P}_{\mathrm{tx}}$ in [dB]},
ymin=0,
ytick={0,1,...,7},
ymax=7.01,
ylabel={Rate},
title={$M=4,8,16,32,64$},
ylabel shift = 0pt
]

\begin{pgfonlayer}{foreground}

\addplot[PAM,name path global=pam4,restrict x to domain=-4:13] table [x=Ptxdb, y=IqXY, col sep=comma]{plots/V1.5/Q4_L30/v1.5,4-PAM,64_128_64_4,lr=5E-04,Tr=120,B=128,S=6,Ls=32,I=2.0E+04,V=1.0E+03,n=6.0E+04,P=151,a=0,L=3.0E+04,Rs=3.5E+10,C=3.1E+04.txt};

\addplot[ASK,restrict x to domain=-4:13,name path global=ask4] table [x=Ptxdb, y=IqXY, col sep=comma,name path global=ask4]{plots/V1.5/Q4_L30/v1.5,4-ASK,64_128_64_4,lr=5E-04,Tr=120,B=128,S=6,Ls=32,I=2.0E+04,V=1.0E+03,n=6.0E+04,P=151,a=0,L=3.0E+04,Rs=3.5E+10,C=3.1E+04.txt};

\addplot[PAM,forget plot,name path global=pam8,restrict x to domain=-4:17] table [x=Ptxdb, y=IqXY, col sep=comma]{plots/V1.5/Q8_L30/v1.5,8-PAM,64_128_128_8,lr=3E-04,Tr=120,B=128,S=6,Ls=32,I=5.0E+04,V=1.0E+03,n=6.0E+04,P=151,a=0,L=3.0E+04,Rs=3.5E+10,C=4.6E+04.txt};

\addplot[ASK,forget plot,name path global=ask8,restrict x to domain=-4:17] table [x=Ptxdb, y=IqXY, col sep=comma]{plots/V1.5/Q8_L30/v1.5,8-ASK,64_128_128_8,lr=3E-04,Tr=120,B=128,S=6,Ls=32,I=5.0E+04,V=1.0E+03,n=6.0E+04,P=151,a=0,L=3.0E+04,Rs=3.5E+10,C=4.6E+04.txt};

\addplot[PAM,forget plot,name path global=pam16,restrict x to domain=-4:20] table [x=Ptxdb, y=IqXY, col sep=comma]{plots/V1.5/Q16_L30/v1.5,16-PAM,84_200_128_128_16,lr=5E-05,Tr=120,B=64,S=6,Ls=64,I=8.0E+04,V=3.0E+03,n=8.0E+04,P=151,a=0,L=3.0E+04,Rs=3.5E+10,C=1.1E+05.txt};

\addplot[ASK,forget plot,name path global=ask16,restrict x to domain=-4:20] table [x=Ptxdb, y=IqXY, col sep=comma]{plots/V1.5/Q16_L30/v1.5,16-ASK,84_200_128_128_16,lr=5E-05,Tr=120,B=64,S=6,Ls=64,I=8.0E+04,V=3.0E+03,n=8.0E+04,P=151,a=0,L=3.0E+04,Rs=3.5E+10,C=1.1E+05.txt};

\addplot[PAM,forget plot,name path global=pam32,restrict x to domain=-4:24] table [x=Ptxdb, y=IqXY, col sep=comma,]{plots/V1.5/Q32_L30/v1.5,32-PAM,100_200_200_200_168_32,lr=4E-05,Tr=120,B=64,S=6,Ls=64,I=1.0E+05,V=7.0E+03,n=8.0E+04,P=151,a=0,L=3.0E+04,Rs=3.5E+10,C=2.3E+05.txt};

\addplot[ASK,forget plot,name path global=ask32,restrict x to domain=-4:24] table [x=Ptxdb, y=IqXY, col sep=comma]{plots/V1.5/Q32_L30/v1.5,32-ASK,100_200_200_200_168_32,lr=4E-05,Tr=120,B=64,S=6,Ls=64,I=1.0E+05,V=7.0E+03,n=8.0E+04,P=151,a=0,L=3.0E+04,Rs=3.5E+10,C=2.3E+05.txt};

\addplot[PAM,forget plot,name path global=pam64,restrict x to domain=-4:27] table [x=Ptxdb, y=IqXY, col sep=comma,]{plots/V1.5/Q64_L30/v1.5,64-PAM,100_300_300_300_240_64,lr=4E-05,Tr=120,B=64,S=6,Ls=100,I=1.0E+05,V=7.0E+03,n=8.0E+04,P=151,a=0,L=3.0E+04,Rs=3.5E+10,C=4.9E+05.txt}; 

\addplot[ASK,forget plot,name path global=ask64,restrict x to domain=-4:27] table [x=Ptxdb, y=IqXY, col sep=comma,]{plots/V1.5/Q64_L30/v1.5,64-ASK,100_300_300_300_240_64,lr=4E-05,Tr=120,B=64,S=6,Ls=100,I=1.0E+05,V=7.0E+03,n=8.0E+04,P=151,a=0,L=3.0E+04,Rs=3.5E+10,C=4.9E+05.txt};

\end{pgfonlayer}

\addlegendentry{PAM}; 
\addlegendentry{ASK}; 

\begin{pgfonlayer}{main}
\foreach \i in {2,3,4,5,6} {
    \edef\temp{
    \noexpand\draw[solid,line width=0.1pt] (axis cs: -4,\i) node[below right,yshift=0.05cm]{} -- (axis cs: 30,\i); %
    }\temp}
\end{pgfonlayer}

\begin{pgfonlayer}{foreground}

\path[name path global=line] (axis cs:\pgfkeysvalueof{/pgfplots/xmin},1.6) -- (axis cs: \pgfkeysvalueof{/pgfplots/xmax},1.6);
\path[name intersections={of=line and ask4, name=p1}, name intersections={of=line and pam4, name=p2}];
\draw[arr] let \p1=(p1-1), \p2=(p2-1) in (p1-1) -- ([xshift=-0.0cm]p2-1) node [left,midway,opacitylabel,font=\footnotesize,yshift=0.1cm,xshift=-0.18cm] {%
	\pgfplotsconvertunittocoordinate{x}{\x1}%
	\pgfplotscoordmath{x}{datascaletrafo inverse to fixed}{\pgfmathresult}%
	\edef\valueA{\pgfmathresult}%
	\pgfplotsconvertunittocoordinate{x}{\x2}%
	\pgfplotscoordmath{x}{datascaletrafo inverse to fixed}{\pgfmathresult}%
	\pgfmathparse{\pgfmathresult - \valueA}%
	\pgfmathprintnumber{\pgfmathresult} dB
};

\path[name path global=line] (axis cs:\pgfkeysvalueof{/pgfplots/xmin},2.4) -- (axis cs: \pgfkeysvalueof{/pgfplots/xmax},2.4);
\path[name intersections={of=line and ask8, name=p1}, name intersections={of=line and pam8, name=p2}];
\draw[arr] let \p1=(p1-1), \p2=(p2-1) in (p1-1) -- ([xshift=-0.0cm]p2-1) node [left,midway,opacitylabel,font=\footnotesize,yshift=0.1cm,xshift=-0.18cm] {%
	\pgfplotsconvertunittocoordinate{x}{\x1}%
	\pgfplotscoordmath{x}{datascaletrafo inverse to fixed}{\pgfmathresult}%
	\edef\valueA{\pgfmathresult}%
	\pgfplotsconvertunittocoordinate{x}{\x2}%
	\pgfplotscoordmath{x}{datascaletrafo inverse to fixed}{\pgfmathresult}%
	\pgfmathparse{\pgfmathresult - \valueA}%
	\pgfmathprintnumber{\pgfmathresult} dB
};

\path[name path global=line] (axis cs:\pgfkeysvalueof{/pgfplots/xmin},3.2) -- (axis cs: \pgfkeysvalueof{/pgfplots/xmax},3.2);
\path[name intersections={of=line and ask16, name=p1}, name intersections={of=line and pam16, name=p2}];
\draw[arr] let \p1=(p1-1), \p2=(p2-1) in (p1-1) -- ([xshift=-0.0cm]p2-1) node [left,midway,opacitylabel,font=\footnotesize,yshift=0.1cm,xshift=-0.18cm] {%
	\pgfplotsconvertunittocoordinate{x}{\x1}%
	\pgfplotscoordmath{x}{datascaletrafo inverse to fixed}{\pgfmathresult}%
	\edef\valueA{\pgfmathresult}%
	\pgfplotsconvertunittocoordinate{x}{\x2}%
	\pgfplotscoordmath{x}{datascaletrafo inverse to fixed}{\pgfmathresult}%
	\pgfmathparse{\pgfmathresult - \valueA}%
	\pgfmathprintnumber{\pgfmathresult} dB
};

\path[name path global=line] (axis cs:\pgfkeysvalueof{/pgfplots/xmin},4) -- (axis cs: \pgfkeysvalueof{/pgfplots/xmax},4);
\path[name intersections={of=line and ask32, name=p1}, name intersections={of=line and pam32, name=p2}];
\draw[arr] let \p1=(p1-1), \p2=(p2-1) in (p1-1) -- ([xshift=-0.0cm]p2-1) node [left,midway,opacitylabel,font=\footnotesize,yshift=0.1cm,xshift=-0.18cm] {%
	\pgfplotsconvertunittocoordinate{x}{\x1}%
	\pgfplotscoordmath{x}{datascaletrafo inverse to fixed}{\pgfmathresult}%
	\edef\valueA{\pgfmathresult}%
	\pgfplotsconvertunittocoordinate{x}{\x2}%
	\pgfplotscoordmath{x}{datascaletrafo inverse to fixed}{\pgfmathresult}%
	\pgfmathparse{\pgfmathresult - \valueA}%
	\pgfmathprintnumber{\pgfmathresult} dB
};

\path[name path global=line] (axis cs:\pgfkeysvalueof{/pgfplots/xmin},4.8) -- (axis cs: \pgfkeysvalueof{/pgfplots/xmax},4.8);
\path[name intersections={of=line and ask64, name=p1}, name intersections={of=line and pam64, name=p2}];
\draw[arr] let \p1=(p1-1), \p2=(p2-1) in (p1-1) -- ([xshift=-0.0cm]p2-1) node [left,midway,opacitylabel,font=\footnotesize,yshift=0.1cm,xshift=-0.18cm] {%
	\pgfplotsconvertunittocoordinate{x}{\x1}%
	\pgfplotscoordmath{x}{datascaletrafo inverse to fixed}{\pgfmathresult}%
	\edef\valueA{\pgfmathresult}%
	\pgfplotsconvertunittocoordinate{x}{\x2}%
	\pgfplotscoordmath{x}{datascaletrafo inverse to fixed}{\pgfmathresult}%
	\pgfmathparse{\pgfmathresult - \valueA}%
	\pgfmathprintnumber{\pgfmathresult} dB
};

\end{pgfonlayer}

\end{axis}
\end{tikzpicture}%
\end{minipage}
\begin{minipage}[t]{0.32\textwidth}
\input{plots/plot_settings}
\begin{tikzpicture}[]
\pgfdeclarelayer{background}
\pgfdeclarelayer{foreground}
\pgfsetlayers{background,main,foreground}
\begin{axis}[%
yminorticks=true,
xmajorgrids,
ymajorgrids,
yminorgrids,
minor x tick num=1,
minor y tick num=4,
grid=both,
legend style={legend cell align=left,  draw=white!15!black, font=\footnotesize,  %
legend pos=south east
},
xlabel style={font=\color{white!15!black}},
ylabel style={font=\color{white!15!black}},
axis background/.style={fill=white},
scale only axis,
width=1*\gridwidth,
height=1*\gridheight,
xmin=-4,
xmax=27,
xlabel={$\mathrm{P}_{\mathrm{tx}}$ in [dB]},
ymin=0,
ytick={0,1,...,7},
ymax=7.01,
ylabel={Rate},
title={$M=4, 8,16,32,64$},
ylabel shift = 0pt
]

\begin{pgfonlayer}{foreground}

\addplot[QAM,name path global=sqam4,restrict x to domain=-4:13] table [x=Ptxdb, y=IqXY, col sep=comma]{plots/V1.5/Q4_L30/v1.5,4-SQAM,64_256_128_128_4,lr=5E-04,Tr=120,B=128,S=6,Ls=32,I=2.0E+04,V=1.0E+03,n=6.0E+04,P=151,a=0,L=3.0E+04,Rs=3.5E+10,C=1.3E+05.txt};

\addplot[QAM,forget plot,name path global=sqam8,restrict x to domain=-4:17] table [x=Ptxdb, y=IqXY, col sep=comma]{plots/V1.5/Q8_L30/v1.5,8-SQAM,64_256_128_128_8,lr=3E-04,Tr=120,B=128,S=6,Ls=32,I=5.0E+04,V=1.0E+03,n=6.0E+04,P=151,a=0,L=3.0E+04,Rs=3.5E+10,C=1.3E+05.txt};

\addplot[QAM,forget plot,name path global=sqam16,restrict x to domain=-4:20] table [x=Ptxdb, y=IqXY, col sep=comma]{plots/V1.5/Q16_L30/v1.5,16-SQAM,84_200_128_128_16,lr=5E-05,Tr=120,B=64,S=6,Ls=64,I=8.0E+04,V=3.0E+03,n=8.0E+04,P=151,a=0,L=3.0E+04,Rs=3.5E+10,C=1.2E+05.txt};

\addplot[QAM,forget plot,name path global=sqam32,restrict x to domain=-4:24] table [x=Ptxdb, y=IqXY, col sep=comma]{plots/V1.5/Q32_L30/v1.5,32-SQAM,100_200_200_200_168_32,lr=4E-05,Tr=120,B=64,S=6,Ls=64,I=1.0E+05,V=7.0E+03,n=8.0E+04,P=151,a=0,L=3.0E+04,Rs=3.5E+10,C=2.4E+05.txt};

\addplot[QAM,forget plot,name path global=sqam64,restrict x to domain=-4:27] table [x=Ptxdb, y=IqXY, col sep=comma,]{plots/V1.5/Q64_L30/v1.5,64-SQAM,100_300_300_300_240_64,lr=4E-05,Tr=120,B=64,S=6,Ls=100,I=1.0E+05,V=7.0E+03,n=8.0E+04,P=151,a=0,L=3.0E+04,Rs=3.5E+10,C=5.2E+05.txt};

\end{pgfonlayer}

\addlegendentry{SQAM}; 

\begin{pgfonlayer}{main}
\foreach \i in {2,3,4,5,6} {
    \edef\temp{
    \noexpand\draw[solid,line width=0.1pt] (axis cs: -4,\i) node[below right,yshift=0.05cm]{} -- (axis cs: 30,\i); %
    }\temp}
\end{pgfonlayer}

\node[font=\footnotesize] at (axis cs: 12,1.8){$M=4$}; 
\node[font=\footnotesize] at (axis cs: 15,2.63){$M=8$}; 
\node[font=\footnotesize] at (axis cs: 18,3.7){$M=16$}; 
\node[font=\footnotesize] at (axis cs: 21,4.5){$M=32$}; 
\node[font=\footnotesize] at (axis cs: 23.2,5.3){$M=64$};

\end{axis}
\end{tikzpicture}%
\end{minipage}
}%
\caption{SIC rates for (a) $L_\text{fib}=0$ and $S=2$ and (b) $L_\text{fib}=\SI{30}{\kilo\meter}$ and $S=6$.}
\label{fig:Q8-Q128}
\end{figure*}

\subsection{Periodically Time-Varying RNNs}
Fig.~\ref{fig:Q8_L30km_ASK_TVRNN_RNN} and~\ref{fig:Q8_L30km_SQAM_TVRNN_RNN} compare the rates of the periodically time-varying RNNs to classic RNNs that use time-invariant recurrent cells $(\Cfw{}{i},\Cbw{}{i})_{i=1}^{L-1}$~\cite[Chap.~10]{goodfellow2016deep}. The latter RNNs use the same configuration as their time-varying counterparts; see Tab.~\ref{tab:nnparams}. Thus, both structures have the same computational complexity. To illustrate the results, we consider 8-ASK and 8-SQAM for $L_\text{fib}=\SI{30}{\kilo\meter}$. 

Fig.~\ref{fig:Q8_L30km_ASK_TVRNN_RNN_AVG} compares the SIC rates for $S=1,2,6$. For SDD ($S=1$) and $S=2$, the proposed RNNs per SIC stage are the same as classic RNNs~\cite[Chap.~10.3]{goodfellow2016deep}. SIC with $S=2$ stages gains $\approx \SI{1.2}{dB}$ over SDD for medium to high SNRs. Using $S=6$ stages almost doubles the energy gains to $\approx\SI{2.2}{dB}$ and $\approx\SI{2.4}{dB}$ at $\SI{2.6}{bpcu}$ and $\SI{2.8}{bpcu}$, respectively. In contrast, operating classic RNNs with $S=6$ is suboptimal and the gains reduce to $\SI{1.6}{dB}$ and $\SI{1.4}{dB}$ over SDD at $\SI{2.6}{bpcu}$ and $\SI{2.8}{bpcu}$ respectively. 

Fig.~\ref{fig:Q8_L30km_ASK_TVRNN_RNN_STAGE} compares the individual stage rates for $S=6$. For $s=1$ and $s=S$, our RNNs are the same as classic RNNs because the time-varying period is $\Gamma = 1$; cf. Sec.~\ref{sec:app_nn}. Therefore, the rates in these stages are equal.
For time-varying RNNs, the rates $I_{q,N\!,\text{SIC}}^s$ are non-decreasing in $s$. This is in contrast to classic RNNs, where for $P_\text{tx} \geq \SI{8}{dB}$ the stage $s=2$ rate is smaller than the stage $s=1$ rate. In other words, classic RNNs are suboptimal because they cannot process cyclostationary inputs correctly and cannot account for the decoded a-priori information from previous SIC stages. The periodically time-varying RNNs are tailored to cyclostationary processing and thus achieve significant energy gains in stages $s=2,\ldots,5$ compared to classical RNNs. 

Fig.~\ref{fig:Q8_L30km_SQAM_TVRNN_RNN} shows similar results for 8-SQAM, where periodically time-varying RNNs gain \SI{1.1}{dB} over classic RNNs for $S=6$ and rates around \SI{2.5}{bpcu}. More SIC stages may be required to approach JDD performance for higher modulation orders, channels with more memory, or stronger ISI. Periodically time-varying RNNs should be even more advantageous for such cases.
\begin{figure*}[!t]
\centering
\begin{minipage}[t]{0.49\textwidth}
\subfloat[\label{fig:Q8_L30km_ASK_TVRNN_RNN_AVG}]{%
\hspace*{-25pt}
\input{plots/plot_settings}

\pgfdeclarelayer{fg}    %
\pgfsetlayers{main,fg}  %

\begin{tikzpicture}[spy using outlines={magnification=1.8, connect spies}]

\begin{axis}[%
yminorticks=true,
xmajorgrids,
ymajorgrids,
yminorgrids,
minor y tick num=4,
minor x tick num=4,
minor y tick num=4,
xtick distance=2,
grid=both,
legend style={legend cell align=left,  draw=white!15!black, font=\footnotesize,  
legend pos=south east
},
xlabel style={font=\color{white!15!black}},
ylabel style={font=\color{white!15!black}},
axis background/.style={fill=white},
scale only axis,
width=1.5*\gridwidth,
height=\gridheight,
xmin=-5,
xmax=17,
xlabel={$\mathrm{P}_{\mathrm{tx}}$ in [dB]},
ymin=0,
ymax=3.001,
ylabel={Rate},
title={8-ASK},
ylabel shift = -4pt
]

\addplot[ASK,solid, forget plot, black,postaction={decorate,decoration={raise=-1.5ex,text along path,text color={black},text align={left, left indent=4.0cm},text={|\footnotesize|SDD}}}] table [x=Ptxdb, y=SIC1, col sep=comma]{plots/V1.5/Q8_L30/v1.5,8-ASK,64_128_128_8,lr=3E-04,Tr=120,B=128,S=6,Ls=32,I=5.0E+04,V=1.0E+03,n=6.0E+04,P=151,a=0,L=3.0E+04,Rs=3.5E+10,C=4.6E+04.txt};

\addplot[ASK,densely dash dot,orange,line width=0.6pt,forget plot,postaction={decorate,decoration={raise=-1.4ex,text along path,text color={orange},text align={left, left indent=3.8cm},text={|\footnotesize|{$S=2$}}}}] table [x=Ptxdb, y=IqXY, col sep=comma]{plots/V1.5/Q8_L30/v1.5,8-ASK,64_128_128_8,lr=3E-04,Tr=84,B=128,S=2,Ls=32,I=5.0E+04,V=1.0E+03,n=6.0E+04,P=151,a=0,L=3.0E+04,Rs=3.5E+10,C=4.6E+04.txt};

\addplot[ASK,line width=0.7pt] table [x=Ptxdb, y=IqXY, col sep=comma]{plots/V1.5/Q8_L30/v1.5,8-ASK,64_128_128_8,lr=3E-04,Tr=120,B=128,S=6,Ls=32,I=5.0E+04,V=1.0E+03,n=6.0E+04,P=151,a=0,L=3.0E+04,Rs=3.5E+10,C=4.6E+04.txt};
\addlegendentry{Time-varying RNNs $S=6$};

\addplot[ASK,black,densely dashed] table [x=Ptxdb, y=IqXY, col sep=comma]{plots/V1.5/Q8_L30/notv_v1.5,8-ASK,64_128_128_8,lr=3E-04,Tr=120,B=128,S=6,Ls=32,I=5.0E+04,V=1.0E+03,n=6.0E+04,P=151,a=0,L=3.0E+04,Rs=3.5E+10,C=4.6E+04.txt};
\addlegendentry{Classic RNNs $S=6$};

\coordinate (spypoint) at (axis cs:9.65,2.72);
\coordinate (magnifyglass) at (axis cs:-0.2,2.45);

\draw[arrmagn,shadowedb,shadowed,gray!40!black] (axis cs: 7.45,2.6) -- (9.6,2.6); %
\draw[arrmagn,shadowedb,shadowed,gray!40!black] (axis cs: 8.3,2.65) -- (9.9,2.65);%

\draw[arrmagn,shadowedb,shadowed,gray!40!black] (axis cs: 8.9,2.8) -- (11.3,2.8); %
\draw[arrmagn,shadowedb,shadowed,gray!40!black] (axis cs: 10.5,2.85) -- (11.90,2.85); %

\end{axis}

\spy [blue,size=1.4cm, width=3cm,] on (spypoint) in node[fill=white] at (magnifyglass);
 \begin{pgfonlayer}{fg}    %
     \node[opacitylabel,font=\footnotesize]  at (2.3,3.42) {$\SI{1.6}{dB}$};
     \node[opacitylabel,font=\footnotesize]  at (2.1,3.16) {$\SI{2.2}{dB}$};

      \node[opacitylabel,font=\footnotesize]  at (1.4,4.1) {$\SI{1.4}{dB}$};
      \node[opacitylabel,font=\footnotesize]  at (0.7,3.8) {$\SI{2.4}{dB}$};
  \end{pgfonlayer}
  
\end{tikzpicture}%
}
\end{minipage}
\hfill
\begin{minipage}[t]{0.49\textwidth}
\subfloat[\label{fig:Q8_L30km_ASK_TVRNN_RNN_STAGE}]{%
\hspace{-20pt}
\pgfdeclarelayer{background}
\pgfdeclarelayer{foreground}
\pgfsetlayers{background,main,foreground}
\input{plots/plot_settings}

\begin{tikzpicture}[]

\begin{axis}[%
yminorticks=true,
xmajorgrids,
ymajorgrids,
yminorgrids,
minor y tick num=4,
minor x tick num=4,
minor y tick num=4,
xtick distance=2,
grid=both,
legend style={legend cell align=left,  draw=white!15!black, font=\footnotesize,  
legend pos=south east
},
xlabel style={font=\color{white!15!black}},
ylabel style={font=\color{white!15!black}},
axis background/.style={fill=white},
scale only axis,
width=1.5*\gridwidth,
height=\gridheight,
xmin=-5,
xmax=17,
xlabel={$\mathrm{P}_{\mathrm{tx}}$ in [dB]},
ymin=0,
ymax=3.001,
ylabel={Rate},
title={8-ASK},
ylabel shift = -4pt
]

\addplot[ASK,line width=0.7pt] table [x=Ptxdb, y=SIC1, col sep=comma]{plots/V1.5/Q8_L30/v1.5,8-ASK,64_128_128_8,lr=3E-04,Tr=120,B=128,S=6,Ls=32,I=5.0E+04,V=1.0E+03,n=6.0E+04,P=151,a=0,L=3.0E+04,Rs=3.5E+10,C=4.6E+04.txt};
\addlegendentry{Time-varying RNNs $S=6$};

\addplot[ASK,forget plot,line width=0.7pt] table [x=Ptxdb, y=SIC2, col sep=comma]{plots/V1.5/Q8_L30/v1.5,8-ASK,64_128_128_8,lr=3E-04,Tr=120,B=128,S=6,Ls=32,I=5.0E+04,V=1.0E+03,n=6.0E+04,P=151,a=0,L=3.0E+04,Rs=3.5E+10,C=4.6E+04.txt};

\addplot[ASK,forget plot,line width=0.7pt] table [x=Ptxdb, y=SIC3, col sep=comma]{plots/V1.5/Q8_L30/v1.5,8-ASK,64_128_128_8,lr=3E-04,Tr=120,B=128,S=6,Ls=32,I=5.0E+04,V=1.0E+03,n=6.0E+04,P=151,a=0,L=3.0E+04,Rs=3.5E+10,C=4.6E+04.txt};

\addplot[ASK,forget plot,line width=0.7pt] table [x=Ptxdb, y=SIC4, col sep=comma]{plots/V1.5/Q8_L30/v1.5,8-ASK,64_128_128_8,lr=3E-04,Tr=120,B=128,S=6,Ls=32,I=5.0E+04,V=1.0E+03,n=6.0E+04,P=151,a=0,L=3.0E+04,Rs=3.5E+10,C=4.6E+04.txt};

\addplot[ASK,forget plot,line width=0.7pt] table [x=Ptxdb, y=SIC5, col sep=comma]{plots/V1.5/Q8_L30/v1.5,8-ASK,64_128_128_8,lr=3E-04,Tr=120,B=128,S=6,Ls=32,I=5.0E+04,V=1.0E+03,n=6.0E+04,P=151,a=0,L=3.0E+04,Rs=3.5E+10,C=4.6E+04.txt};

\addplot[ASK,forget plot,line width=0.7pt] table [x=Ptxdb, y=SIC6, col sep=comma]{plots/V1.5/Q8_L30/v1.5,8-ASK,64_128_128_8,lr=3E-04,Tr=120,B=128,S=6,Ls=32,I=5.0E+04,V=1.0E+03,n=6.0E+04,P=151,a=0,L=3.0E+04,Rs=3.5E+10,C=4.6E+04.txt};

\addplot[ASK,black,densely dashed,line width=0.8pt] table [x=Ptxdb, y=SIC1, col sep=comma]{plots/V1.5/Q8_L30/notv_v1.5,8-ASK,64_128_128_8,lr=3E-04,Tr=120,B=128,S=6,Ls=32,I=5.0E+04,V=1.0E+03,n=6.0E+04,P=151,a=0,L=3.0E+04,Rs=3.5E+10,C=4.6E+04.txt};

\addplot[ASK,forget plot,black,densely dashed,line width=0.8pt] table [x=Ptxdb, y=SIC2, col sep=comma]{plots/V1.5/Q8_L30/notv_v1.5,8-ASK,64_128_128_8,lr=3E-04,Tr=120,B=128,S=6,Ls=32,I=5.0E+04,V=1.0E+03,n=6.0E+04,P=151,a=0,L=3.0E+04,Rs=3.5E+10,C=4.6E+04.txt};

\addplot[ASK,forget plot,black,densely dashed,line width=0.8pt] table [x=Ptxdb, y=SIC3, col sep=comma]{plots/V1.5/Q8_L30/notv_v1.5,8-ASK,64_128_128_8,lr=3E-04,Tr=120,B=128,S=6,Ls=32,I=5.0E+04,V=1.0E+03,n=6.0E+04,P=151,a=0,L=3.0E+04,Rs=3.5E+10,C=4.6E+04.txt};

\addplot[ASK,forget plot,black,densely dashed,line width=0.8pt] table [x=Ptxdb, y=SIC4, col sep=comma]{plots/V1.5/Q8_L30/notv_v1.5,8-ASK,64_128_128_8,lr=3E-04,Tr=120,B=128,S=6,Ls=32,I=5.0E+04,V=1.0E+03,n=6.0E+04,P=151,a=0,L=3.0E+04,Rs=3.5E+10,C=4.6E+04.txt};

\addplot[ASK,forget plot,black,densely dashed,line width=0.8pt] table [x=Ptxdb, y=SIC5, col sep=comma]{plots/V1.5/Q8_L30/notv_v1.5,8-ASK,64_128_128_8,lr=3E-04,Tr=120,B=128,S=6,Ls=32,I=5.0E+04,V=1.0E+03,n=6.0E+04,P=151,a=0,L=3.0E+04,Rs=3.5E+10,C=4.6E+04.txt};

\addplot[ASK,forget plot,black,densely dashed,line width=0.8pt] table [x=Ptxdb, y=SIC6, col sep=comma]{plots/V1.5/Q8_L30/notv_v1.5,8-ASK,64_128_128_8,lr=3E-04,Tr=120,B=128,S=6,Ls=32,I=5.0E+04,V=1.0E+03,n=6.0E+04,P=151,a=0,L=3.0E+04,Rs=3.5E+10,C=4.6E+04.txt};
\addlegendentry{Classic RNNs $S=6$};

\draw[arrR] (axis cs:7,1.0)node[right,opacitylabel]{$s=1,2,\ldots,6$} -- (axis cs: 1,1.9); 
\end{axis}

\end{tikzpicture}%
}
\end{minipage}
\caption{Time-varying RNNs vs. classic RNNs for 8-ASK and $L_\text{fib}=\SI{30}{\kilo\meter}$: (a) SIC rates for $S=1,2,6$; (b) individual stage rates $I_{q,N\!,\text{SIC}}^s$ for $S=6$.}
\label{fig:Q8_L30km_ASK_TVRNN_RNN}
\end{figure*}
\begin{figure*}[!t]
\centering
\begin{minipage}[t]{0.49\textwidth}
\subfloat[\label{fig:Q8_L30km_SQAM_TVRNN_RNN_AVG}]{%
\hspace*{-25pt}
\input{plots/plot_settings}

\pgfdeclarelayer{background}
\pgfdeclarelayer{foreground}
\pgfsetlayers{background,main,foreground}

\begin{tikzpicture}[spy using outlines={magnification=1.8, connect spies}]

\begin{axis}[%
yminorticks=true,
xmajorgrids,
ymajorgrids,
yminorgrids,
minor y tick num=4,
minor x tick num=4,
minor y tick num=4,
xtick distance=2,
grid=both,
legend style={legend cell align=left,  draw=white!15!black, font=\footnotesize,  
legend pos=south east
},
xlabel style={font=\color{white!15!black}},
ylabel style={font=\color{white!15!black}},
axis background/.style={fill=white},
scale only axis,
width=1.5*\gridwidth,
height=\gridheight,
xmin=-5,
xmax=17,
xlabel={$\mathrm{P}_{\mathrm{tx}}$ in [dB]},
ymin=0,
ymax=3.001,
ylabel={Rate},
title={8-SQAM},
ylabel shift = -4pt
]

\addplot[QAM,solid, forget plot, black,postaction={decorate,decoration={raise=-1.5ex,text along path,text color={black},text align={left, left indent=4.0cm},text={|\footnotesize|SDD}}}] table [x=Ptxdb, y=SIC1, col sep=comma]{plots/V1.5/Q8_L30/v1.5,8-SQAM,64_256_128_128_8,lr=3E-04,Tr=84,B=128,S=2,Ls=32,I=5.0E+04,V=1.0E+03,n=6.0E+04,P=151,a=0,L=3.0E+04,Rs=3.5E+10,C=1.3E+05.txt};

\addplot[QAM,densely dash dot,orange,line width=0.6pt,forget plot,postaction={decorate,decoration={raise=-1.4ex,text along path,text color={orange},text align={left, left indent=3.8cm},text={|\footnotesize|{$S=2$}}}}] table [x=Ptxdb, y=IqXY, col sep=comma]{plots/V1.5/Q8_L30/v1.5,8-SQAM,64_256_128_128_8,lr=3E-04,Tr=84,B=128,S=2,Ls=32,I=5.0E+04,V=1.0E+03,n=6.0E+04,P=151,a=0,L=3.0E+04,Rs=3.5E+10,C=1.3E+05.txt};

\addplot[QAM,line width=0.7pt,name path global=TVS6] table [x=Ptxdb, y=IqXY, col sep=comma]{plots/V1.5/Q8_L30/v1.5,8-SQAM,64_256_128_128_8,lr=3E-04,Tr=120,B=128,S=6,Ls=32,I=5.0E+04,V=1.0E+03,n=6.0E+04,P=151,a=0,L=3.0E+04,Rs=3.5E+10,C=1.3E+05.txt};
\addlegendentry{Time-varying RNNs $S=6$};

\addplot[QAM,black,densely dashed,name path global=NOTVS6] table [x=Ptxdb, y=IqXY, col sep=comma]{plots/V1.5/Q8_L30/notv_v1.5,8-SQAM,64_256_128_128_8,lr=3E-04,Tr=120,B=128,S=6,Ls=32,I=5.0E+04,V=1.0E+03,n=6.0E+04,P=151,a=0,L=3.0E+04,Rs=3.5E+10,C=1.3E+05.txt};
\addlegendentry{Classic RNNs $S=6$};

\coordinate (spypoint) at (axis cs:9.65,2.7);
\coordinate (magnifyglass) at (axis cs:-0.2,2.45);

\begin{pgfonlayer}{foreground}
\path[name path global=line] (axis cs:\pgfkeysvalueof{/pgfplots/xmin},2.5) -- (axis cs: \pgfkeysvalueof{/pgfplots/xmax},2.5);
\path[name intersections={of=line and TVS6, name=p1}, name intersections={of=line and NOTVS6, name=p2}];
\draw[arr] let \p1=(p1-1), \p2=(p2-1) in (p1-1) -- ([xshift=-0.0cm]p2-1) node [left,midway,opacitylabel,font=\footnotesize,yshift=0.0cm,xshift=-0.3cm] {%
	\pgfplotsconvertunittocoordinate{x}{\x1}%
	\pgfplotscoordmath{x}{datascaletrafo inverse to fixed}{\pgfmathresult}%
	\edef\valueA{\pgfmathresult}%
	\pgfplotsconvertunittocoordinate{x}{\x2}%
	\pgfplotscoordmath{x}{datascaletrafo inverse to fixed}{\pgfmathresult}%
	\pgfmathparse{\pgfmathresult - \valueA}%
	\pgfmathprintnumber{\pgfmathresult} dB
};
\end{pgfonlayer}

\end{axis}

\end{tikzpicture}%
}
\end{minipage}
\hfill
\begin{minipage}[t]{0.49\textwidth}
\subfloat[\label{fig:Q8_L30km_SQAM_TVRNN_RNN_STAGE}]{%
\hspace{-20pt}
\pgfdeclarelayer{background}
\pgfdeclarelayer{foreground}
\pgfsetlayers{background,main,foreground}
\input{plots/plot_settings}

\begin{tikzpicture}[]

\begin{axis}[%
yminorticks=true,
xmajorgrids,
ymajorgrids,
yminorgrids,
minor y tick num=4,
minor x tick num=4,
minor y tick num=4,
xtick distance=2,
grid=both,
legend style={legend cell align=left,  draw=white!15!black, font=\footnotesize,  
legend pos=south east
},
xlabel style={font=\color{white!15!black}},
ylabel style={font=\color{white!15!black}},
axis background/.style={fill=white},
scale only axis,
width=1.5*\gridwidth,
height=\gridheight,
xmin=-5,
xmax=17,
xlabel={$\mathrm{P}_{\mathrm{tx}}$ in [dB]},
ymin=0,
ymax=3.001,
ylabel={Rate},
title={8-SQAM},
ylabel shift = -4pt
]

\addplot[QAM,line width=0.7pt] table [x=Ptxdb, y=SIC1, col sep=comma]{plots/V1.5/Q8_L30/v1.5,8-SQAM,64_256_128_128_8,lr=3E-04,Tr=120,B=128,S=6,Ls=32,I=5.0E+04,V=1.0E+03,n=6.0E+04,P=151,a=0,L=3.0E+04,Rs=3.5E+10,C=1.3E+05.txt};
\addlegendentry{Time-varying RNNs $S=6$};

\addplot[QAM,forget plot,line width=0.7pt] table [x=Ptxdb, y=SIC2, col sep=comma]{plots/V1.5/Q8_L30/v1.5,8-SQAM,64_256_128_128_8,lr=3E-04,Tr=120,B=128,S=6,Ls=32,I=5.0E+04,V=1.0E+03,n=6.0E+04,P=151,a=0,L=3.0E+04,Rs=3.5E+10,C=1.3E+05.txt};

\addplot[QAM,forget plot,line width=0.7pt] table [x=Ptxdb, y=SIC3, col sep=comma]{plots/V1.5/Q8_L30/v1.5,8-SQAM,64_256_128_128_8,lr=3E-04,Tr=120,B=128,S=6,Ls=32,I=5.0E+04,V=1.0E+03,n=6.0E+04,P=151,a=0,L=3.0E+04,Rs=3.5E+10,C=1.3E+05.txt};

\addplot[QAM,forget plot,line width=0.7pt] table [x=Ptxdb, y=SIC4, col sep=comma]{plots/V1.5/Q8_L30/v1.5,8-SQAM,64_256_128_128_8,lr=3E-04,Tr=120,B=128,S=6,Ls=32,I=5.0E+04,V=1.0E+03,n=6.0E+04,P=151,a=0,L=3.0E+04,Rs=3.5E+10,C=1.3E+05.txt};

\addplot[QAM,forget plot,line width=0.7pt] table [x=Ptxdb, y=SIC5, col sep=comma]{plots/V1.5/Q8_L30/v1.5,8-SQAM,64_256_128_128_8,lr=3E-04,Tr=120,B=128,S=6,Ls=32,I=5.0E+04,V=1.0E+03,n=6.0E+04,P=151,a=0,L=3.0E+04,Rs=3.5E+10,C=1.3E+05.txt};

\addplot[QAM,forget plot,line width=0.7pt] table [x=Ptxdb, y=SIC6, col sep=comma]{plots/V1.5/Q8_L30/v1.5,8-SQAM,64_256_128_128_8,lr=3E-04,Tr=120,B=128,S=6,Ls=32,I=5.0E+04,V=1.0E+03,n=6.0E+04,P=151,a=0,L=3.0E+04,Rs=3.5E+10,C=1.3E+05.txt};

\addplot[QAM,black,densely dashed,line width=0.8pt] table [x=Ptxdb, y=SIC1, col sep=comma]{plots/V1.5/Q8_L30/notv_v1.5,8-SQAM,64_256_128_128_8,lr=3E-04,Tr=120,B=128,S=6,Ls=32,I=5.0E+04,V=1.0E+03,n=6.0E+04,P=151,a=0,L=3.0E+04,Rs=3.5E+10,C=1.3E+05.txt};

\addplot[QAM,forget plot,black,densely dashed,line width=0.8pt] table [x=Ptxdb, y=SIC2, col sep=comma]{plots/V1.5/Q8_L30/notv_v1.5,8-SQAM,64_256_128_128_8,lr=3E-04,Tr=120,B=128,S=6,Ls=32,I=5.0E+04,V=1.0E+03,n=6.0E+04,P=151,a=0,L=3.0E+04,Rs=3.5E+10,C=1.3E+05.txt};

\addplot[QAM,forget plot,black,densely dashed,line width=0.8pt] table [x=Ptxdb, y=SIC3, col sep=comma]{plots/V1.5/Q8_L30/notv_v1.5,8-SQAM,64_256_128_128_8,lr=3E-04,Tr=120,B=128,S=6,Ls=32,I=5.0E+04,V=1.0E+03,n=6.0E+04,P=151,a=0,L=3.0E+04,Rs=3.5E+10,C=1.3E+05.txt};

\addplot[QAM,forget plot,black,densely dashed,line width=0.8pt] table [x=Ptxdb, y=SIC4, col sep=comma]{plots/V1.5/Q8_L30/notv_v1.5,8-SQAM,64_256_128_128_8,lr=3E-04,Tr=120,B=128,S=6,Ls=32,I=5.0E+04,V=1.0E+03,n=6.0E+04,P=151,a=0,L=3.0E+04,Rs=3.5E+10,C=1.3E+05.txt};

\addplot[QAM,forget plot,black,densely dashed,line width=0.8pt] table [x=Ptxdb, y=SIC5, col sep=comma]{plots/V1.5/Q8_L30/notv_v1.5,8-SQAM,64_256_128_128_8,lr=3E-04,Tr=120,B=128,S=6,Ls=32,I=5.0E+04,V=1.0E+03,n=6.0E+04,P=151,a=0,L=3.0E+04,Rs=3.5E+10,C=1.3E+05.txt};

\addplot[QAM,forget plot,black,densely dashed,line width=0.8pt] table [x=Ptxdb, y=SIC6, col sep=comma]{plots/V1.5/Q8_L30/notv_v1.5,8-SQAM,64_256_128_128_8,lr=3E-04,Tr=120,B=128,S=6,Ls=32,I=5.0E+04,V=1.0E+03,n=6.0E+04,P=151,a=0,L=3.0E+04,Rs=3.5E+10,C=1.3E+05.txt};
\addlegendentry{Classic RNNs $S=6$};

\draw[arrR] (axis cs:7,1.0)node[right,opacitylabel]{$s=1,2,\ldots,6$} -- (axis cs: 1,1.9); 
\end{axis}

\end{tikzpicture}%
}
\end{minipage}
\caption{Time-varying RNNs vs. classic RNNs for 8-SQAM and $L_\text{fib}=\SI{30}{\kilo\meter}$: (a) SIC rates for $S=1,2,6$; (b) individual stage rates $I_{q,N\!,\text{SIC}}^s$ for $S=6$.}
\label{fig:Q8_L30km_SQAM_TVRNN_RNN}
\end{figure*}
\section{Conclusion}
\label{sec:conclusion}
We designed time-varying RNN equalizers for SIC and showed that they outperform FBA-SIC and GS-SIC with substantially less complexity. We simulated NN-SIC rates up to \SI{7}{bpcu} with 128-PAM/ASK/SQAM for short-reach fiber-optic links with a SLD. Moreover, bipolar ASK and complex modulations gain up to $\SI{3}{dB}$ over classic unipolar PAM.

For future work, we plan to verify the computed rates with realistic coded modulations as in~\cite{prinz2023successive}; we also plan to include probabilistic shaping. We further plan to investigate the robustness and generalization capabilities of NN-SIC with data from hardware experiments; see~\cite{karanov2018end}. The NN-SIC receiver may be a candidate for optical short-range links with direct detection if high-speed optical modulators~\cite{moor2024plasmonic} are available.
One may further explore NN-compression methods~\cite{oneil2020compression} to reduce the number of NN parameters; we expect that the NN sizes for the higher SIC stages can be reduced because much of the interference is known. Finally, one can mitigate phase symmetries by precoding; see~\cite[Appendix]{prinz2023successive}.

\section*{Acknowledgment}
\noindent The authors wish to thank M. Schädler, S. Calabr\`{o}, D. Lentner and the reviewers for helpful discussions.

\bibliographystyle{IEEEtran}
\bibliography{IEEEabrv,nn_dd}

\end{document}